%% file: nci_survey.tex
\documentclass[10pt,journal,compsoc]{IEEEtran}

\usepackage{epsfig}
\usepackage{graphicx}
\usepackage{amsmath}
\usepackage{amssymb}
\usepackage{multirow}
\usepackage{amsthm}
\usepackage{subfigure}
\usepackage{subcaption}
\usepackage{xcolor}
\usepackage{csquotes}
\usepackage{soul}
\usepackage{enumerate}
\usepackage[normalem]{ulem}
\usepackage{algorithm}
\usepackage{algorithmicx}
\usepackage{algpseudocode}
\usepackage{booktabs}
\usepackage{marvosym}
\usepackage{enumitem}
\usepackage{rotating}
\usepackage{arydshln}
\usepackage{ragged2e}
\usepackage{wrapfig}
\usepackage{pdflscape}
\usepackage{adjustbox}

\definecolor{citecolor}{RGB}{34,139,34} 
\definecolor{veronica-red}{RGB}{196,30,58}
\definecolor{cvprblue}{RGB}{98,149,202}
\definecolor{darkblue}{rgb}{0, 0, 0.5}

\newcommand{\coloredblock}[1]{\textcolor{#1}{\blacksquare}}
\definecolor{code1}{RGB}{197,224,180}
\definecolor{code2}{RGB}{226,240,217}
\definecolor{code3}{RGB}{112,173,71}
\definecolor{reason}{RGB}{91,155,213}
\definecolor{color1}{RGB}{199,176,61}
\definecolor{color2}{RGB}{218,81,58}
\definecolor{color3}{RGB}{50,157,156}
\definecolor{color4}{RGB}{32,80,114}
\definecolor{timecolor1}{RGB}{187,219,173}
\definecolor{timecolor2}{RGB}{255,226,148}
\definecolor{timecolor3}{RGB}{211,223,240}
\definecolor{takeaway-blue}{RGB}{218,227,243}
\definecolor{takeaway-title-blue}{RGB}{51,74,133}
\usepackage[most,skins,theorems]{tcolorbox}

\tcbset{
  takeawaybox/.style={
    width=\linewidth,
    top=10pt,
    bottom=4pt,
    left=2pt,
    right=2pt,
    colback=takeaway-blue,
    colframe=black,
    colbacktitle=takeaway-title-blue,
    boxrule=0.5pt,
    enhanced,
    center,
    attach boxed title to top left={yshift=-0.1in,xshift=0.15in},
    boxed title style={boxrule=0pt,colframe=white,},
    breakable,
    after={\vspace{-1em}}
  }
}
\newtcolorbox{TakeawayBox}[2][]{takeawaybox,title=#2,#1}

\usepackage[pagebackref=true, 
            breaklinks=false, 
            colorlinks, 
            citecolor=citecolor, 
            linkcolor=veronica-red, 
            urlcolor=cvprblue,
            bookmarks=false]{hyperref}

\usepackage{amssymb}
\usepackage{pifont}
\usepackage{url}
\usepackage{diagbox}
\usepackage{makecell}
\usepackage[numbers]{natbib}
\usepackage[normalem]{ulem}

\newcommand{\czr}[1]{{\color{color4}{[(zhirui): #1]}}} %

\usepackage{xspace}

\newcommand{\pak}{\text{pass@$k$}\xspace}

\definecolor{ForestGreen}{RGB}{34,139,34}
\definecolor{BrickRed}{rgb}{.72,0,0}
\definecolor{LakeBlue}{RGB}{0,61,153}
\usepackage{pifont}
\newcommand{\greencheck}{\textcolor{ForestGreen}{\ding{51}}}
\newcommand{\redcross}{\textcolor{BrickRed}{\ding{55}}}
    
\ifCLASSOPTIONcompsoc
  \usepackage[nocompress]{cite}
\else
  \usepackage{cite}
\fi
\ifCLASSINFOpdf
\else
\fi
\begin{document}
%

\title{A Survey of Neural Code Intelligence: Paradigms, Advances and Beyond}

%
%


\author{Qiushi~Sun,
        Zhirui~Chen, 
        Fangzhi~Xu, 
        Kanzhi~Cheng, 
        Chang~Ma, 
        Zhangyue~Yin,
        Jianing~Wang, 
        Chengcheng~Han, 
        Renyu~Zhu, 
        Shuai~Yuan, 
        Qipeng~Guo, 
        Xipeng~Qiu,
        Pengcheng~Yin, 
        Xiaoli~Li,~\IEEEmembership{Fellow,~IEEE,}
        Fei~Yuan, 
        Lingpeng~Kong,
        Xiang~Li,
        and~Zhiyong~Wu 
\IEEEcompsocitemizethanks{


\IEEEcompsocthanksitem Qiushi Sun (\url{qiushisun@connect.hku.hk}), Fangzhi Xu, Kanzhi Cheng, Chang Ma, Shuai Yuan, Fei Yuan, and Zhiyong Wu (\url{wuzhiyong@pjlab.org.cn}) are with Shanghai AI Laboratory, Shanghai, China.
\IEEEcompsocthanksitem Zhirui Chen, Jianing Wang, Chengcheng Han, and Xiang Li (\url{xiangli@dase.ecnu.edu.cn}) are with the School of Data Science and Engineering, East China Normal University, Shanghai, China.
\IEEEcompsocthanksitem Zhangyue Yin, Qipeng Guo, and Xipeng Qiu are with the School of Computer Science, Fudan University, Shanghai, China.
\IEEEcompsocthanksitem Renyu Zhu is with NetEase Fuxi AI Lab, Zhejiang, China.
\IEEEcompsocthanksitem Pengcheng Yin is with Google DeepMind, Mountain View, CA, USA.
\IEEEcompsocthanksitem Xiaoli Li is with the Institute for Infocomm Research (I\textsuperscript{2}R), Agency for Science, Technology and Research (A*STAR), Singapore, and also with the School of Computer Science and Engineering at Nanyang Technological University, Singapore.
\IEEEcompsocthanksitem Lingpeng Kong is with the Department of Computer Science, The University of Hong Kong, Hong Kong, China.
}
\thanks{Version 1.6} 
}

%
%

\markboth{S\lowercase{un et al.}}
{Shell \MakeLowercase{\textit{et al.}}: Bare Advanced Demo of IEEEtran.cls for IEEE Computer Society Journals}
%




\IEEEtitleabstractindextext{%
\begin{abstract}
\justifying
Neural Code Intelligence -- leveraging deep learning to understand, generate, 
and optimize code -- 
holds immense potential for transformative impacts on the whole society.
Bridging the gap between Natural Language and Programming Language, 
this domain has drawn significant attention from researchers in both research communities over the past few years.
This survey presents a systematic and chronological review of the advancements in code intelligence, 
encompassing over 50 representative models and their variants, 
more than 20 categories of tasks, 
and over 700 related works.
We follow the historical progression to trace the paradigm shifts across different research phases (\textit{e.g.}, from modeling code with recurrent neural networks to the era of Large Language Models). 
Concurrently,
we highlight the major technical transitions in models, tasks, and evaluations spanning through different stages.
For applications, 
we also observe a co-evolving shift.
It spans from initial endeavors to tackling specific scenarios, 
through exploring a diverse array of tasks during its rapid expansion, 
to currently focusing on tackling increasingly complex and varied real-world challenges.
Building on our examination of the developmental trajectories, 
we further investigate the emerging synergies between code intelligence and broader machine intelligence,
uncovering new cross-domain opportunities and illustrating the substantial influence of code intelligence across various domains.
Finally, 
we delve into both the opportunities and challenges associated with this field, 
alongside elucidating our insights on the most {promising research directions}.
An ongoing, dynamically updated project and resources associated with this survey have been released at 
\url{https://github.com/QiushiSun/Awesome-Code-Intelligence}. 
\end{abstract}
\begin{IEEEkeywords}
Code Intelligence; Natural Language Processing; Language Models; Software Engineering
\end{IEEEkeywords}}

\maketitle

\IEEEdisplaynontitleabstractindextext

%
\IEEEpeerreviewmaketitle

\input{sections/1-introduction}


\input{sections/2-code-embeddings}

\input{sections/3-CodePTMs.tex}

\input{sections/4-LLM.tex}

\input{sections/5-reasoning}

\input{sections/6-applications}

\input{sections/7-future}

\input{sections/8-conclusion}


%

\appendices

\input{appendices/resources}

\input{appendices/more-benchmarks}
\input{appendices/codeptm-objects}
\input{appendices/eval-details}
\ifCLASSOPTIONcaptionsoff
  \newpage
\fi



%

\clearpage
\bibliographystyle{IEEEtranN}
\bibliography{custom,bibliographies/models,bibliographies/rel-surveys,bibliographies/datasets-benchmark,bibliographies/methods,bibliographies/code-embd,bibliographies/softwares,bibliographies/analysis,bibliographies/applications,bibliographies/preference}


%

\end{document}

%% file: sections/1-introduction.tex
\IEEEraisesectionheading{\section{Introduction}\label{sec:introduction}}




\IEEEPARstart{C}{ode} is one of the elegant languages created by humans, 
which replaces the diverse forms of natural language (NL) through a high degree of abstraction~\citep{pierce2002types}.
As a conduit between humans and machines,
it is ultimately transformed into specific programs\footnote{We use \textit{code} and \textit{program} interchangeably in this paper.} 
that substitute human effort in accomplishing various tasks, 
characterized by advantages such as precision, logic, modularity, and executability.

\begin{figure}[t]
    \centering
    \includegraphics[width=0.45\textwidth]{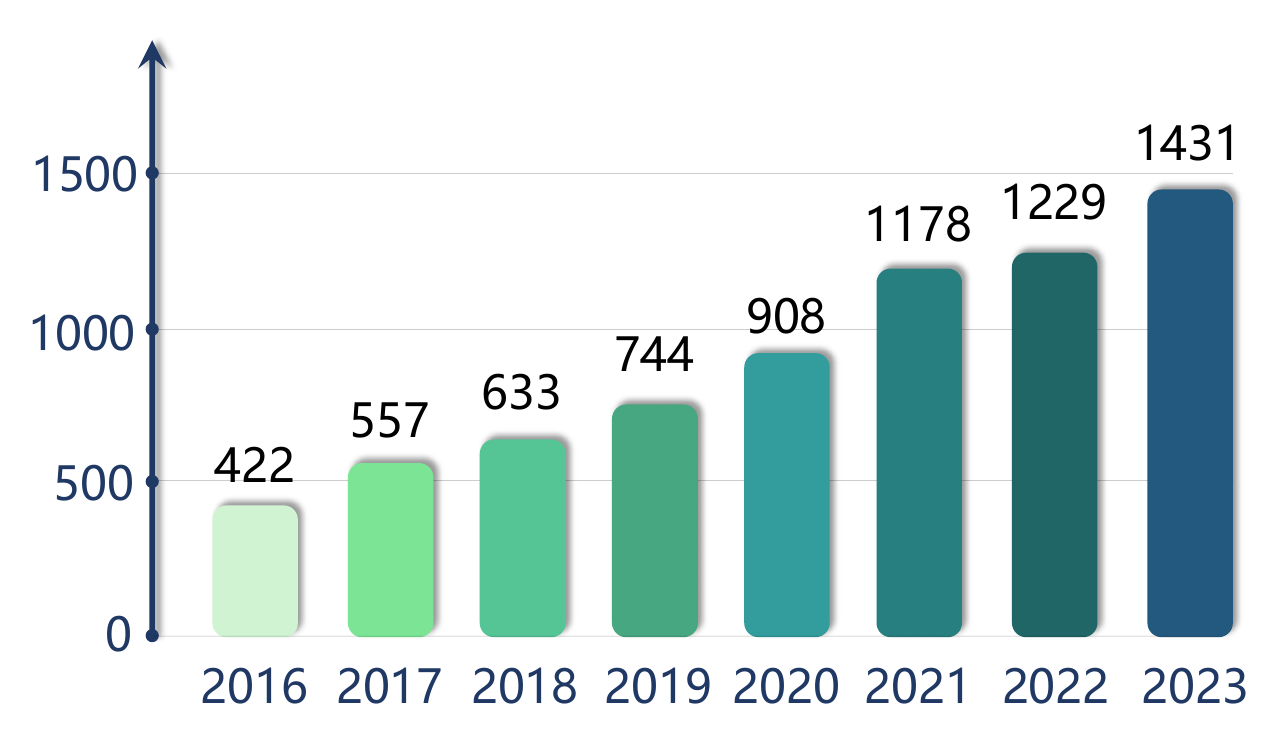}
    \caption{Cumulative number of publications/preprints related to neural code intelligence (from arXiv). 
    Over the past few years, the number of articles has been steadily increasing.
    }
    \label{fig:trend}
    \vspace{-0.5em}
\end{figure}

\begin{figure*}[t]
   \begin{center}
   {\includegraphics[width=\linewidth]{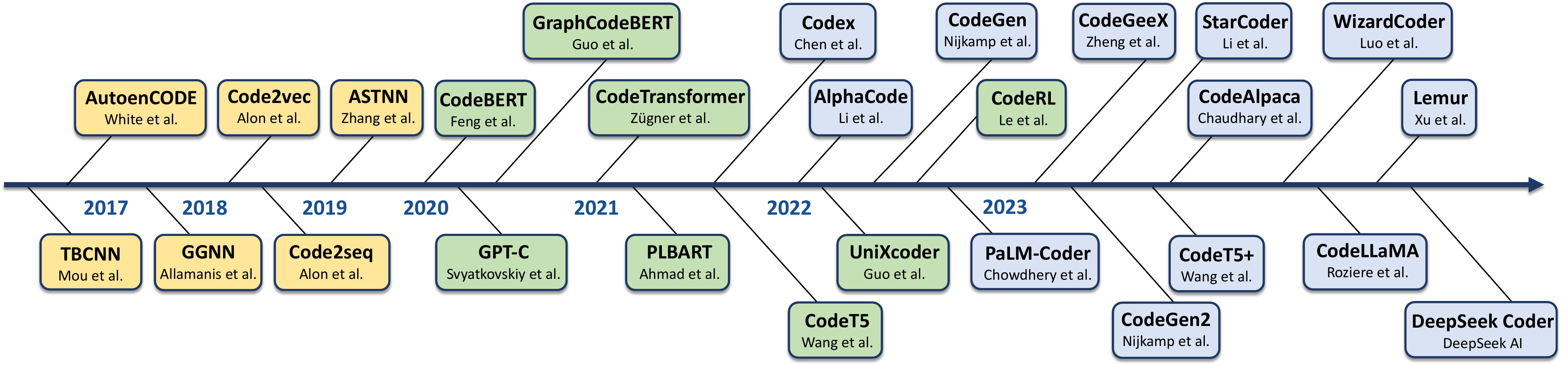}} 
   \end{center}
\caption{A chronological overview of representative works in neural code intelligence over recent years. Works are differentiated by background colors to represent distinct evolutionary phases: $\coloredblock{timecolor2}$ represents milestones of neural language models using code structures, 
$\coloredblock{timecolor1}$ denotes code pre-trained models with typical architectures, 
and $\coloredblock{timecolor3}$ signifies some influential large language models for code.
The timeline is established mainly according to the release date of the paper or model.
}
\label{fig:timeline}
\end{figure*}

The fusion of rapidly advancing deep learning techniques with the availability of ``Big Code''~\citep{allamanis2013mining,allamanis2018survey} has led to the emergence of neural code intelligence. 
This domain,
which applies neural approaches to understand, generate, 
and manipulate code, 
has garnered significant attention from the research community.
Figure~\ref{fig:trend} illustrates the cumulative
publication statistics\footnote{
The statistics are derived by querying a set of specific keywords (\textit{e.g.}, code representation, code generation, code intelligence) through an exact match search in the titles or abstracts of documents.
} 
over recent years, 
showcasing the growing interest and significant efforts being dedicated to this area.
Notably, this domain transcends disciplinary boundaries, spanning Natural Language Processing (NLP)~\citep{zan2023large}, Software Engineering (SE)~\citep{zhang2023unifying}, Robotics~\citep{hu2023generalpurpose}, and beyond.
Moreover,
the unique duality of code, which combines human-readable semantics with executability, establishes research in this area as a critical bridge between artificial intelligence and the real world, 
laying a cornerstone on the path toward artificial general intelligence~\citep{bubeck2023sparks}.
At the heart of these studies lies the foundational concept of ``Software Naturalness''~\citep{hindle2016natural}, 
which posits that programming language (PL), much like human languages, is characterized by predictable and repetitive patterns that can be effectively modeled.
From a macroscopic perspective,
the progression of techniques for processing these artificial languages has largely mirrored the evolution observed in NLP~\citep{sun2022paradigm}.

Specifically,
after a brief initial period characterized by statistical~\citep{nguyen2013statistical,allamanis2014naturalize,allamanis2015bimodel} and probabilistic~\citep{bielik2016phog} modeling,
the learning paradigms of 
neural code intelligence
has transitioned from the earliest word embedding techniques to Large Language Models (LLMs), 
broadly categorized into three main phases of evolution:

$\bullet$ \emph{Neural Language Models for Code.} 
This era witnessed the early yet fruitful efforts of applying deep learning to process code. 
The methods designed during this period primarily relied on well-developed recurrent~\citep{schmidhuber1997lstm} or convolutional~\citep{kim2014conv} structures to model code. 
Notably,
they not only leverage the textual information of the code but also intricately incorporate structural information extracted from code structures like Abstract Syntax Trees (ASTs) into the modeling process~\citep{mou2016convolutional,zhang2019astnn}, 
aligned closely with the principles of neural semantic parsing~\citep{liang2017neural,yinneubig2018tranx,yin2021learning}.

Meanwhile, 
as code snippets can be  
represented as continuous vectors, 
the techniques evolved during this period were also known as code embeddings~\citep{han2021comp}.
The most representative techniques, 
code2vec~\citep{alon2019code2vec} and code2seq~\citep{alon2018codeseq},
captured the semantic and structural information of code by embedding paths from ASTs into a vector space, 
enabling the application of neural methods to diverse scenarios.

$\bullet$ \emph{Code Pre-trained Models (CodePTMs).} Pre-trained language models~\citep{devlin2018bert, radford2018improving} with multi-layer Transformer architecture~\citep{vaswani2017attention} have established a ``pre-train'' and ``fine-tune'' learning paradigm~\citep{qiu2020pre}.
Following this, various studies have emerged on how to build CodePTMs.
These models, exemplified by CodeBERT~\citep{feng2020codebert}, CodeT5~\citep{wang2021codet5}, and PLBART~\citep{ahmad2021unified},
have long dominated the mainstream approaches in code intelligence and initiated a surge of research interest -- a sharp increase in arXiv papers can be observed in Figure~\ref{fig:trend} after the release of CodeBERT in 2020. 
During the pre-training phase, 
they learn general-purpose context-aware representations from massive GitHub code data, 
as well as their structural information of code (\textit{e.g.}, ASTs). 
Subsequently, 
they are fine-tuned on task-specific code data or NL-PL pairs,
significantly improving the performance of various code-related tasks. 
These approaches mark a shift from previous learning paradigms, 
where CodePTMs no longer require individual modeling for each task,
but adapt to different scenarios by fine-tuning on relatively smaller labeled datasets.

$\bullet$ \emph{Large Language Models for Code (CodeLLMs).}  Recent studies have indicated that scaling language models through increasing their parameters or the volume of training data~\citep{kaplan2020scaling,hoffmann2022an} consistently results in an enhancement of the model's capacity to perform effectively on downstream tasks. 
Following the success of general LLMs such as GPT-3~\citep{brown2020GPT3} and PaLM~\citep{chowdhery2022palm} in both academia and industry~\citep{zhao2023survey},
models like Codex~\citep{chen2021evaluating} have sparked a new wave of research in code intelligence -- a notable increase of related papers can be observed in Figure~\ref{fig:trend} after the debut of \textit{ChatGPT}\footnote{\url{https://openai.com/blog/chatgpt}} in late 2022. 
This phase has also seen a shift in the learning paradigm from task-specific fine-tuning to prompting~\citep{liu2023prompt} and in-context learning~\citep{dong2023survey},
as well as expanding the application of code intelligence from solely code-related tasks to a broader array of real-world scenarios.

Aware that the development of code intelligence is significantly reflected in the evolution of language models designed for code, 
we present in Figure~\ref{fig:timeline} a chronological summary of representative works that outline the developmental trajectory of neural code intelligence.
It provides a framework for this paper by outlining an overview of the technical advances within the domain.
Building upon this timeline, 
we embrace a new perspective characterized by the \textit{paradigm shifts} in models, applications, evaluations, and beyond, 
to explore these exciting advancements in code intelligence.
We thoroughly investigate the literature and distill the key findings, techniques, and interconnections between research across various epochs.
Moreover, we broaden our scope to explore the integration of other domains with code intelligence, 
discussing how code generation assists in machine reasoning, 
how code training enhances models' mathematics capabilities,
and how code serves as a medium to provide new approaches for solving typical NLP tasks. 
Furthermore,
we explore a wide array of real-world cross-domain applications, spanning coding assistants, data science, autonomous agents, and AI4science.
A GitHub project associated with this survey is actively maintained at \url{https://github.com/QiushiSun/Awesome-Code-Intelligence}, 
which contains the resources used to construct this paper, 
as well as comprehensive reading lists to support
further exploration. 
We hope that this will facilitate the continued development of the field.

The following parts of this survey are organized
as follows.
Section~\ref{sec:code-embd} begins by reviewing the preliminaries. 
We then examine classic yet crucial methods for processing code using neural language models, 
as well as an introduction to quintessential code-related tasks.
Section~\ref{sec:codeptms} delves into the evolution of code intelligence within the pre-train and fine-tune paradigm, 
offering a comprehensive review and discussion of the techniques of this era and their implications for future research.
Section~\ref{sec:codellms} explores the research advancements in the era of LLMs, 
alongside a review of progress in NL2Code,
and conducts thorough discussions of representative models and benchmarks. 
Section~\ref{sec:reasoning} investigates
the synergies between code intelligence and other domains of machine intelligence.
Section~\ref{sec:app} is on the application of code intelligence in real-world scenarios, demonstrating its practical utility.
Section~\ref{sec:discuss} engages in a dual-faceted discussion of open issues, 
both from the perspectives of model architecture and practical application,
and shares what we believe are worthy of future research directions.
Finally, 
Section~\ref{sec:conclusion} summarizes the insights and findings of the paper.

%% file: sections/2-code-embeddings.tex
\section{The Spark of Code Intelligence}
\label{sec:code-embd}
In recent years, 
the vast availability of source code from public repositories has significantly boosted the application of deep learning techniques to source codes~\citep{allamanis2018survey}. 
Under the concept of ``software naturalness''~\citep{hindle2016natural},
neural language models designed for processing text can be naturally applied to code. 
Combined with the evolving demands of software engineering for automated code processing, 
neural code intelligence has experienced a prosperous development.

By conceptualizing code snippets as language sequences,
sequential neural architectures, 
such as LSTM~\citep{schmidhuber1997lstm}, 
are naturally adaptable for code understanding and generation~\citep{dam2016deep,liang2019seml}.
However, 
it is imperative to recognize that, unlike natural language sentences, 
programs contain  explicit and complicated structures, 
which introduces more opportunities and possibilities for modeling code.
Subsequently, 
code embeddings have emerged, 
which can be defined as numerical sequences representing the inherent concepts found in codes. 
This development has had a pioneering and long-lasting impact on future research in code intelligence.


In this section, we first provide readers with preliminaries regarding code structure, followed by a review of some classic methods of modeling based on it. Subsequently, we introduce a series of classic, significant, and continuously explored code-related tasks. 




\subsection{Code Features Through Structural Views}
\label{sec:code-embd:struct}

Viewing source code merely as a token sequence overlooks their inherent structures,
a characteristics can greatly enhance the model's ability to comprehend code.
To illustrate this, let's consider a straightforward example.
Consider the expression \texttt{s} = \texttt{min\_value} + \texttt{max\_value},
\texttt{s} is evidently derived from the maximum and minimum values.
However, since programmers do not always adhere to naming conventions, 
it is challenging for models to grasp the semantics of \texttt{s} solely from its variable name.
Nevertheless, 
by leveraging the dependency relation between variables, 
it becomes possible to facilitate the comprehension of the semantics of the \texttt{s} and to predict program properties~\citep{alon2018predict}.
Structural information represented by such dependency relations plays a crucial role in modeling code~\citep{maddison2014structured,allamanis2015bimodal}.
Therefore, in this part, we will briefly cover three typical carriers of structural information, providing readers with some background knowledge.


\noindent$\bullet$~\textbf{Abstract Syntax Tree.}
AST stands as a quintessential intermediate representation during code compilation, 
where a program is parsed into a tree structure of operations and their operands. 
Serving as a syntactic-level structure, 
it encapsulates both the syntax and structural information of a program, 
while its components also embody distinct semantics~\citep{wang2020modular}.
It can be obtained by applying parsers (\textit{e.g.}, Tree-sitter\footnote{\url{https://github.com/tree-sitter}}, pycparser\footnote{\url{https://github.com/eliben/pycparser}} and javalang\footnote{\url{https://github.com/c2nes/javalang}}) on source codes.

\noindent$\bullet$~\textbf{Data flow.}
Unlike syntactic-level code structures like AST,
Data flow (Graph) represents a semantic-level structure within code. 
Its nodes represent variables,
and the edges reflect the relationships and origins among these variables. 
It can be extracted from AST and characterized by reduced complexity and does not entail a deep hierarchy,
resulting in relatively lower costs for modeling and analysis~\citep{cummins2021programl}.

\noindent$\bullet$~\textbf{Control flow.}
Contrasting with data flow's semantics, 
Control Flow (Graph) provides a structural view of code executive information. 
Here, 
nodes represent executable blocks, and edges indicate control transitions between them. This emphasizes the sequence and potential paths of program execution rather than variable interactions~\citep{allen1970control}. 
Control Flow is key for understanding program dynamics,
and offering insights into program logic. 
It can be constructed through the use of static analyzers~\citep{li2022scalpel}.




Building on the above code features, 
in processing codes with neural methods,
one can consider not only the plain text of the source code but also leverage code structural information.
This process can be typically divided into three main strategies:
(1) Directly Encoding AST: A representative method is TBCNN~\citep{mou2016convolutional},
which utilizes tree-based convolution kernels on ASTs to capture information from subtrees. 
The features of subtrees will be aggregated through 
pooling to formulate the embedding of the program.
Following this,
subsequent work has increasingly integrated ASTs with convolution to capture local code features~\citep{li2017soft,bui2018cross}.
(2) Utilizing AST Paths: This approach is exemplified by code2vec~\citep{alon2019code2vec} and code2seq~\citep{alon2018codeseq}. 
Code2vec integrates the representations of AST leaf nodes
and aggregates their path representation to build combined context vectors.  
Following this idea, 
code2seq extracts more fine-grained information from the AST paths and leverages an LSTM to encode the entire path to suit generation tasks.
(3) Transforming AST, 
represented by: 
AutoenCODE~\citep{white2016deep} converts ASTs into binary trees and utilizes autoencoder to learn code embedding from it.
GGNN~\citep{allamanis2018learning} introduces additional edges to explicitly represent data/control flows and employs graph neural networks to learn nodes' representations. 
To address long-term dependency issues,
ASTNN~\citep{zhang2019astnn} breaks each AST into a sequence of statement subtrees and encodes them into vectors by capturing both the lexical and syntactical knowledge of the statements. The approach has been further extended for industrial applications~\citep{xu2023xastnn}.
InferCode~\citep{bui2021infercode} employs self-supervised learning by exploiting the structural similarities within code to automatically generate labels for training.



Moreover, 
beyond utilizing these features in their original forms, 
researchers have adapted them to apply various deep learning approaches.
\citet{wang2020detecting,wang2021completion} initially augment AST with explicit control and data flow edges to facilitate the application of graph algorithms~\citep{kipf2017semisupervised,veličković2018graph,brockschmidt2018generative}.
Later, 
issues related to the low connectivity of ASTs and the out-of-vocabulary problem~\citep{karampatsis2020vocab} during modeling are identified~\citep{cvitkovic2019open}.
To mitigate these issues, 
researchers connecting adjacent leaf nodes, 
which aids in the graph partitioning~\citep{zhu2022neural} and analysis~\citep{zhu2022catprobing}.


\subsection{Overview of Core Tasks in Code Intelligence}
\label{sec:code-embd:task-overview}

This part provides an overview of the most important tasks in code intelligence and the challenges they face, 
categorized based on the form of their inputs and outputs.

\subsubsection{Code-Code Tasks}

Code-code tasks refer to a series of tasks that involve operations on source code with the aim of understanding, generating, or transforming code.

\noindent\textbf{Clone Detection.}
\label{sec:code-embd:clone}
Clone detection is widely studied in SE research~\citep{wei2017supervised,zhang2019astnn,wang2020detecting},
which predicts whether two code snippets are clones of each other, 
and can be conceptualized as a binary sentence classification task. 
Code clones refer to pairs of code snippets that display notable similarities, occurring within or across different software systems~\citep{svajlenko2020clone}.
Programmers often create clones by reusing code through copy and paste.
While cloning can offer advantages, 
such as accelerated software development, it presents significant drawbacks.
When buggy code is cloned, 
the bug is duplicated throughout the system, exacerbating the complexity of debugging and maintenance~\citep{zhang2021clone}.
Furthermore, 
clones may introduce new bugs if updates to a code fragment are not uniformly applied to its clones. 
Such practices can adversely impact software by unnecessarily inflating the system's size and consequently increasing the expenses related to re-engineering~\citep{allamanis2019duplication}.
Beyond finding suitable matching algorithms or metric~\citep{eghbali2203crystal},
The primary challenge in developing automated approaches for clone detection lies in equipping the detector with the ability to fully comprehend syntactic~\citep{roy2008nicad,michel2009clone,sajnani2016sourcerercc} or semantic~\citep{roy2007aso,wu2020scdetector,wang2020blended} similarities, thereby minimizing the risk of false positives~\citep{krinke2022bcb}.


\noindent\textbf{Defect Detection.}
\label{sec:code-embd:defect}
The incidence of source code defects and vulnerabilities has been increasing rapidly,
as evidenced by public reports through CVE (Common Vulnerabilities \& Exposures), 
as well as identified within proprietary code bases.
This trend poses a crucial yet complex challenge in security.
In response to this, 
defect (vulnerability) detection has emerged as a solution,
aiming to liberate human programmers from the extensive demands of manual code inspection~\citep{ray2016on,wang2016auto,li2019improving,zhou2019devign,li2024bug}.
The task can be formalized as binary classification,
\textit{i.e.}, learning to determine whether a given code snippet contains a defect or not.
As a non-generative task, it shares similar challenges with clone detection, 
and further, the calibration of models~\citep{zhou2024on,spiess2024calibration} plays a vital role in ensuring reliability as well.


\noindent\textbf{Code Repair.}
\label{sec:code-embd:repair}
Writing codes often involves errors, 
a common experience for programmers.
Often, 
these errors are minor, 
necessitating only limited modifications to the original program.
Such errors can interrupt the workflow of experienced developers and may pose significant challenges to beginners while localizing and rectifying them is known to be effort-prone and time-consuming.
Code repair aims to refine the code by automatically localizing~\citep{zhou2012should}  and fixing these bugs, 
which can be modeled as a seq2seq task~\citep{gupta2017deepfix,allamanis2018learning,bhatia2018neuro,hellendoorn2019global,tufano2019bugfix,han2023errorclr}. 
By integrating code repair with defect detection, it becomes possible to streamline the processes of identifying issues and implementing fixes~\citep{allamanis2021self}.

\noindent\textbf{Code Completion.}
\label{sec:code-embd:completion}
Code completion is one of the most common application scenarios for coding assistants like Copilot\footnote{\url{https://github.com/features/copilot}}, 
and its usage is bifurcated into two subcategories: token-level completion and line-level completion. 
The former involves predicting a single code token, 
while the latter entails completing an entire, yet unfinished line. 
The objective of the task is to predict subsequent token(s) within a given code context~\citep{robbes2008completion,bruch2009completion,sangmok2009completion,hou2011completion,svyatkovskiy2021fast}. 
It can also be viewed as a seq2seq task, 
but the target needs to be a continuation of the input. 
With the evolution of code intelligence, 
code completion has also begun to encompass infilling tasks~\citep{fried2023incoder,bavarian2022fim}, 
which entails not only left-to-right completion but also filling in code before or in the middle of a given context.


\noindent\textbf{Code Translation.}
\label{sec:code-embd:trans}
Code translation, 
also known as transpilation, 
involves translating a code snippet from one PL to another.
It has many use cases, 
such as modernizing artifacts~\citep{malia2021mono2micro} implemented in PLs like COBOL or Python 2, 
and migrating legacy software in proprietary PLs to applications written in general-purpose PLs~\citep{nitin2023cargo}.
Over the past decades, 
the paradigm of code translation has undergone a significant transformation,
shifting from labor-intensive 
rewriting methods to more efficient and reliable automated solutions. 
While it can also be described as a seq2seq task, 
it presents more difficulty compared to previous tasks.
The greatest challenges 
include 
(1) the need to faithfully preserve the original functionality, 
and 
(2) the requirement to generate syntactically correct code without introducing bugs~\citep{pan2024lost}.
Existing research includes strategies that utilize annotated PL pairs and their syntactic structures for training~\citep{nguyen2015divide,chen2018tree}, 
as well as unsupervised methods that learn from monolingual source code without parallel data~\citep{roziere2020unsupervised,wen2022babeltower,liu2023syntax}.
Additionally, 
the scope of this area also includes pseudocode-to-code translation~\citep{oda2015learn,kulal2019pseudocode}.
In comparison to NL machine translation, the functional correctness of all translated code is more critical than its similarity~\citep{lin2004rouge,denkowski2014meteor} to the reference.





\input{tables/task-wo-nl2code}

\subsubsection{Code-Text Tasks}
\label{sec:code-embd:codesum}

Code-text tasks refer to the challenge of generating natural language from source code.

\noindent\textbf{Code Summarization.}
Code summarization represents a prominent task in the field of code intelligence,
which entails generating concise and descriptive comments for codes, 
derived from analyzing its semantics~\citep{zhu2019automatic}.
It is vital for updating and maintaining software systems~\citep{hu2023maintainability}, 
particularly those with collaboration among multiple developers.
Code summarization can be modeled in a seq2seq format, 
aiming to take a code snippet (and its structure) as input and produce an NL description~\citep{iyer2016summarizing,alon2018codeseq,fernandes2018structured,wan2018improve,leclair2019neural} or the function/method’s name as output~\citep{allamanis2016convatt,ahmad2020transformer,peng2021integrating,pian2023metatptrans}.
Beyond directly synthesizing summaries, strategies include retrieving keywords from the source code~\citep{haiduc2010on,de2012using} or employing clone detection to find comments from similar code snippets~\citep{wong2015clocom}. 
Additionally, 
leveraging the API knowledge can further enhance the relevance of generated content~\citep{ijcai2018TLCodeSum}.


\noindent\textbf{Commit Message Generation.} 
Developers may frequently edit their code for bug fixing, 
adding new features, etc. 
Version control systems like Git often track these edits,
which utilizes commit to document the changes. 
When code is updated frequently, manually writing commit messages becomes a laborious task. 
Fortunately, 
code embeddings can also be employed to represent these edits~\citep{yin2018learning}. 
Commit message generation is an emerging task aimed at automating the creation of commit messages for code changes. 
It involves taking two versions of the code, 
before and after the edits, 
as input and generating summaries that describe their differences~\citep{tao2021cmgmodel,tao2022cmg}.
In practice, 
the methods employed for commit message generation span a range of techniques, 
including the use of predefined rules or templates~\citep{coy2014on}, 
leveraging commit messages from similar code changes~\citep{Liu2018NNGen},
employing seq2seq modeling~\citep{Jiang_McMillan2017,Xu2019CoDiSum,jung2021commitbert},
and incorporating retrieval-augmented approaches~\citep{shi2022race}. 


\subsubsection{Text-Code Tasks}

Text-code tasks involve finding or generating executable source code aligned with NL descriptions.

\noindent\textbf{Code Retrieval.}
\label{sec:code-embd:retrieve}
To enhance coding productivity, seeking ready-made solutions that closely match their requirements serves as a shortcut. The objective of code retrieval, also known as NL code search, is to identify and retrieve functionally relevant codes in response to NL queries~\citep{husain2019codesearchnet,parvez2021ret,xie2023survey} for both developers and models.
A common practice involves using specially designed metrics to measure the similarity between the contextual embeddings of the given query and the candidate code snippets~\citep{Gu2018DeepCS,wang2020trans3}.
Additionally, 
there is a parallel task to code retrieval known as code search~\citep{grazia2023codesearch,kim2023bigcode}, 
where the key difference lies in the query: here, the query is also a code snippet. 
This task can be viewed as searching for clones within a candidate pool, allowing developers to find code snippets that perform similar functions or have similar implementations based on code-based queries. 
The code obtained can serve as a reference for generating more complex yet related code~\citep{hayati2018retrieval}.
Retrieval has long been an active research area, and modern approaches (\textit{e.g.}, CodeXEmbed~\citep{liu2024codexembed}) demonstrate that effective retrieval enables models to better generalize across diverse code-related tasks.

\begin{figure*}[ht]
   \begin{center}
   {\includegraphics[width=0.995\linewidth]{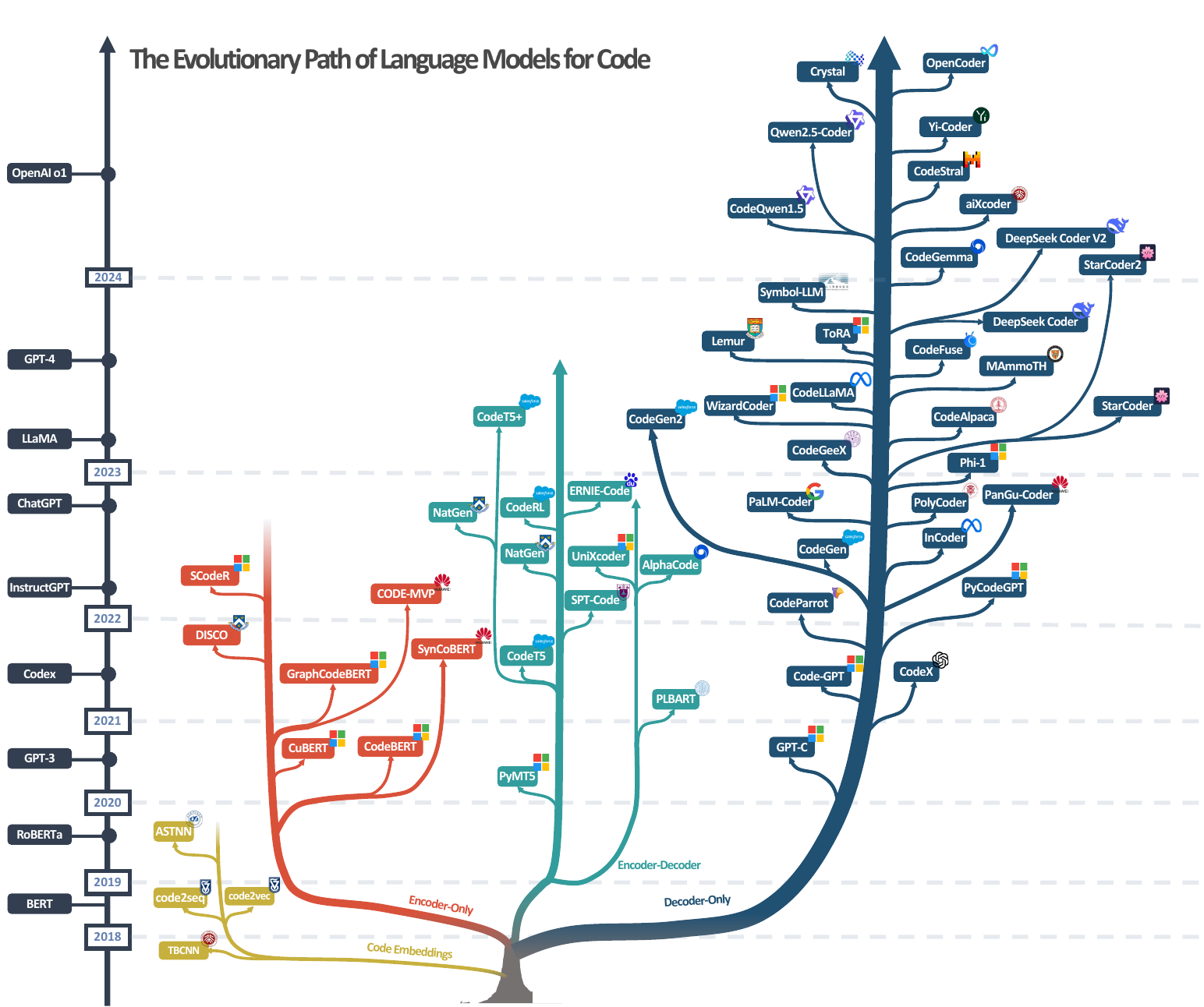}} 
   \end{center}
\caption{The trajectory of neural code intelligence's evolution is encapsulated through the development of language models for code.
This is delineated by four principal branches,
each representing a distinct category of models. 
The first branch showcases models based on \textcolor{color1}{code embedding techniques}, 
while the subsequent three branches feature Transformer-based models, each exemplifying unique architectures: 
\textcolor{color2}{Encoder-only}, \textcolor{color3}{Encoder-Decoder}, and \textcolor{color4}{Decoder-only}. 
Models on the same sub-branch have closer relationships. Additionally, the vertical axis chronicles the timeline of these models' release dates, paralleled by some seminal NLP models. The details of creating this figure are listed in Appendix~\ref{app:trees}.}
\label{fig:codelm-tree}
\end{figure*}

\noindent\textbf{Code Generation.}
\label{sec:code-embd:generation}
Code generation, also known as program synthesis,
broadly refers to the use of NL to generate code (NL2Code). This long-standing task aims to lower the barriers associated with coding, streamline some routine tasks through automation, and empower non-programmers to obtain solutions tailored to their intentions~\citep{robert1985automatic}.
Beyond merely treating it as another form of text generation, early research primarily relied on the guidance of code syntax to generate small code snippets for specific scenarios,   
and view it as a semantic parsing task~\citep{ben2018neural,sun2019grammar,shin2019program}. 
Over time, 
this focus gradually shifted towards general-purpose code generation~\citep{yin2017syntactic,lin2017program,hayati2018retrieval,sun2019cnndecoder} for a single programming language. 
Examples such as Hearthstone~\citep{ling2016latent}, CONCODE~\citep{iyer2018mapping}, and NL2Bash~\citep{lin2018nl2bash} respectively represent efforts to convert natural language into Python, Java, and Bash.
Subsequently, with advancements in the field, 
especially with the help of LLMs,
this domain has experienced prosperous growth~\citep{zan2023large}.
The scope has gradually covered the generation of code in multiple PLs~\citep{cassano2022multiple},
data science notebooks~\citep{agashe2019juice,chandel2022jupyt5}, 
and other more complex scenarios. 
We will delve into more detailed discussions of these developments in Section~\ref{sec:codellm:nl2code}.

Besides, Text-to-SQL can be viewed as a special case of code generation,
which translates NL requirements into SQL statements~\citep{yu2018spider,yu2019sparc,yu2019cosql}. 
This has been a long-studied topic and plays a significant role in bridging the gap between humans and relational database management systems~\citep{qin2022survey,deng2022sql,george2023sql}.
As for their more recent developments, 
we will delve into this topic in Section~\ref{sec:app:ds}.

For the tasks mentioned above, we have compiled a list of representative benchmarks along with their brief descriptions in Table~\ref{tab:code_tasks}.
Furthermore, we expand on this list in Table~\ref{tab:code_tasks_full_p1}, \ref{tab:code_tasks_full_p2} and \ref{tab:code_tasks_full_p3}, 
which include additional benchmarks and cover some tasks that are not detailed extensively, such as question answering~\citep{liu2021codeqa}, comments generation~\citep{yu2020towards}, log analytics research~\citep{yadav2020log}, document translation~\citep{lu2021codexglue}, and programming learning~\citep{miao2020pipe,puri2021codenet,zhu2022programming}.

The neural modeling for the code-related tasks discussed above was primarily developed before 2020, 
as illustrated on the left branch of Figure~\ref{fig:codelm-tree}. 
Although most models were designed ad hoc for specific tasks and may seem relatively simple by today's standards, they signified the rise of code intelligence and laid the foundation for subsequent research.



\begin{TakeawayBox}
{\textit{Takeaways}}
\begin{itemize}[itemsep=2pt,topsep=3pt,parsep=0pt]
    \item[(1)] {The application of neural networks to code marked a foundation for neural code intelligence.}
    \item[(2)] {Compared to modeling NL, incorporating code structures represents a critical divergence. These code features can be extracted and encoded in diverse ways, a practice that subsequent research has continued to leverage.}
    \item[(3)] {A wide range of code-related tasks were introduced and formalized during this period. Beyond their practical value, they also became testbeds for future advancements.}
    \item[(4)] {Despite these earlier techniques becoming overshadowed in the subsequent transformer era, they remain crucial as lightweight, interpretable, and practical solutions for developing code representations.}
\end{itemize}
\end{TakeawayBox}

%% file: tables/task-wo-nl2code.tex
\begin{table*}[ht]
\footnotesize
    \centering
    \caption{Representative benchmarks for different types of code-related downstream tasks, including the number of programming languages they cover and brief descriptions. Complete benchmarks are listed in Table~\ref{tab:code_tasks_full_p1}, Table~\ref{tab:code_tasks_full_p2} and Table~\ref{tab:code_tasks_full_p3}.} 
    \begin{tabular}{l|l|c|c|l}
        \toprule[1.2pt]
        \textbf{Task} & \textbf{Dataset} & \textbf{Date} & \textbf{\# PLs.} & \textbf{Description}  \\
        \midrule
        \multirow{3}{*}{Clone Detection}
        & POJ-104~\citep{mou2016convolutional}~\href{https://sites.google.com/site/treebasedcnn/}{[link]} & 2014 & 2 & a program classification dataset of 52K C/C++ programs  \\
        & BigCloneBench~\citep{svajlenko2014towards}~\href{https://github.com/clonebench/BigCloneBench}{[link]} & 2015 & 1 & a clone detection dataset of eight million Java validated clones  \\
        & CLCDSA~\citep{nafi2019CLCDSA}~\href{https://github.com/Kawser-nerd/CLCDSA}{[link]} & 2019 & 3 & a cross-language clone dataset of more than 78K solutions\\
        \midrule
        \multirow{3}{*}{Defect Detection}
        & Devign~\citep{zhou2019devign}~\href{https://shannon.cs.illinois.edu/DenotationGraph/}{[link]} & 2019 & 1 & a dataset of vulnerable C functions\\
        & CrossVul~\citep{nikitopoulos2021crossvul}~\href{https://zenodo.org/records/4734050}{[link]} & 2021 & $>$40 & a dataset of 13K/27K (vulnerable/non-vulnerable) files \\
        & DiverseVul~\citep{chen2023diversevul}~\href{https://zenodo.org/records/4734050}{[link]} & 2023 & 2 & a dataset of 18K/330K (vulnerable/non-vulnerable) functions\\
        \midrule
        \multirow{3}{*}{Code Repair}
        & Defects4J~\href{https://github.com/rjust/defects4j}{[link]} & 2014 & 1 & a database of real Java bugs \\
        & DeepFix~\citep{gupta2017deepfix}~\href{https://bitbucket.org/iiscseal/deepfix/src/master/}{[link]} & 2017 & 1 & a dataset of 7K erroneous C programs for 93 programming tasks \\
        & QuixBugs~\citep{ye2021quixbugs}~\href{https://github.com/jkoppel/QuixBugs}{[link]} & 2017 & 2 & a multilingual benchmark of similar buggy programs  \\
        \midrule
        \multirow{3}{*}{Code Search}
        & CodeSearchNet~\citep{husain2019codesearchnet}~\href{https://github.com/github/CodeSearchNet}{[link]} & 2019 & 6 & a dataset of 6M functions and natural language queries \\
        & AdvTest~\citep{lu2021codexglue}~\href{https://github.com/microsoft/CodeXGLUE/tree/main/Text-Code/NL-code-search-Adv}{[link]} & 2021 & 1 & a Python code search dataset filtered from CodeSearchNet \\
        & WebQueryTest~\citep{lu2021codexglue}~\href{https://github.com/microsoft/CodeXGLUE/tree/main/Text-Code/NL-code-search-WebQuery}{[link]} & 2021 & 1 & a testing set of Python code search of 1K query-code pairs \\
        \midrule
        \multirow{3}{*}{Code Translation}
        & CodeTrans~\citep{lu2021codexglue}~\href{https://github.com/microsoft/CodeXGLUE/tree/main/Code-Code/code-to-code-trans}{[link]} & 2021 & 2 & a C\#/Java dataset collected from several repos  \\
        & CoST~\citep{zhu2022cost}~\href{https://github.com/reddy-lab-code-research/MuST-CoST}{[link]} & 2022 & 7 & a dataset containing parallel data from 7 programming languages  \\
        & CodeTransOcean~\citep{yan2023codetransocean}~\href{https://github.com/WeixiangYAN/CodeTransOcean}{[link]} & 2023 & 45 & a large-scale comprehensive benchmark for code translation  \\
        \midrule
        \multirow{3}{*}{Code Completion}
        & GitHub Java Corpus~\citep{allamanis2013mining}~\href{https://groups.inf.ed.ac.uk/cup/javaGithub/}{[link]} & 2013 & 1 &  a giga-token corpus of Java code from a wide variety of domains   \\
        & Py150~\citep{raychev2016py150}~\href{https://www.sri.inf.ethz.ch/py150}{[link]} & 2016 & 1 & a corpus of Python programs from GitHub  \\
        & LCC~\citep{guo23longcoder}~\href{https://github.com/microsoft/CodeBERT}{[link]} & 2023 & 3 &  a benchmark of code completion with long code context   \\
        \midrule
         \multirow{3}{*}{Code Summarization} 
        & CODE-NN~\citep{iyer2016summarizing}~\href{https://github.com/sriniiyer/codenn}{[link]} & 2016 & 2 & a dataset of (title, query) pairs from StackOverflow  \\
        & TL-CodeSum~\citep{ijcai2018TLCodeSum}~\href{https://github.com/xing-Approaches}{[link]} & 2018 & 1 &  a dataset containing 69K pairs of (API sequence, code, summary)   \\
        & CodeSearchNet~\citep{husain2019codesearchnet}~\href{https://github.com/github/CodeSearchNet}{[link]} & 2019 & 6 & a dataset of 6M functions and natural language queries \\
        \midrule
        \multirow{3}{*}{GitHub} 
        & CommitGen~\citep{loyola2017commitgen}~\href{https://github.com/epochx/commitgen}{[link]} & 2017 & 4 & a multilingual dataset collected from open source projects \\
        & CommitBERT~\citep{jung2021commitbert}~\href{https://github.com/graykode/commit-autosuggestions}{[link]} & 2021 & 6 & a multilingual dataset of code modification and commit messages \\
        & SWE-bench~\citep{jimenez2023swebench}~\href{https://www.swebench.com/}{[link]} & 2023 & 1 & a benchmark of 2K SE problems and corresponding PRs \\
        \bottomrule[1.2pt]
    \end{tabular}
    \label{tab:code_tasks}
\end{table*}

%% file: sections/3-CodePTMs.tex
\section{An Odyssey of Pre-train and Fine-tune}
\label{sec:codeptms}




Following the remarkable success that pre-trained language models~\citep{qiu2020pre} have achieved in NLP,
the code intelligence community rapidly integrated their architecture and learning paradigms, 
leading to the proliferation of CodePTMs.
Coming after the era of neural language modeling,
this marks a flourishing period for code intelligence which retains code structural insights while incorporating transformer-based models.
The construction of language models for code has undergone a major paradigm shift, characterized by the following features:

\begin{enumerate}
    \item Architecture: Multi-layer transformer~\citep{lin2022transformer} has become the de facto choice for model backbone,
    moving away from building models from scratch for each task or relying on tailored feature engineering.
    \item Training Data: Pre-training is primarily conducted on large volumes of unlabeled data harvested from GitHub~\citep{buratti2020exploring}, 
    with a smaller portion of labeled data typically used for adapting the model to various downstream tasks.
    \item Learning objectives: While optimized through self-supervised objectives, 
    the approach still retains the utilization of structural information to varying degrees to enable more effective learning of code representations.
\end{enumerate}

The development of this stage is reflected in the three main branches shown in Figure~\ref{fig:codelm-tree}.
In this section, we will first conduct a systematic review of representative CodePTMs and their variants, followed by an in-depth discussion of their other aspects.




\subsection{Pre-trained Language Models for Code}
\label{sec:codeptm_models}

In this part,
we discuss a wide range of CodePTMs,
categorizing them based on their architectures.

\subsubsection{Encoder-only}

Existing CodePTMs with encoder-only architecture can be classified based on their use of structural information into two distinct categories: \textit{structure-free} and \textit{structure-based}.
The former only utilizes raw code texts, 
whereas the latter incorporates code structure during pre-training to more effectively grasp the inherent structure of code.

\noindent$\bullet$~\textbf{Structure-free Models}.
CuBERT~\citep{kanade2020learning} marks a pioneering endeavor in the integration of transformer architecture into the realm of code intelligence.
It is trained on a corpus of Python data collected from GitHub, 
employing the same training objectives as BERT~\citep{devlin2018bert} and replicating its training pipeline.
Another milestone is CodeBERT~\citep{feng2020codebert},
which distinguishes itself from CuBERT by adopting a cross-modal training strategy that utilizes both bimodal NL-PL data and unimodal data.
The pre-training of CodeBERT is centered around two objectives: Masked Language Modeling (MLM) and Replaced Token Detection (RTD)~\citep{clark2020electra}.
For implementation, 
it is initialized through RoBERTa~\citep{liu2019roberta} and trained on CodeSearchNet~\citep{husain2019codesearchnet},
a pioneering and influential corpus encompassing six PLs constructed by scraping open-source GitHub repositories.

Regardless of whether task-specific fine-tuning is applied,
both models have achieved performance far surpassing previous word2vec models and multi-layered bidirectional LSTMs (discussed in Section~\ref{sec:code-embd}) across a wide range of code-related tasks, 
paving the way for the successful application of the pre-train and fine-tune paradigm on code.

\noindent$\bullet$~\textbf{Structure-based Models}.
After the success of CodePTMs that solely rely on code tokens for training, 
researchers revisit earlier strategies centered on code features,
innovatively incorporating code structural information in other modalities (\textit{e.g.}, data flow) into the training process of transformer-based models.
GraphCodeBERT~\citep{guo2021graphcodebert} represents one of the earliest endeavors, 
which leverages data flow in the pre-training stage.
Beyond MLM,
it innovatively introduces two tasks: predicting code structure edges and aligning code with its structure. These tasks collectively aim to enable the model to understand the relationships between variables as well as between variables and tokens.
Trained on CodeSearchNet, 
this structure-aware training approach allows GraphCodeBERT to outperform previous models (\textit{e.g.}, CodeBERT) on a range of code-related tasks, 
pioneeringly demonstrating the importance of structural information in code understanding.

Contrastive pre-training~\citep{neelakantan2022text} emerges as another pathway.
SynCoBERT~\citep{wang2022syncobert} extends structure-aware training further by not only considering the structural information of code but also synchronizing the embeddings of code and its corresponding comments through contrastive learning, 
aiming to bridge the gap between code semantics and NL comments.
Similarly employing contrastive objectives, 
CODE-MVP~\citep{wang2022codemvp} explores multi-view learning of code. 
It processes different code structures, such as AST, data flow, and control flow in parallel. The contrastive objects come into play when it compares these multiple views of the same code snippet in training,
thus identifying and reinforcing the commonalities and differences across these representations.
DISCO~\citep{ding2022disco} leverages code transformation algorithms to generate synthetic code clones and inject real-world security bugs, utilized respectively to construct positive and negative samples. 
This approach enables models to discern subtle differences in functionalities. Later, \citet{li2022soft} observe that positive samples created through transformation algorithms, such as variable renaming~\citep{jain2021contrastive} or injecting non-functional code~\citep{bui2021contra}, 
could lead the model to prioritize learning superficial code structures over significant code semantics.
To prevent the model from being misled by superficial content,
SCodeR further employs code comments and subtrees of ASTs to build positive samples, 
compelling the model to deeply understand code semantics and learn to infer code based on its context.

Additionally, in the context of training with program transformations, the concept of identifier deobfuscation in SE has also been employed.
DOBF~\citep{roziere2021dobf} objective begins by concealing the names of functions and variables using placeholder tokens,
then trains the CodePTM to restore the original names through dictionary mapping. 


\subsubsection{Encoder-Decoder}
In contrast with encoder-only models,
which excel in code understanding,
their encoder-decoder counterparts possess inherent advantages in the realm of controllable text or code generation.
Yet, their evolution mirrors that of encoder-only models in significant ways.
Initially, 
code was treated purely as text and applied to encoder-decoder transformers~\citep{elnaggar2021codetrans}, 
followed by the integration of various structural information to enhance the learned code representations~\citep{jiang2021treebert}. 
This evolution has gradually led to the development of three distinct categories of models equipped with classic architectures, which will be discussed as follows:

\noindent$\bullet$~\textbf{BART}. \citet{ahmad2021unified} propose PLBART.
Following the training objectives of BART~\citep{lewis2020bart}, 
it is pre-trained on a corpora constructed by Java and Python functions (from GitHub) and NL documents from StackOverflow\footnote{\url{https://stackoverflow.com/}} via denoising autoencoding.
PLBART is distinctive for its unified training on code and NL, 
aiming to learn the alignment between semantic spaces across different PLs.

\input{tables/codeptm-comparisons}

\noindent$\bullet$~\textbf{T5}. 
The initial exploration of the T5~\citep{raffel2019exploring} architecture's potential on source code is initiated by \citet{mastropaolo2021study}, 
who drew inspiration from the concept of multitask learning. 
This approach commenced with the presentation of a series of code-related tasks as text-to-text transformations.
Similarly, PyMT5~\citep{clement2020PyMT5} replicates this approach by leveraging Python methods and method-docstring data.

CodeT5~\citep{wang2021codet5} is the first structure-aware encoder-decoder model and is among the most influential models today. 
It follows the T5-learning~\citep{raffel2019exploring} pipeline and, in addition to the original span corruption training objective, 
it incorporates:
(1) identifier tagging, which informs the model about whether a code token is an identifier or not;
(2) masked identifier prediction, similar to the deobfuscation mentioned earlier, a variant of span corruption where all identifiers tokens are masked; and
(3) text $\leftrightarrow$ code generation.
For pre-training data, it not only utilizes the CodeSearchNet but also extends to include C/C\# data to accommodate a wider task range. 
In fine-tuning, 
it is capable of performing task-specific transfer learning as well as multi-task learning to address both code generation and code understanding simultaneously.
Building on this,
CodeRL~\citep{le2022coderl} innovatively combines code generation with deep reinforcement learning (using Unit Test Signals), 
and enhances CodeT5 in terms of learning objectives, model sizes, and pretraining data, to better adapt to the NL2Code task.
In light of the outstanding performance, 
CodeT5 has seen further development in the future,
which will be discussed in Section~\ref{sec:codellm:model}.

Meanwhile,
SPT-Code~\citep{niu2022sptcode} enhances its input by integrating linearized ASTs, 
thereby enabling the use of both natural language and code structures as inputs during the pre-training phase.
To improve the code generation ability of T5-based models, 
researchers also explore strategies to enable the decoder part to learn syntax and data flow~\citep{tipirneni2024structcoder}.
NatGen~\citep{chakraborty2022natgen} represents an extension of CodeT5 that leverages the bimodal and dual-channel nature of structural information.
Like DOBF, 
it is trained by ``Naturalizing'' source code to exploit the codes' naturalness and semantics.
It requires the model to receive ``unnatural'' synthetic code as input and produce semantically equivalent code,
mirroring the quality and style a human developer would prefer to write.
CodeT5Mix~\citep{wang2023codetmix} is composed of a mixture of encoders and decoders, 
each with specific code functionality. 
They can be flexibly combined to suit different scenarios and enjoy mutual benefits from joint pretraining on various targets. 
Further, weight-sharing strategies in decoders are used to act as task-specific experts to reduce interference across code-related tasks.
Very recently, AST-T5~\citep{gong2024astt5} employs a structure-aware code segmentation method during its training process, 
enabling the model to reconstruct code structures at various granularities.

Beyond the aforementioned general models,
specialized models also emerged for the first time during this stage.
For instance,
JuPyT5~\citep{chandel2022jupyt5} emerges as a CodePTM tailored for the data science domain.
It is trained on Jupyter Notebook repositories from GitHub, 
with each cell in each notebook considered as a target during the pre-training process,
aiming to serve as a data science assistant.
After that,
leveraging code intelligence to address data science problems has seen substantial advancement,
which will be discussed in subsequent sections.

\noindent$\bullet$~\textbf{UniLMs}. 
It is noteworthy that the UniLM~\citep{dong2019unified, bao2020unilmv2} architecture has also been adopted by researchers to develop its successors trained on source code.
One such model, 
CugLM~\citep{liu2020cuglm}, adopts BERT-like training objectives, 
focusing on code completion tasks.

Another essential UniLM-style CodePTM is UniXcoder~\citep{guo2022unixcoder}, 
which integrates various novel training objectives (\textit{e.g.}, code fragment representation learning) and utilizes cross-modal content such as AST and code comments.
Interestingly, 
it has developed a lossless method to convert ASTs into sequences, 
incorporating these alongside code comments as cross-modal content for pre-training.
The model also expands its training dataset by utilizing both the C4 dataset and the CodeSearchNet data.
Further, 
it employs a prefix mechanism to determine whether the model functions as an encoder-decoder model, a decoder-only model, or an encoder-only model.

\subsubsection{Decoder-only}
During the era where pre-training and fine-tuning are the primary paradigms for code learning, 
CodePTMs with a decoder-only architecture are essentially replicas of GPT~\citep{radford2018improving} models applied to code, 
mostly adhering to the original GPT architectures and employing Causal Language Modeling.
GPT-C~\citep{Svyatkovskiy2020gptc} is a variant of the GPT-2~\citep{radford2019language} trained from scratch on multilingual source code corpora. 
Its purpose is to serve as the \textit{Intellicode} extension in the Visual Studio IDE,
representing one of the initial attempts to utilize language models for code as coding assistants. 
Subsequently, 
GPT-CC~\citep{codedotai2021gptcc}, 
derived through fine-tuning GPT-Neo~\citep{black2021gptneo} on The Pile~\citep{gao2020pile}, 
has been used to build an open-source version of GitHub Copilot.
Additionally, CodeGPT, 
a GPT-style pre-trained model is released alongside CodeXGLUE~\citep{lu2021codexglue}.
It shares a similar parameter size with CodeBERT and is utilized to help solve completion and generation problems during benchmarking machine learning research for program generation.

For task-specific variants,
apart from adapting for specific PL~\citep{su2023java},
there is also a model: PyCodeGPT~\citep{zan2022cert}, 
designed to generate library-oriented codes,
which share similar code sketches (the code structure after anonymizing the user-defined terms). 
Innovatively, 
it employs a specialized trained tokenizer for Python and judges the data quality during the training process through each file's star count and unit test function rate, 
prioritizing the high-quality portions.
These efforts enable PyCodeGPT to achieve excellent code generation capabilities, and the corresponding techniques have been adopted in subsequent research.

In comparison to the previous two types of CodePTMs, 
the development of decoder-only models at this stage is somewhat constrained, 
predominantly revolving around developing the ``code-version'' of GPT-2. 
Owing to their autoregressive characteristics, 
these models find it hard to incorporate structural information of code into their training process.
Nonetheless, 
they will demonstrate remarkable achievements in later research endeavors,
which will be comprehensively investigated in Section~\ref{sec:codellm:model}.



\subsection{Task-specific Adaptation of CodePTMs}
\label{sec:codeptms:adapt}

Unlike the practices in the pre-transformer era where each task requires individualized modeling,
CodePTMs, similar to their counterparts in NLP,
can adapt to the required scenarios through task-specific fine-tuning or by adding new training objectives without significant modifications to the architecture. 
Additionally, they benefit from readily available end-to-end pipelines~\citep{wolf2020tf,wang2023hugnlp,bui2023codetf} and toolkits~\citep{wan2023deep}.

We first revisit the task-enhanced variants of CodePTMs mentioned in Section~\ref{sec:codeptm_models}.
Multiple variants have opted to build upon GraphCodeBERT~\citep{guo2021graphcodebert} for their developments. 
To enhance the capability of code search, CodeRetriever~\citep{li2022coderetriever} incorporates additional Uni/Bimodal training objectives, 
making the model more aware of the semantic similarities between code-code or code-text pairs.
In efforts to utilize external information to bolster code generation and summarization, 
both REDCODER~\citep{parvez2021ret} and ReACC~\citep{lu2022reacc} extend GraphCodeBERT by integrating an additional dense retriever.
Additionally,
to automate code review activities, CodeReviewer~\citep{li2022codereviewer} is constructed by training real-world code changes and code reviews on CodeT5~\citep{wang2021codet5}.
Later, 
CodeExecutor~\citep{liu2023codeexecutor} utilizes UniXcoder~\citep{guo2022unixcoder} as its basis, 
further learning to execute programs and predict their execution traces for improved code generation capabilities.

We then consider models retrained from scratch for specific scenarios.
In terms of further leveraging code features for enhanced generation, CodeTransformer~\citep{zugner2021codetransformer} leverages both the context and structure of code to extract language-agnostic features, aiming to build a multilingual code summarization model. 
GrammarFormer~\citep{guo2022learning} utilizes code syntax to guide the generation of code completions, enabling it to refrain from making predictions in instances where the context is ambiguous. 
Regarding long-range modeling,
\citet{clement2021eWASH} enhance transformer models by prioritizing higher-level syntactic elements to extend context windows,
considering a broader range of contextual information.
Conversely,
to facilitate code completions with longer contexts,
LongCoder~\citep{guo23longcoder} employ sparse attention.
It utilizes a sliding window to attend only the local information,
enabling models to maintain high performance even when dealing with extensive code segments.
Moreover, ERNIE-Code~\citep{chai2023erniecode} represents a specialized case, 
being among the first to recognize the importance of moving beyond English-centric texts. 
It advocates for multilingual text-to-code generation and summarization, 
considering both NL and PL,
increasing the accessibility for a global user base.



\subsection{Understanding and Analyzing CodePTMs}
\label{sec:ptm:interpretable}

As the field progresses, 
in addition to exceptional performance, 
explorations into why these models work and what features they can capture have gradually begun. 
Furthermore, researchers have also started to pay attention to the security threats to these models.

\subsubsection{Understanding the Inner Mechanisms of CodePTMs}
It is crucial for us to understand the inner mechanisms of language models for code and their differences from parallels in NL. Drawing from the experiences in explainable deep learning over the past few years~\citep{raganato2018analysis, serrano2019attention, wiegreffe2019attention, hewitt2019structural}, research into their interpretability has primarily focused on two main areas: (1) task-level inspection and (2) internal mechanisms exploration in conjunction with code structure.


\citet{karmakar2021pre,karmakar2023inspect} first construct diagnostic tasks to discover to what extent CodePTMs learn about specific aspects of source code. 
\citet{troshin2022probing} also employ probing tasks to verify models are aware of code syntactic structure.
The follow-up research delves into the inner workings,
drawing inspiration from previous research targeting NLP models~\citep{clark2019does,jawahar2019bert,yenicelik2020bert} and protein models~\citep{vig2021bertology}.
A primary focus is analyzing models' attention within Transformer layers,
utilizing the structural information of code to provide additional signals for analysis.
Specifically, 
\citet{wan2022capture} conduct qualitative analyses to evaluate how CodePTMs interpret code structure, 
discovering that attention aligns strongly with the code's syntax.
Subsequently, quantitative characterization of the code structure learned by models is also established by linking attention weights to AST nodes~\citep{zhu2022catprobing}.
Probing experiments also point out that CodePTMs can induce entire ASTs~\citep{jose2023astprobe}.

Attention analysis also highlights CodePTMs' propensity to prioritize specific types of tokens and statements, 
notably keywords and data-relevant statements.
Based on these findings, input codes can be simplified for the model's lightweight application~\citep{zhang2022diet}.
More recent research has highlighted how the lexical, 
syntactic, and structural properties of code are distributed across different model layers,
which could pave the way for more efficient fine-tuning strategies for code models via layer freezing~\citep{shi2023telly}.

\subsubsection{Evaluating the Robustness and Safety of CodePTMs}

Just like their NL counterparts, 
CodePTMs might not be resistant to changes in the input and, 
thus, are potentially susceptible to adversarial attacks~\citep{wang2022measure} and perturbations~\citep{sun2024rethink}. 
In such scenarios, 
the robustness of these models requires careful investigation~\citep{wang2023recode}. 
\citet{yefet2020adv} utilize gradient-based methods to slightly perturb the input code to force a given trained model to make an incorrect prediction.
For robust training, 
program transformations with preserved semantics are also employed~\citep{henkel2022semantic}.
Considering the ``naturalness'' of textual perturbation,
\citet{zhou2022attack} pioneer an example generation strategy that adversarially transforms inputs to make victim models produce wrong outputs, 
balancing both natural semantic and operational semantics.

Beyond exploiting misleading instructions~\citep{wu2023deceptprompt},
Attacks on CodePTMs can also be based on the structural information of code. CodeAttack~\citep{jha2023codeattack} is a representative black-box attack method that leverages code structure to generate imperceptible adversarial code samples, achieving higher attack success rates than direct applications of adversarial attacks in NLP~\citep{li2020bertattack}.
It has also been used to explore vulnerabilities in less common PLs, such as code entities in R~\citep{zhao2024r}.
\citet{zhang2023black} exploit the uncertainty in CodePTMs' outputs, using it to guide the search for adversarial examples through variable name substitution.
Later on, traversing the ASTs in different ways to construct adversarial samples has also been adopted as a strategy to test models' sensitivity to input variations~\citep{sun2023evaluating}.

In parallel,
backdoor attacks have also gained attention alongside the advancement of neural code intelligence. 
A backdoor-attacked CodePTM can behave as usual on benign examples but will generate pre-defined malicious outputs when injected with inputs embedded with backdoor triggers~\citep{schuster2021poison,li2023poison}.
Subsequent developments include the creation of more stealthy backdoors~\citep{yang2023stealthy},
as well as the proposal of multi-target backdoors that simultaneously aim at code understanding and generation tasks~\citep{li2023multitarget}.

\begin{TakeawayBox}
{\textit{Takeaways}}
\begin{itemize}[itemsep=2pt,topsep=3pt,parsep=0pt]
    \item[(1)] {Applying pre-trained transformers to code represents a groundbreaking initiative,
    addressing the previously encountered dilemma of having to model each task from scratch. Moreover, it demonstrates that pre-training on a vast corpus of unlabeled code followed by task-specific fine-tuning can boost performance across all downstream tasks.}
    \item[(2)] {Various code features have been leveraged during pre-training to enhance models' perception of code structure. However, this explicit modeling of structures is not a free lunch; alterations to the model's inner mechanisms due to explicit structural modeling make it challenging for CodePTMs to generalize across different tasks.}
    \item[(3)] {Leveraging code structure offers an additional perspective for interpreting and analyzing CodePTMs. Compared to their NL counterparts, researchers are becoming increasingly aware of utilizing these structures to analyze model behavior. As we step into the era of LLMs, such research is still in its nascent stage.}
\end{itemize}
\end{TakeawayBox}

\begin{figure*}[t]
   \begin{center}
   {\includegraphics[width=0.995\linewidth]{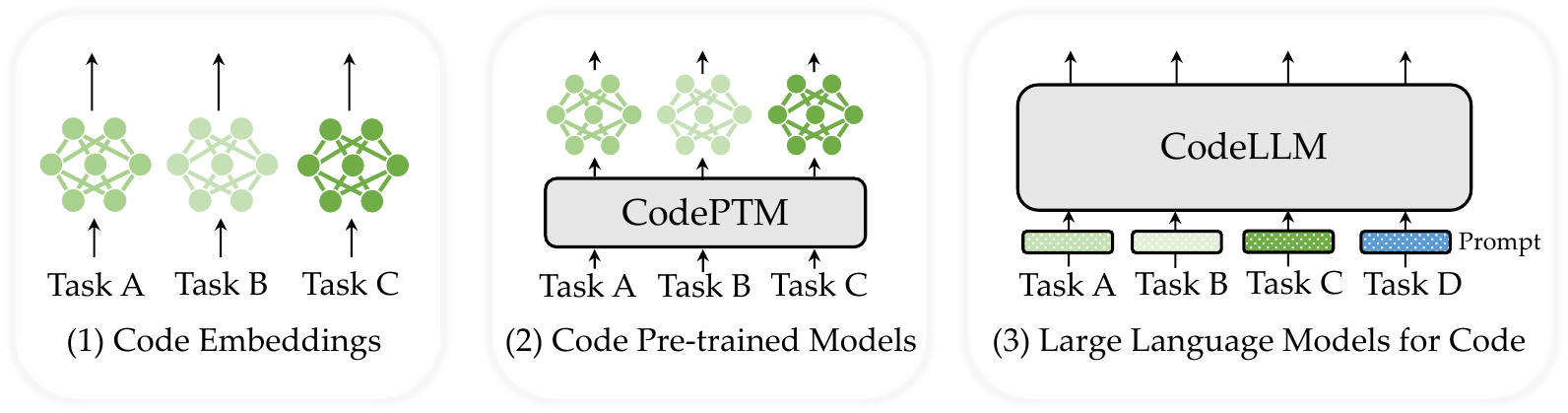}} 
   \end{center}
\caption{Schematic illustration of different paradigms of applying language models for code to downstream applications. $\coloredblock{code1}$ $\coloredblock{code2}$ $\coloredblock{code3}$ indicate different code downstream tasks (\textit{e.g.}, Defect Detection, Code Translation, NL2Code) and $\coloredblock{reason}$ indicates tasks that can be addressed by employing code-based solutions (\textit{e.g.}, mathematical reasoning).}
\label{fig:app_paradigm}
\end{figure*}

%% file: tables/codeptm-comparisons.tex
\begin{table*}[ht]
	\centering
 	\caption{An overview of Code Pre-trained Models’ architecture and pre-training strategies, along with whether these models leverage code structure information during the pre-training phase. 
    Due to space limitations, we abbreviate some of the training strategies, 
    with detailed descriptions provided in Table~\ref{tab:codeptm_objs}.}
	    \begin{tabular}{l|l|c|c|c|c}
		\toprule[1.2pt]
            Architecture & Models  & Struct. & Base & Strategy & Size\\
	    \midrule
            \multirow{7}{*}{Encoder} 
            & CuBERT~\citep{kanade2020learning}    & \redcross & - & MLM + NSP & 340M\\
            & CodeBERT~\citep{feng2020codebert}    & \redcross & RoBERTa & MLM + RTD & 125M\\
	    & GraphCodeBERT~\citep{guo2021graphcodebert}                    & \greencheck & CodeBERT & MLM + Edge Pred. + Node Align. & 125M\\
            & SynCoBERT~\citep{wang2022syncobert}                           & \greencheck & CodeBERT & MMLM + IP + TEP + MCL & 125M\\
            
            & CODE-MVP~\citep{wang2022codemvp}         & \greencheck & GraphCodeBERT & FGTI + MCL + MMLM & 125M\\
            & SCodeR~\citep{li2022soft}               & \greencheck& UniXcoder & Soft-Labeled Contrastive Pre-training & 125M\\
            & DISCO~\citep{ding2022disco}                                & \greencheck & - & MLM + NT-MLM + CLR & 110M\\
            \midrule
            \multirow{9}{*}{Enc-Dec}                              
            & PLBART~\citep{ahmad2021unified}                               & \redcross & - & Denoising Pre-training & 140M/406M\\
	    & CodeT5~\citep{wang2021codet5}                                 & \greencheck & - & MSP + IP + MIP + Bimodal Generation & 60M/220M/770M\\
            & PyMT5~\citep{clement2020PyMT5}                                & \redcross & - & MSP & 374M\\
	    & UniXcoder~\citep{guo2022unixcoder}                            & \greencheck & -  & MLM + ULM + MSP + MCL + CMG & 125M\\
            & NatGen~\citep{chakraborty2022natgen}                          & \greencheck & CodeT5 & Code Naturalization & 220M\\
            & TreeBERT~\citep{jiang2021treebert}   & \greencheck & - & TMLM + NOP & 210M\\
            & ERNIE-Code~\citep{chai2023erniecode}                          & \redcross & mT5 & SCLM + PTLM & 560M\\
            & CodeExecutor~\citep{liu2023codeexecutor}                      & \redcross & UniXcoder & Code execution + Curriculum Learning & 125M \\
            & LongCoder~\citep{guo23longcoder}                      & \redcross & UniXcoder & CLM & 150M \\
            \midrule
            \multirow{3}{*}{Decoder} 
            & GPT-C~\citep{Svyatkovskiy2020gptc}                            & \redcross & - & CLM & 366M\\
            & CodeGPT~\citep{lu2021codexglue}                               & \redcross & - & CLM & 124M\\
            & PyCodeGPT~\citep{zan2022cert}                               & \redcross & GPT-Neo & CLM & 110M\\
	    \bottomrule[1.2pt]
		\end{tabular}
    \label{table:CodePTMs}
\end{table*}

%% file: sections/4-LLM.tex
\section{The LLM Era: A New Frontier}
\label{sec:codellms}

The domain of code intelligence has been significantly transformed by the swift advancement of Large Language Models (LLMs)~\citep{openai2023gpt4,geminiteam2023gemini,groeneveld2024olmo},
signaling the dawn of a new era and introducing new opportunities~\citep{bommasani2021foundation}.
In addition to displaying emergent abilities~\citep{wei2022emergent},
typical LLMs such as PaLM~\citep{chowdhery2022palm}, 
LaMDA~\citep{chen2022lamda} and BLOOM~\citep{workshop2023bloom},
inherently possess competent coding capabilities. 
This innate ability stems from their pre-training data, 
which is often a diverse mixture containing a considerable amount of code corpus. 
For instance, commonly used datasets like ROOTS~\citep{laurenon2022roots} and the Pile~\citep{gao2020pile} corpora include significant portions of code data; 
Pile contains 95.16GB of GitHub data out of 800GB, 
while ROOTS comprises 163GB out of 1.6TB.
This substantial inclusion of code enables these models to learn and understand programming concepts, syntax, and semantics,
thus equipping them with the ability to generate and interpret code across various PLs and tasks.

\subsection{Large Language Models for Code}
\label{sec:codellm:model}

To harness the power of LLMs to further propel the field of code intelligence, 
CodeLLMs
have emerged.
Benefiting from the advantages of being successors to general LLMs,
CodeLLMs are naturally equipped with: 
\uppercase\expandafter{\romannumeral1}. access to vast and high-quality data for training~\citep{gao2020pile,laurenon2022roots,ms2024redstone}; 
\uppercase\expandafter{\romannumeral2}. modern positional encoding and interpolation techniques~\citep{press2022train,su2023roformer} for tackling longer sequences.
\uppercase\expandafter{\romannumeral3}. efficient strategies for training deployment~\citep{shazeer2019fast,dao2022flashattention,ainslie2023gqa};
\uppercase\expandafter{\romannumeral4}. the resources and weights from mature open-source LLMs~\citep{zhang2022opt, touvron2023llama, vicuna2023, 2023internlm}.
Supported by these technological advancements, 
the design philosophy of CodeLLMs has, at a macro level,
evolved from the CodePTMs era in the following ways: 
\begin{enumerate}
    \item Architecture: Aside from a few cases~\citep{li2022alphacode,wang2023codet5plus} retaining the encoder part, 
    the majority of CodeLLMs have embraced decoder-only autoregressive models to better align with generative tasks.
    \item Training data: Compared to their widely adopted predecessor like CodeSearchNet~\citep{husain2019codesearchnet},
    the new emerging corpora have rapidly grown in size and the number of PLs covered, 
    as listed in Table~\ref{tab:pre-training-data}.
    \item Learning objectives: There is a shift away from explicitly learning code structural information. 
    Moreover, beyond left-to-right generation, some models are designed to learn infilling tasks~\citep{fried2023incoder} to support scenarios such as code completion.
\end{enumerate}
\input{tables/pre-training-data}


In terms of application paradigms, 
there is no longer a necessity to meticulously select annotated data for training from scratch or to engage in task-specific fine-tuning.
As illustrated in Figure~\ref{fig:app_paradigm}, 
the primary approach now leans towards leveraging prompt learning and providing relevant in-context demonstrations~\citep{li2023largelanguagemodelawareincontext}, 
which are usually non-invasive to models~\citep{zan2023large,sun2023promptcs}. 
Consequently, the approach to handling various code-related tasks has gradually evolved from the diversified forms mentioned in Section~\ref{sec:code-embd:task-overview} to predominantly generative methods~\citep{joshi2023repair}.



Within this ongoing evolutionary path, 
the contemporary CodeLLMs that have emerged can mainly be observed on the branches on the right side of Figure~\ref{fig:codelm-tree}.
In the ongoing discussion,
we will concentrate on some of the most representative models (and their derivatives),
subsequently offering a comparatively concise review of other models and the hallmark techniques they employ. 
As for all available CodeLLMs and their properties,
we present them in Table~\ref{table:CodeLLM}, 
aiming to offer a comprehensive overview of the current landscape.



\noindent\raisebox{-.185\height}{\includegraphics[height=1em]{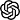}}~\textbf{Codex Series}.
The initial version of Codex is a GPT model~\citep{brown2020GPT3} fine-tuned on publicly available code from GitHub. 
Literature~\citep{chen2021evaluating} describes a 12B version of Codex, 
fine-tuned on a 159 GB corpus of deduplicated, filtered Python code, 
which later presumably evolved to be the \textit{code-cushman-001} within the OpenAI APIs.
In 2022, 
OpenAI initiated the development of a new Codex variant, termed \textit{code-davinci-002},
which is perceived as a larger model (175B) further trained on a blend of text and code data~\citep{fu2022gptroadmap},
and subsequently fine-tuned on instructions~\citep{ouyang2022training}.

The advent of Codex has had a profound impact on the development of code intelligence, 
pioneering and demonstrating the potential of building large-scale language models specialized for code.
However, it is slightly regrettable that the exact sizes, corpora, 
and certain training details of the Codex series remain undisclosed to the public.
Nevertheless, 
it is undeniable that Codex is a landmark,
and researchers also believe that it has played an essential role in the evolution of other seminal OpenAI models, 
such as \textit{text-davinci-003} and \textit{ChatGPT}~\citep{fu2022gptroadmap}.
After Mar, 2023,
The Codex series is no longer publicly available, 
access is limited to API credit applications via the researcher access program\footnote{\url{https://openai.com/form/researcher-access-program}}.

\noindent\raisebox{-.185\height}{\includegraphics[height=1em]{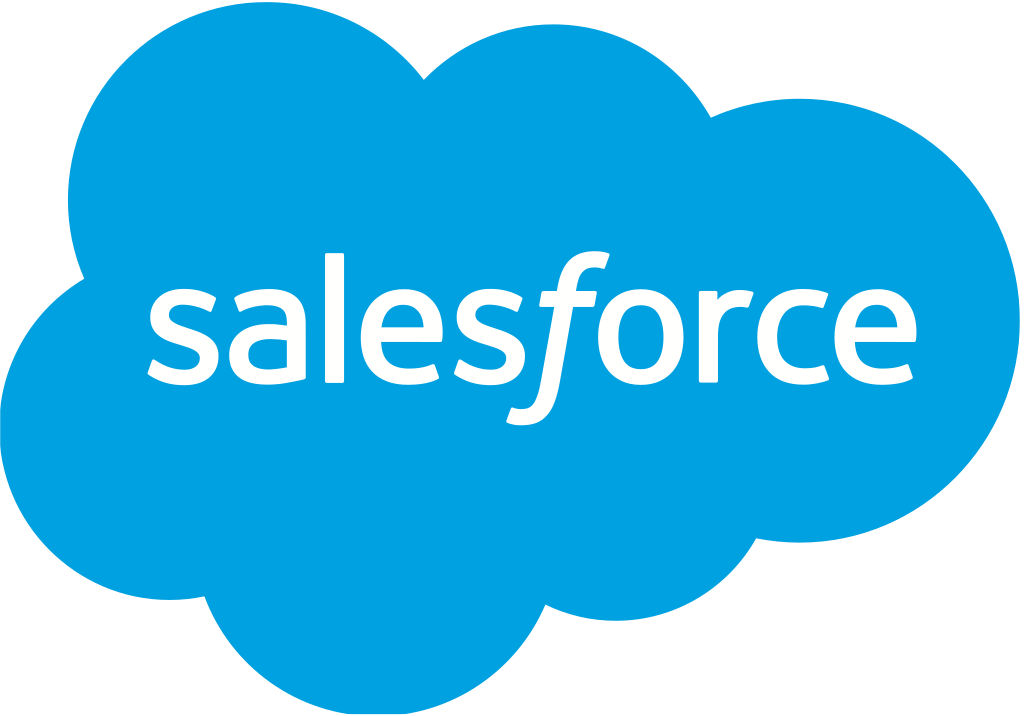}}~\textbf{CodeGen Series}. 
The CodeGen series~\citep{nijkamp2022codegen} is one of the first open-source CodeLLMs for code generation, 
featuring models ranging from 350M to 16.1B.
It stands as one of the initial efforts utilizing autoregressive transformers to learn from both NL and code data,
with next-token prediction language modeling training objective.
CodeGen innovatively proposes a multi-turn code generation approach, 
where a user interacts with the model by progressively providing requirements in natural language and then receives responses in the form of ``subprograms.''
In terms of training, CodeGen first learns general knowledge on The Pile, 
followed by training on a subset of Google BigQuery (which includes 6 PLs) to obtain CodeGen-Multi. 
The model is then further trained on BigPython to acquire a Python-oriented model, termed CodeGen-mono. 
At the time, these models demonstrated capabilities close to those of CodeX. 
Moreover, CodeGen serves to fill a niche between larger and smaller language models for code, 
partially marking a transition from CodePTMs to CodeLLMs.

CodeGen2~\citep{nijkamp2023codegen2} represents a comprehensively upgraded iteration that delves into the training of CodeLLMs from four main aspects: model architectures, learning methods, training objectives, and data distributions. 
Compared to its predecessor, this version imposes stricter control over the quality of training data and is trained on mixed objects of causal language modeling and span corruption. 
The resulting model is capable of infilling and supports a broader range of PLs, marking a significant advancement of contemporary CodeLLMs.

The most recent model release is CodeGen2.5\footnote{\url{https://blog.salesforceairesearch.com/codegen25/}},
which adopts multi-epoch training on StarCoderData~\citep{li2023starcoder} and employs span corruption for training. 
This model exemplifies the principle that with a robust data recipe—specifically, running multiple epochs and utilizing data augmentation—a relatively smaller CodeLLM (7B) can be on par with its larger predecessors (typically greater than 15B),
paving the way for subsequent enhancement of models from the perspective of training data.

Interestingly, 
given the robust performance and flexible size options of the CodeGen series, it has also been used as an initializer for other general-purpose LLMs. 
A case in point is MOSS~\citep{sun2023moss},
which is initialized with CodeGen-mono-16B and then further pre-trained on Chinese tokens, as well as samples drawn from the Pile and BigQuery.

\input{tables/codellms}
\noindent\raisebox{-.185\height}{\includegraphics[height=1em]{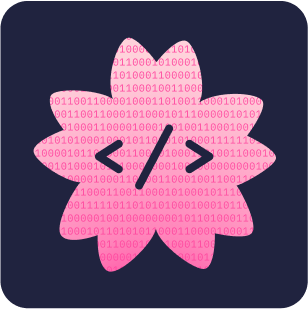}}~\textbf{BigCode Models}. 
BigCode\footnote{\url{https://www.bigcode-project.org/}} represents an open-scientific collaboration focused on
CodeLLMs,
aiming to provide the research community with full insight into the development process. 
It has made significant contributions in various aspects, including data, models, evaluation, and ethics. 
The CodeLLMs developed by BigCode are among the first to adopt the fill-in-the-middle (FIM) objectives~\citep{bavarian2022fim} (also known as causal masking objective utilized by InCoder~\citep{fried2023incoder}, which is a 6.7B model specialized in infilling).
For using CodeLLMs,
it is a common necessity to generate or insert contextually appropriate content based on given snippets of code.
However,
due to the dependency relationship with code,
solely relying on ``predicting next token'' falls short of capturing these complexities.
FIM addresses this challenge by innovatively segmenting the code into three parts, 
then shuffling these segments and reconnecting them with special tokens.
This strategy aims to enhance the model's pre-training by incorporating a ``fill-in-the-blank'' that goes beyond causal modeling, 
enabling an understanding of bidirectional context by considering code dependencies.

One of the most representative models,
StarCoder~\citep{li2023starcoder} has a size of 15.5B, equipped with infilling capabilities and the capacity for efficient generation~\citep{shazeer2019fast}.
In terms of its training data, 
StarCoder employs heuristic filtering, manual inspection, 
and cleaning processes to compile the StarCoderData, 
which includes 86 programming languages.
The format of this dataset encompasses text-code pairs, GitHub issues, Jupyter notebooks, and GitHub commits.
SantaCoder~\citep{allal2023santacoder} is an earlier variant with 1.1B size.
It shares the same architecture as StarCoder but is exclusively trained on Python, Java, and JavaScript.
Additionally, this family of models also includes fine-tuned versions on conversation data to act as coding assistants~\citep{tunstall2023starchatalpha}.

Later developments release OctoCoder along with OctoPack~\citep{muennighoff2023octopack},
an instruction-tuned model created by fine-tuning StarCoder on newly collected commit messages that resemble instructions (CommitPackFT) and the OpenAssistant (OASST) conversations dataset.
Furthermore, the data is also used to construct OctoGeeX based on CodeGeeX~\citep{zheng2023codegeex},
Both models demonstrate performance that rivals non-permissive models, 
showcasing the effectiveness of instruction tuning and specialized training. 
The most recently released StarCoder2~\citep{lozhkov2024starcoder} represents an even more capable version, 
utilizing training data that is 4 x larger and extends beyond to include notebooks from Kaggle,
GitHub pull requests,
and code documentation. 
Moreover, 
it uses data augmentation strategies to boost low-resource language performance by enhancing source code by pairing it with its LLVM~\citep{lattner2004LLVM} intermediate representation.

Beyond the success and impact of the aforementioned models,
FIM has also been widely adopted in subsequent research.
While \citet{bavarian2022fim} propose that FIM could be learned without harming the ability to do left-to-right generation,
some research holds the view that equipping the model with this infilling ability might not be a ``free lunch''~\citep{nijkamp2023codegen2}. 
For instance, models trained through FIM have been observed to sometimes struggle with determining the appropriate moments to cease infilling.
We believe that these conflicting views may arise because models must learn to balance the nuances of generating code linearly with the ability to jump back and forth to fill gaps as needed. 
Further evidence is required to delve deeper into this discussion.

\noindent\raisebox{-.185\height}{\includegraphics[height=1em]{figures/logos/Salesforce.com_logo.svg.png}}~\textbf{CodeT5+}. 
Unlike the previously mentioned models that are designed as decoder-only models,
CodeT5+~\citep{wang2023codet5plus} is a family of open-source CodeLLMs with encoder-decoder architecture, ranging from 220M to 16B.
This model not only scales up from its predecessor, CodeT5~\citep{wang2021codet5},
but also showcases refined architectural design and training objectives.
Architecturally, the encoder is tasked with encoding contextual representations, 
whereas the decoder is adept at generating diverse types of outputs.
Distinctively, 
CodeT5+ adopts a ``shallow encoder and deep decoder'' structure~\citep{li2022alphacode}, 
with both parts initialized using CodeGen and connected by cross-attention.
For pre-training,
the objectives employ a combination of span denoising and CLM~\citep{soltan2022alexatm,tay2023ul}, 
thereby equipping the models with the capability to learn code context representations and to reconstruct missing information at different levels, including code spans, partial programs, and complete programs.

In contrast to previous models that are treated as a single system across all tasks,
CodeT5+ offers the versatility to operate in encoder-only, decoder-only, and encoder-decoder modes to accommodate different downstream applications.
This flexibility significantly mitigates the issue of inter-task interference encountered in UniLM-style models~\citep{guo2022unixcoder,liu2020cuglm}.

For further improvement,
InstructCodeT5+ is created as an instruction-tuned variant to align with NL instructions, 
employing a strategy revolving around using synthetic instruction-following prompts~\citep{taori2023alpaca,codealpaca2023chau,wang2023selfinstruct}.

\noindent\raisebox{-.185\height}{\includegraphics[height=1em]{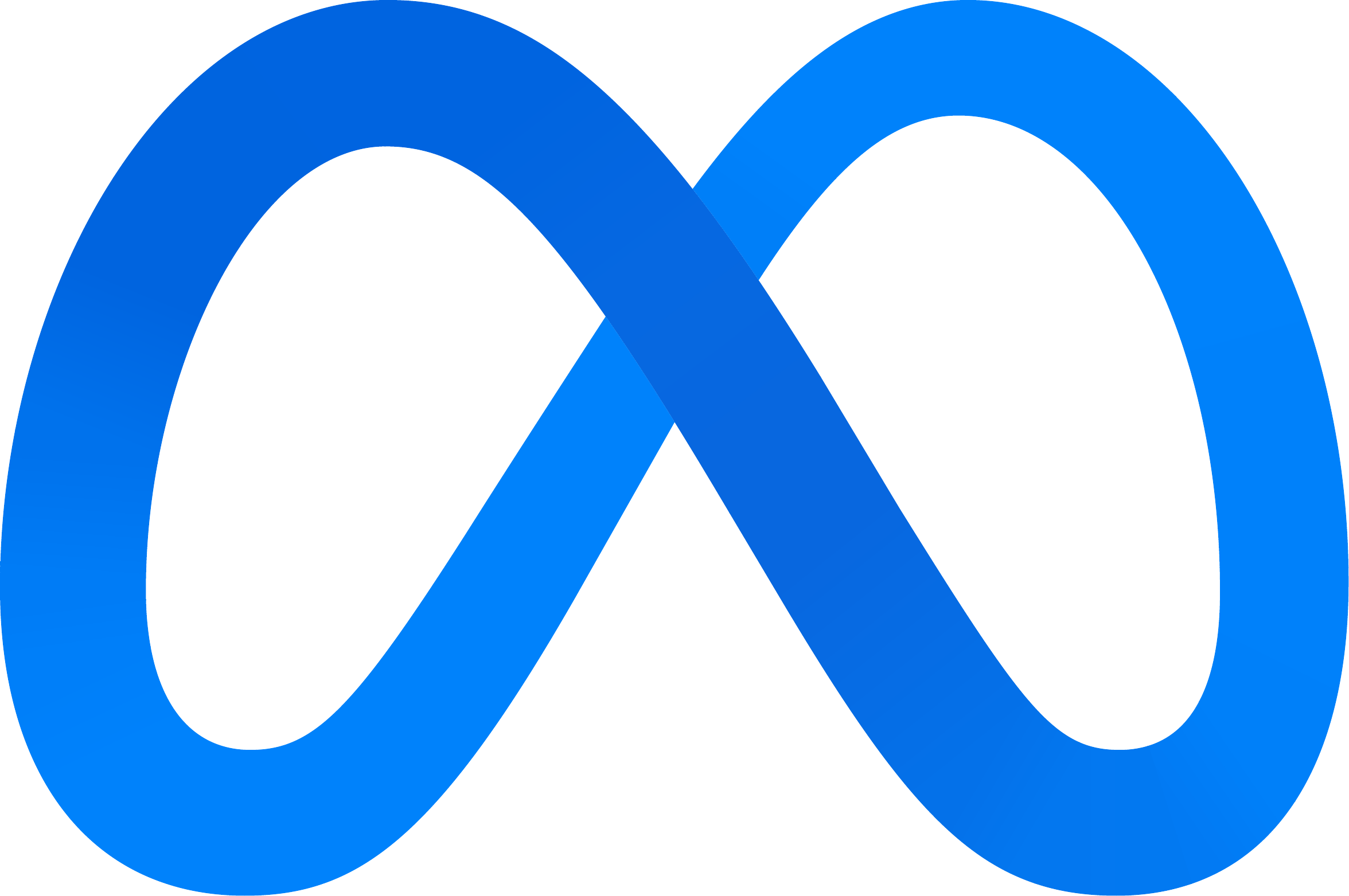}}~\textbf{CodeLLaMA}. 
Shortly after the debut of LLaMA2~\citep{touvron2023llama2} in July 2023,
CodeLLaMA~\citep{roziere2023code} is swiftly released as a family of foundation models for code generation, 
covering model sizes 7B, 13B, and 34B (with 70B later announced in Jan 2024).
Diverging from aforementioned models that are trained exclusively on code from scratch,
CodeLLaMA is derived from LLaMA2 through training on an additional 500B code tokens.
The 7B, 13B, and 34B versions have also been trained with FIM, allowing them to insert code into existing code.
For the 70B version,
an extra stage of long context fine-tuning was introduced, 
leveraging position interpolation~\citep{chen2023extending} to extend the context length from the initial 4K, 
as was standard with the LLaMA2 model, to an extended 16K. 
Experiments have shown that the pursuit of tackling longer sequences might slightly hurt the performance on shorter sequences. 
Still, it boosts the model's ability to generate meaningful content in tasks like long code completion~\citep{guo23longcoder}.

In addition to the foundation models, 
Meta has provided two additional variations, 
namely 
(1) CodeLLaMA - Python, a language-specialized variant of CodeLLaMA that has been further fine-tuned on 100B Python code.
(2) CodeLLaMA - Instruct, 
which is trained on self-instruct~\citep{wang2023selfinstruct} dataset created by prompting LLaMA2 with programming problems, as well as data aimed at improving safety and helpfulness.
These models have exerted a profound and positive impact on the domain, 
being widely applied in the development of derivatives~\citep{gou2023tora,yue2024mammoth,azerbayev2024llemma} and various instruction tuning practices~\citep{liu2024tuning,song2024code}. 

Beyond the pursuit of enhanced capacity, CodeLLaMA takes a forward-looking step toward responsible AI and safety. 
It rigorously compares itself against other CodeLLMs from the perspectives like truthfulness~\citep{lin2022truthfulqa}, 
toxicity~\citep{hartvigsen2022toxigen}, 
and bias~\citep{jwala2021bold} in generated code and text. 
Through empirical evaluations, 
CodeLLaMA demonstrates the possibility of achieving high coding performance without compromising harmlessness.

\noindent\raisebox{-.185\height}{\includegraphics[height=1em]{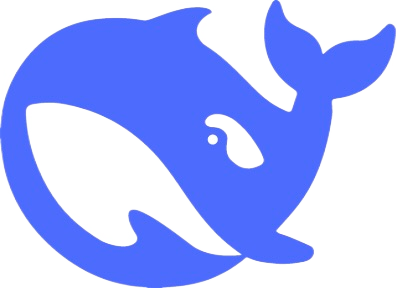}}~\textbf{DeepSeek-Coder}.
DeepSeek Coder~\citep{deepseekcoder2023} represents a range of open-source CodeLLMs with sizes varying from 1.3B to 33B, 
trained from scratch on a curated code corpus.
In terms of data composition, 
besides utilizing a filtering process for GitHub data similar to the StarCoderData rules~\citep{li2023starcoder}, 
it uniquely constructs repository-level code data to enhance the model's capability for cross-file completion within repositories. 
Moreover,
it incorporates a small code-unrelated Chinese corpus to aid the model in understanding instructions in Chinese. 
The model is trained on 2T tokens comprising 87 PLs,
with combining objectives of next token prediction and FIM.
Further,
it employs linear scaling to extend the context window~\citep{chen2023extending} to 16K, 
supporting repository-level code training. 
In terms of performance, it has achieved stunning results on tasks such as NL2Code and code completion, 
being considered one of the strongest open-source CodeLLMs currently available.

Beyond the models built from scratch, 
DeepSeek also provides versions that continue pre-training on general LLMs, termed DeepSeekCoder-v1.5 which can be viewed as a branch of DeepSeek LLM~\citep{deepseekai2024deepseek}.
Compared to its code-exclusive counterpart, 
this variant, 
while witnessing a slight decrease in coding capabilities, 
performs better in tasks for math and natural language.
Later, 
it played a vital role in constructing mathematical models~\citep{shao2024deepseekmath}.


\noindent\raisebox{-.185\height}{\includegraphics[height=1em]{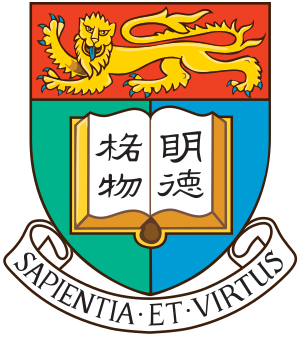}}~\textbf{Lemur}.
Contrasting with the prevailing open-source CodeLLMs that predominantly focus on code-centric optimization,
Lemur~\citep{xu2023lemur} represents a novel endeavor to lay the groundwork for LLM-based agents.
To achieve this, 
LLMs must possess not only robust coding capabilities to ensure precise grounding in the relevant environments~\citep{huang2022language,ahn2022can} but also the capacity to comprehend human intentions, reasoning, and planning.
Therefore, Lemur advocates for harmonious integration of language understanding and coding proficiencies.

Similar to CodeLLaMA, 
the model is built by continually pre-training on LLaMA2.
However, in terms of data composition, 
it employs a 90B corpus with a 10:1 ratio of code to text,
followed by fine-tuning with instructions from diverse sources. 
Lemur can achieve stunning results on both NL benchmarks, \textit{e.g.}, MMLU~\citep{hendrycks2021mmlu}, 
BigBench~\citep{suzgun2023bbh}
as well as on code-related benchmarks, 
including multilingual code generation~\citep{cassano2022multiple}, SQL~\citep{yu2018spider}, and data science~\citep{lai2022ds1000}.
Compared to other LLMs that exhibit a disparity between NL and code capabilities, 
Lemur stands out with a balanced skillset, 
achieving the highest overall performance when averaged across a variety of tasks.
Moreover,
Lemur uniquely excels in practical agent tasks, making significant strides in tool usage, self-debugging, following complex instructions, and navigating partially observable environments~\citep{wang2024mint}.
Balancing NL and code capabilities is gaining increasing attention, and recently released Crystal~\citep{tao2024crystal} excels with a multi-phase pretraining strategy that integrates both domains effectively.

Beyond the representatives above, 
the community has witnessed a blossom of interesting and solid works,
as illustrated by the right branch of Figure~\ref{fig:codelm-tree}. 
These contributions mainly revolve around
(1) additional pre-training on general LLMs,
(2) instruction-tuned variants, 
(3) advanced tool-use capability, 
and (4) efficiency enhancements.

Regarding CodeLLMs that are built from additional pre-training, CodeGeeX2~\citep{zheng2023codegeex} represents the second iteration of the CodeGeeX model lineage.
\czr{Semantic repetition}
Unlike its predecessor, which was trained from scratch, 
Diverging from its predecessor trained from scratch,
it is developed upon ChatGLM2~\citep{du2022glm,zeng2023glmb} with further pre-training on code tokens.
Similarly, Code-Qwen~\citep{bai2023qwen} follows a training approach akin to CodeLLaMA~\citep{roziere2023code}, using base model Qwen trained on a combination of text and code data as initialization
and then continuing to pre-train on code data.
This approach is mirrored in other models such as AquilaCode\footnote{\url{https://huggingface.co/BAAI/AquilaCode-multi}},
which also enhances base models with extra training on code corpora.
Recently released CodeGemma~\citep{team2024codegemma} denotes a compilation of lightweight models crafted through further code infilling training on Gemma~\citep{team2024gemma},
particularly adept at code completion and generation from provided code prefixes or suffixes.

Researchers also apply diverse instruction-tuning strategies to CodeLLMs as well.
CodeAlpaca~\citep{codealpaca2023chau,taori2023alpaca} is initially built as an instruction-following LLaMA model for code generation. 
WizardCoder~\citep{luo2023wizardcoder} is constructed by fine-tuning StarCoder~\citep{li2023starcoder} using Evol-Instruct~\citep{xu2023wizardlm} and ChatGPT feedback seeded by CodeAlpaca dataset~\citep{codealpaca2023chau}
WaveCoder~\citep{yu2023wavecoder} enhances CodeLLMs through an innovative instruction tuning process.
It employs an LLM-based generator-discriminator framework to produce a wide array of high-quality instruction data for multiple code-related tasks, 
focusing on improving data quality and task diversity for fine-tuning.
Recently,
DolphCoder~\citep{wang2024dolphcoder} also adopts diversified tuning strategies.
It begins by leveraging multiple chain-of-thought~\citep{wei2022chain} responses to the same instruction and then combines the tasks of code generation and code evaluation in the form of natural language generation. 
Likewise,
MoTCoder~\citep{li2024motcoder} utilizes a modular approach for instruction tuning.
It segments complex coding tasks into logical sub-modules, 
guiding models to first outline and then implement these sub-modules.




Regarding tool use~\citep{cai2023large,wang2024tools},
ToRA~\citep{gou2023tora} targets building tool-integrated agents, 
addressing complex mathematical reasoning. 
To achieve it, ToRA is trained upon the collected interactive trajectories of invoking tools. 
Nevertheless,
its coding ability is mainly
limited to Python.
In the same vein,
MammoTH~\citep{yue2024mammoth} concentrates on equipping off-the-shelf LLM with Python-integrated reasoning abilities (will be discussed in Section~\ref{sec:pal}). 
It utilizes a hybrid composition of data generated from intermediate steps when reasoning with NL or code for further pre-training on LLaMA.
This approach is expected to unleash both program-aided and NL-centric power in mathematics.
Concurrent with Lemur's success in harmonizing text and code capabilities,
a new foundational model, 
Symbol-LLM~\citep{xu2023symbolllm}, 
expands the scope of code capabilities to encompass the entire range of symbol-centric capabilities, complemented by an external symbolic solver.
This extension broadens the application scope of LLMs to more intricate scenarios beyond code generation, 
such as neuro-symbolic reasoning.


As for efficiency-optimized CodeLLMs,
phi-1~\citep{gunasekar2023phi1} distinguishes itself through its compact size and the unique approach of utilizing high-quality, ``textbook-quality'' data for training. 
It demonstrates the efficacy of quality over quantity in data selection and the potential for smaller code models to compete with or outperform larger counterparts.
Factors influencing the quality of code data have also been explored in subsequent studies~\citep{jain2023llmassisted}.
When it comes to CodeLLMs that can run on consumer-level hardware, 
Stable Code~\citep{stable-code-3b} (based on Stable LM~\citep{stablelm}) and DeciCoder~\citep{DeciFoundationModels} adopt a more compact and concise design. 
These models uphold a certain performance standard while allowing users to deploy them locally, 
enhancing the accessibility of code generation to a broader range of users.

In the realm of application frameworks, 
CodeTF~\citep{bui2023codetf} initially provides an interface for both training and inference,
facilitating the integration of CodeLLMs into practical applications. 
This framework aims to make it easier for developers to leverage the power of CodeLLMs in efficiency and functionality of software development processes. 
MFTCoder (CodeFuse)~\citep{liu2023mftcoder} presents a multi-task fine-tuning framework specifically designed for CodeLLMs. 
It supports the efficient tuning and deployment of a broad spectrum of models, 
enabling developers to adapt and optimize these models for various coding tasks and challenges swiftly.
For example,
CodeFuse-CodeLLaMA is a model created by further training of CodeLLaMA through MFT,
which achieves performance surpassing that of GPT-4 on the HumanEval benchmark. 



\subsection{Learning with Execution Feedback}
\label{sec:codellm:rl}

Another pathway to further enhance CodeLLMs involves integrating Reinforcement Learning (RL), 
which incorporates non-differentiable reward signals into the training process.
Unlike approaches represented by reinforcement learning from human feedback (RLHF) that utilize human preferences~\citep{christiano2017human,ouyang2022training,bai2022training}, 
The inherently compilable and executable nature of codes allows compilers or interpreters to automatically generate precise feedback.
Such endeavors were initiated in the era of CodePTMs,
as COMPCODER~\citep{wang2022compcoder} harnesses the compilability signals to optimize both the generator (\textit{e.g.}, CodeGPT~\citep{lu2021codexglue}) and the discriminator (MLPs) via RL strategies.

As the capability for code generation improves, 
using RL in code training becomes increasingly flexible.
CodeRL~\citep{le2022coderl} exploits the code unit test signals in both training and inference stages and uses RL to optimize the model.
PPOCoder~\citep{shojaee2023ppocoder} combines CodeLLMs with Proximal Policy Optimization~\citep{schulman2017proximal} for code generation.
RLTF~\citep{liu2023rltf} is another novel online RL framework,
which uses unit test feedback of multi-granularity for refinement. 
RLCF~\citep{jain2023coarsetuning} further enhances a CodeLLM by incorporating feedback from a grounding function, 
which assesses the quality of generated codes.
Pangu-Coder2~\citep{shen2023pangucoder2} introduces an RRTF (Rank Responses to align Test \& Teacher Feedback) framework,
aimed at steering the model towards producing higher-quality code achieved by synergistically using test signals and human preferences as combined feedback.
ExeDec~\citep{shi2023exedec} innovates in decomposing tasks into execution subgoals, 
improving compositional generalization through tackling complex tasks step-by-step.
In a similar vein,
recently released StepCoder~\citep{dou2024stepcoder} innovates by breaking complex tasks into a curriculum of subtasks, tackling code generation's exploration and optimization challenges.
RLEF~\citep{gehring2024rlef} further expands RL approaches by grounding LLM generations in iterative execution feedback, enabling multi-turn self-correction and optimization.

\input{tables/task-nl2code}

\subsection{Advancements in NL2Code}
\label{sec:codellm:nl2code}

In the era of LLMs,
the ability of NL2Code has leaped forward, 
with machine learning models now truly capable of assisting professional developers through crafting accurate code snippets based on human intent.
This holds a tantalizing promise of ``programming in natural language''.
Moreover, the role of NL2Code has also transitioned; 
it has transcended the initial function as merely a downstream coding task and has become a pivotal metric for evaluating the capabilities of LLMs.
Here we first discuss the shift in evaluation paradigms and then focus on extensively utilized benchmarks and their derivatives.

\subsubsection{The Shift in Evaluation Metrics}

\noindent\textbf{Limitations of Match-based Approaches}.
In the Pre-LLM era, 
code generation capabilities were primarily benchmarked by matching samples against a reference solution, 
using metrics like (smoothed) BLEU scores~\citep{papineni2002bleu,lin2004orange, eghbali2203crystal}.
Nevertheless, 
in addition to the lingering issues already identified in NLG systems~\citep{post2018call,sai2022eval},
BLEU-based evaluations struggle to capture semantic features specific to code~\citep{mikhail2023out}. 
Although variants like CodeBLEU~\citep{ren2020codebleu} propose several semantic modifications based on code structure,
a fundamental problem remains: match-based metrics are unable to fully represent the broad and complex space of programs that are functionally equivalent to the reference solutions.
This dilemma extends to other evaluation metrics~\citep{lin2004rouge,denkowski2014meteor,popovic2015chrf,tran2019ruby} as well.
Consequently, \citet{chen2021evaluating} advocate the use of execution-based evaluation to measure functional correctness for NL2Code tasks instead.

\noindent\textbf{The Rise of Execution-based Evaluation}.
For evaluating the functional correctness of a generated code snippet,
the most reliable approach is to examine if it can be successfully executed and passes a set of unit tests, 
a method commonly employed in software engineering's test-driven development.
\pak is initially designed for assessing pseudocode-to-code translations~\citep{kulal2019pseudocode}.
Generating $k$ code samples for each problem,
a problem deemed solved if any of the samples pass, and it reports the total fraction of problems solved. 
However, due to the high variance, \citet{chen2021evaluating} refine it into a more stable metric as delineated in Equation~\ref{eq:pak}.

\begin{align}
\pak &:= \mathop{\mathbb{E}}_{\text{Problems}} \left[ 1 - \frac{{\binom{n-c}{k}}} {\binom{n}{k}} \right]
\label{eq:pak}
\end{align}

This revised method involves producing $n \geq k$ samples for each task, 
counting correct samples $c \leq n$ that pass unit tests successfully,
thereby deriving an unbiased estimator.
\pak represents a milestone in NL2Code evaluation,
which transcends the limitations of earlier metrics incapable of accurately reflecting functional correctness and has become a standard practice in all subsequent benchmarks.
Building on this, contemporary work has also started incorporating elements like self-debugging and candidate generation into the evaluation pipeline~\citep{li2024doce}.

\subsubsection{Commonly Used Benchmarks and Evaluations}


Herein,
we offer detailed descriptions and discussions for HumanEval~\citep{chen2021evaluating} and MBPP~\citep{austin2021program}, 
two of the most widely utilized contemporary benchmarks, 
along with MultiPL-E~\citep{cassano2022multiple} benchmark derived from them.
Subsequently,
we provide a bird’s eye view of other datasets and the new benchmarks based on them.
Subsequently, we present an overview of additional datasets and introduce benchmarks developed for specific purposes.

\noindent$\bullet$~\textbf{HumanEval}.
HumanEval was initially released in conjunction with Codex~\citep{chen2021evaluating},
comprising 164 manually crafted Python coding problems.
These problems are validated using test cases~(with an average of 7.7 tests per problem) to evaluate the code generated by CodeLLMs,
typically in a zero-shot setting.
The necessity for these problems to be manually crafted stems from the fact that the majority of contemporary CodeLLMs are trained on a large fraction of GitHub data,
which already contains solutions to some ready-made programming challenges, such as those collected from Codeforces\footnote{\url{https://codeforces.com/}}, LeetCode\footnote{\url{https://leetcode.com/}} and CodeChef\footnote{\url{https://www.codechef.com/}}.
HumanEval includes problems of varying difficulty levels, ranging from basic string and array manipulations to relatively complex data structures and algorithms.
Each problem provides a function signature, clarifying the input/output format to aid the model in understanding the requirements.
Since its release, 
HumanEval has been widely adopted by the community as a primary tool for evaluating code generation capabilities, and subsequently, it has been extended with the goals of (1) covering more programming languages and (2) constructing more robust evaluation.

Multilingual HumanEval~\citep{athiwaratkun2023mbxp} introduces a scalable automated framework capable of converting datasets from Python into variants for 12 different languages.
In contrast to the automated approach, 
HumanEval-X~\citep{zheng2023codegeex} is another multilingual extension constructed by the authors of CodeGeeX.
Similar to the original HumanEval, it is manually crafted, 
involving a rewrite of prompts, canonical solutions, and test cases for C++, Java, JavaScript, and Go.

Unlike the strategies of expanding the number of languages to achieve comprehensive evaluation, 
researchers are aware that the quantity and quality of test cases in problems can be inadequate for fully assessing functional correctness~\citep{liu2023code}.
This gap may inadvertently lead to flawed code being considered correct due to the paucity of testing.
To address this issue, 
HumanEval\textsuperscript{+} is developed by extending the unit tests of HumanEval, augmenting the scale of test cases by 80 times. 
This substantial increase in testing has proven to catch significant amounts of previously undetected incorrect code.

\noindent$\bullet$~\textbf{MBPP}. \textbf{M}ostly \textbf{B}asic \textbf{P}rogramming \textbf{P}roblems~\citep{austin2021program} stands as another widely recognized benchmark for assessing the performance of CodeLLMs in Python programming, especially under few-shot scenarios.
Unlike HumanEval, 
which features varying levels of difficulty, 
MBPP is tailored to be solvable by entry-level programmers.
It contains 974 short Python functions, each accompanied by an English description, 
a predefined function signature, 
and three manually written test cases for validation.
As for its compilation,
MBPP amalgamates a vast collection of crowd-sourced questions and a smaller manually edited and verified subset.
The difficulty of these questions ranges from simple numerical manipulations to those requiring some external knowledge (\textit{e.g.}, the definition of the Fibonacci sequence).

In a similar vein,
MBBP has also been expanded to accommodate multilingual scenarios, 
aiming to offer a more comprehensive and diverse evaluation.
One of the most notable extensions is MBXP (Most Basic X Programming Problems, where X = Java, Go, Ruby, etc)~\citep{athiwaratkun2023mbxp},
constructed in a parallel to Multilingual HumanEval.
Moreover, it has been extended to cover other code-related scenarios,
encompassing zero-shot code translation, prompt perturbation test, code insertion, and code summarization.

\begin{figure*}[t]
   \begin{center}
   {\includegraphics[width=\linewidth]{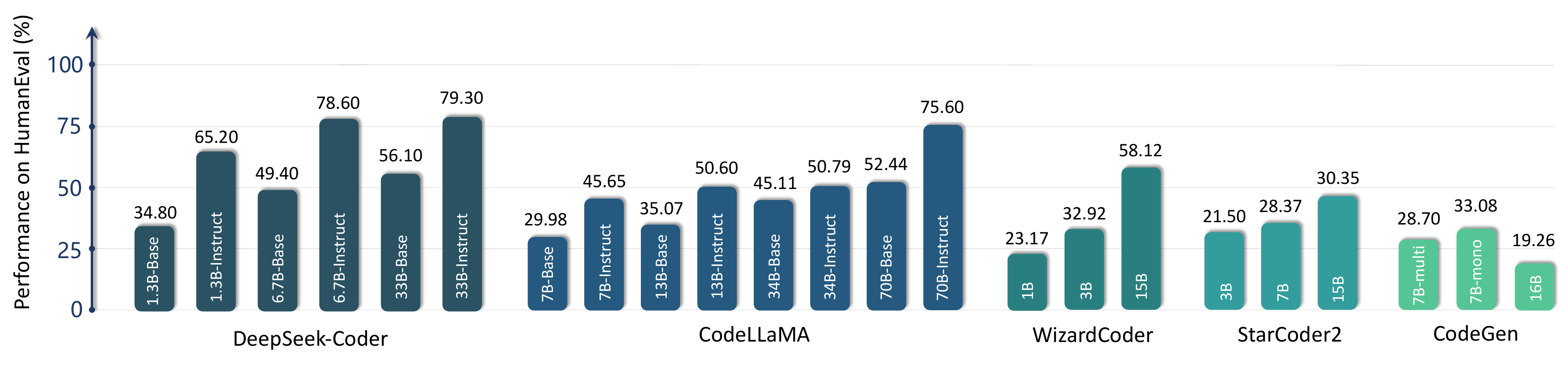}} 
   \end{center}
\caption{The Pass@1 performance of selected open-source CodeLLMs on HumanEval, showing the result for each model across different sizes and versions.}
\label{fig:nl2code-eval}
\end{figure*}

\noindent$\bullet$~\textbf{MultiPL-E}. 
MultiPL-E~\citep{cassano2022multiple} emerges as a new benchmark system tailored for multilingual scenarios, 
constructed upon the two benchmarks discussed previously. 
It extends their scope by translating them into 18 additional programming languages chosen according to TIOBE\footnote{\url{https://www.tiobe.com/tiobe-index/}} rankings and GitHub usage frequencies.
To accommodate the diverse languages,
it features a containerized sandbox environment,
which compiles programs (if necessary), runs them with test cases and appropriate timeouts, 
and categorizes each output as successful, syntax error, etc.
Furthermore, 
MultiPL-E stands out for its scalability and extensibility, 
offering a streamlined process for incorporating new benchmarks and languages, thereby minimizing the need for manual intervention.

Beyond the aforementioned benchmarks, 
more works regarding evaluation are emerging alongside the surge in interest in CodeLLMs,
broadly categorized into three types: 
(1) Those tailored for specific NL2Code scenarios, 
like Code-Contests~\citep{li2022alphacode} and APPS~\citep{hendrycks2021apps} designed for programming competitions. 
(2) The conversion of general benchmarks into testbeds for code generation,
such as MathQA-Python~\citep{austin2021program} and selected BigBench~\citep{srivastava2023bb} problems.
(3) Those that consolidate, refine, and repurpose existing benchmarks, 
such as L2CEval~\citep{ni2023l2ceval},
which has been developed as a standardized benchmark combining semantic parsing, math reasoning, and Python programming. 
Representative benchmarks for various purposes are shown in Table~\ref{tab:nl2code_full}, 
and a comprehensive list of benchmarks currently available is presented in Table~\ref{tab:nl2code_tasks}.
We will further discuss the opportunities and challenges of benchmarking CodeLLMs in Section~\ref{sec:future:eval}.
Additionally, theoretical frameworks for evaluating CodeLLMs are also emerging, such as applying category theory~\citep{lin2024catcode} to express the structural aspects of NL and code.

To provide readers with an intuitive understanding of the current CodeLLMs' capabilities,
we display the HumanEval performance of some open-source representatives in Figure~\ref{fig:nl2code-eval}.
It is evident that, 
in addition to the positive correlation between model size and performance, 
models of the same size that have been fine-tuned with instructions significantly outperform their base versions. The benefits that instruction fine-tuning brings to CodeLLMs are more pronounced than those observed in general LLMs. 
We hold the view that this gap is likely because human intentions, compared to code comments or documents, are more intricate.
Research has shown that models like \textit{code-davinci-002} achieve top performance in benchmarks, 
yet their outputs may not always align with human expectations~\citep{chung2022scaling,fu2022gptroadmap}.
Therefore, for developing practical CodeLLMs,
dedicating more effort to instruction fine-tuning could be more cost-effective than merely expanding the model size. 

\subsubsection{Strategies for Enhanced Code Generation}
Building upon our discussions above,
here we explore a variety of innovative strategies that have significantly advanced the capabilities of code generation.

\noindent\textbf{Decoding-Enhanced}.
In terms of decoding strategies,
\citet{zhang2023planning} propose that conventional decoding algorithms may not be the best choice for code generation, 
and utilize lookahead search to guide future token choices.
Self-planning~\citep{jiang2023selfplanning} improves code generation by first creating a high-level plan to break down complex tasks into simpler steps, and then generating code for each step. 
Self-infilling~\citep{zheng2023selfinfilling} enhances code generation by combining sequential context generation with infilling, 
which allows for more controlled output.
\citet{yang2023chainofthought} break down the problem into multiple steps and utilize programming hints as support, allowing lightweight models to generate better programs.
Considering the reuse of existing code,
AceCoder~\citep{li2023acecoder} searches for programs with similar
requirements as guidance generation for improved code generation.
Taking the structural information into account,
\citet{li2023structured} ask the model to first generate the program's structure (\textit{e.g.}, loops and branches) and then implement the code.

\noindent\textbf{Feedback-Driven}.
As discussed in Section~\ref{sec:codellm:rl},
execution feedback represents an external signal that code can innately utilize. 
Employing/building test cases~\citep{arteca2022nessie,shi2022natural,bareiß2022code} within code generation processes emerges as a viable means to enhance the reliability of the generated code. 
CodeT~\citep{chen2023codet} exemplifies this category of methods by unleashing the model's inherent capability to automatically generate unit tests,
ensuring the consistency of outputs through broader tests. 
Similarly, TiCoder~\citep{lahiri2023interactive} enhances code reliability by constructing an augmented set of examples to cover a wider range of possible user inputs. 
LEVER~\citep{ni2023lever} enhances NL2Code generation by integrating execution feedback provided by trained verifiers, harnessing both the generative and discerning capability.
Further,
ALGO~\citep{zhang2023algo} advances in using LLM-generated oracles for algorithmic program synthesis.
Beyond NL2Code tasks, 
this concept of leveraging unit test feedback can also be applied to other tasks,
such as multilingual code translation or code search,
ensuring the functional equivalence of translated content~\citep{roziere2022leveraging} or retrieved code~\citep{han2023intervention}.

\noindent\textbf{NL-Directed}.
Utilizing NL information represents another pathway. 
Given an instruction, 
\citet{zhang2023coder} evaluate the likelihood of the given instruction given the generated programs, 
thereby improving code quality through reranking. 
DocPrompting~\citep{zhou23iclr} enhances the code generation process by explicitly retrieving relevant document pieces from a pool based on user intent. 
Very recently,
AlphaCodium~\citep{ridnik2024code} proposes a test-based, iterative process combining understanding NL and code generation.


In addition to the insights gleaned thus far,
there are ongoing studies within the field that may hold significant implications for building CodeLLMs in the future,
such as better tokenization and efficiency improvements.
These topics will be elaborated upon in Section~\ref{sec:future:tokenizer} and Section~\ref{sec:future:efficient}.

\subsection{Preference Learning for CodeLLMs}
\label{sec:preference}
Supervised fine-tuning on pairs of NL instructions and code snippets has been shown to enhance a model’s coding abilities effectively. 
To further enable CodeLLMs to prioritize stronger outputs over weaker ones consistently, 
preference learning~\citep{rafailov2023dpo} plays a crucial role.
PLUM~\citep{zhang2024PLUM} leverages automated test case generation to evaluate the functional correctness of model outputs, dynamically collecting preference data during training to guide models toward syntactically and functionally correct solutions. 
Similarly, Code-Optimise~\citep{gee2024codeopt} incorporates both functional correctness and runtime efficiency into its preference signals, using self-generated data to optimize multi-objective performance.

Later, CodeDPO~\citep{zhang2024codedpo} combines correctness and execution efficiency into a unified preference learning framework. It employs a novel self-generation-and-validation mechanism to iteratively refine the ranking of code snippets and test cases, enabling large-scale, automatic dataset construction without relying on external resources. 
CodeLutra~\citep{tao2024codelutra} learns from both successful and failed code attempts to iteratively improve model performance. By constructing preference pairs from self-generated outputs and leveraging a dual-loss function that integrates preference learning with SFT, CodeLutra enhances correctness while reducing non-executable outputs.
Recently, DSTC~\citep{liu2024dstc} introduces a framework that relies solely on self-generated code and tests to construct preference pairs. By employing a minimax selection mechanism and code-test concatenation, DSTC reduces the impact of low-quality tests while leveraging direct preference learning algorithms like DPO and KTO~\citep{ethayarajh2024kto}.




\begin{TakeawayBox}
{\textit{Takeaways}}
\begin{itemize}[itemsep=2pt,topsep=3pt,parsep=0pt]
    \item[(1)] {The advent of CodeLLMs has been revolutionary, indicating a new learning paradigm.
Whether as variants of general LLMs or built from scratch, 
they demonstrate exceptional capabilities not seen in their predecessors, achieved through larger sizes, premium code data, 
task-oriented training objectives, and intricately designed tuning strategies.}
    \item[(2)] {The integration of RL with CodeLLMs offers a promising avenue for enhancing code generation through the use of non-differentiable reward signals, 
such as compiler feedback and unit test results. This allows for the precise and automatic generation of feedback, 
overcoming the high costs associated with learning from human preferences.}
    \item[(3)] {Whether in terms of model capabilities or the diversity of tasks,
NL2Code has experienced unparalleled expansion during this period.
Further,
despite a multitude of efforts to devise intricate matching-based evaluation metrics, 
the LLM era has seen a shift in evaluating code generation tasks toward reliance on execution to assess functional correctness.}
    \item[(4)] Compared to the various ingenious methods mentioned in Section~\ref{sec:codeptms:adapt}, aimed at enhancing code-related tasks through specific training, the approaches for handling downstream tasks have converged towards generative methods. Essentially, these are predominantly based on prompting, which is non-invasive in nature.
\end{itemize}
\end{TakeawayBox}

%% file: tables/pre-training-data.tex
\begin{table}[ht]
	\centering
            \caption{Representative open-source corpora for pre-training CodeLLMs with their sizes (measured by GB/TB for disk size, or measured by M in number of files) and the number of PLs they cover.} 
	    \begin{tabular}{l|c|c}
		    \toprule
	        \textbf{Dataset}  & \textbf{Size} & \textbf{\# PLs.} \\
	        \midrule
                BigQuery (GitHub data)~\href{https://cloud.google.com/blog/topics/public-datasets/github-on-bigquery-analyze-all-the-open-source-code}{[link]} & 2.8M & 6\\ 
	        CodeSearchNet~\citep{husain2019codesearchnet}~\href{https://github.com/github/CodeSearchNet}{[link]} & 20GB / 6.5M & 6\\
                GitHub Code~\href{https://huggingface.co/datasets/codeparrot/github-code}{[link]}  & 1TB / 115M & 32\\
                BigPython~\citep{nijkamp2022codegen} & 217GB & 1 \\
                The Stack~\citep{kocetkov2022stack}~\href{https://www.bigcode-project.org/docs/about/the-stack/}{[link]} & 6.4TB / 546M & 358\\
                StarCoderData~\citep{li2023starcoder}~\href{https://huggingface.co/datasets/bigcode/starcoderdata}{[link]} & 783GB / 207M & 86 \\
                The Stack v2~\citep{lozhkov2024starcoder}~\href{https://huggingface.co/datasets/bigcode/the-stack-v2}{[link]} & 67.5TB & $>$ 600 \\
			\bottomrule
		\end{tabular}
    \label{tab:pre-training-data}
\end{table}

%% file: tables/codellms.tex
\begin{table*}[ht]
	\centering
  \caption{An overview of CodeLLMs categorized based on their architecture, along with their parameter size, base model (if any), vocabulary size, context length, training objectives, data scale used for training (measured by K/B/T in number of tokens, or measured by GB for disk size), and their public availability. Due to space limitations, we do not differentiate between various versions of CodeLLMs and their bases.
  For models built upon a base, the data scale refers to the size of the corpora used during additional pre-training.}
	    \begin{tabular}{l|l|c|c|c|c|c|c|c}
		\toprule
            \textbf{Arch.} & \textbf{Model Name} & \textbf{Size} & \textbf{Base} &\textbf{ Vocab.} & \textbf{Context} & \textbf{Training Objs.} & \textbf{Data Scale} & \textbf{Public} \\
	    \midrule
            \multirow{3}{*}{Enc-Dec} 
            & AlphaCode~\citep{li2022alphacode} & \begin{tabular}[c]{@{}c@{}}284M/1.1B/\\2.8B/8.7B/\\41.1B\end{tabular} & - & 8.0K & 1536+768 & MLM+CLM & \begin{tabular}[c]{@{}c@{}}354B/590B/\\826B/1250B/\\967B\end{tabular} & \redcross \\  
     & CodeT5+~\citep{wang2023codet5plus}   & \begin{tabular}[c]{@{}c@{}}220M/770M/\\2B/6B/16B\end{tabular} & CodeGen & 50.0K & 2048+2048 & MSP+CLM+CL & 51.5B & \greencheck\\  
            \midrule
            \multirow{32}{*}{\begin{tabular}[c]{@{}c@{}}Decoder / \\CodeLLMs\end{tabular}}
            & Codex~\citep{chen2021evaluating} & 2.5B/12B & - & 50.3K & 4K & CLM & 100B/159GB & \greencheck\\
            & CodeParrot~\citep{codeparrot} & 125M/1.5B & - & 32.8K & 1K & CLM & 26B/50GB &      \greencheck\\ 
            & PolyCoder~\citep{xu2022systematic} & 160M/0.4B/2.7B & - & 50.3K & 2K & CLM &  39B/254GB & \greencheck\\ 
            & CodeGen~\citep{nijkamp2022codegen} & \begin{tabular}[c]{@{}c@{}}350M/2.7B/\\6.1B/16.1B\end{tabular} & - & 50.0K & 2K & CLM & 1.2T & \greencheck\\
            & PaLM-Coder~\citep{chowdhery2022palm} & 8B/62B/540B & PaLM & 256K & 2K & CLM & 7.75B &      \redcross\\  
            & InCoder~\citep{fried2023incoder} & 1.3B/6.7B & - & 50.3K & 2K & FIM & 52B/159GB & \greencheck\\  
            & PanGu-Coder~\citep{christopoulou2022pangucoder} & 317M/2.6B & - & 42K & 1K & CLM+MLM & 387B/147GB & \redcross\\ 
            & SantaCoder~\citep{allal2023santacoder} & 1.1B & - & 49.2K & 2K & FIM & 236B/268GB & \greencheck\\ 
            & phi-1~\citep{gunasekar2023phi1} & 350M/1.3B & - & 50.0K & 2K & CLM & 7B & \greencheck\\  
            & CodeGeeX~\citep{zheng2023codegeex} & 13B & - & 52.2K & 2K & CLM & 850B & \greencheck\\  
            & CodeGen2~\citep{nijkamp2023codegen2} & 1B/3.7B/7B/16B & - & 50.0K & 2K & MLM+CLM & 400B & \greencheck\\  
            & StarCoder~\citep{li2023starcoder} & 15.5B  & - & 49.2K & 8K & FIM & 1T/815GB & \greencheck\\  
            & CodeAlpaca~\citep{codealpaca2023chau} & 7B/13B & LLaMA & 32.0K & 4K & CLM & 20K  & \greencheck\\ 
            & WizardCoder~\citep{luo2023wizardcoder} & \begin{tabular}[c]{@{}c@{}}1B/3B/7B/\\13B/15B/34B\end{tabular} &  StarCoder & 32.0K & 2K & CLM & 78k & \greencheck\\
            & AquilaCode~\citep{AquilaCode2023} & 7B & Aquila & 100.0K & 2K & CLM & - & \greencheck\\  
            & CodeGeeX2~\citep{zheng2023codegeex} & 6B & ChatGLM2 & 65.0K & 8K & CLM & 600B & \greencheck\\  
            & CodeLLaMA~\citep{roziere2023code} & 7B/13B/34B/70B & LLaMA2 & 32.0K & 4K & FIM & 500B & \greencheck\\ 
            & ToRA-Code~\citep{gou2023tora} & \begin{tabular}[c]{@{}c@{}}7B/13B/34B\end{tabular} & CodeLLaMA & 32.0K & 2K & Min. NLL & 223K & \greencheck\\
            & MAmmoTH-Coder~\citep{yue2024mammoth} & 7B/13B/34B & CodeLLaMA & 32.0K & 2K & CLM & 260K & \greencheck\\
            & Code-Qwen~\citep{bai2023qwen} & 7B/14B &  Qwen & 152.0K & 8K & CLM & 90B & \greencheck\\ 
            & CodeFuse~\citep{liu2023mftcoder} & \begin{tabular}[c]{@{}c@{}}1.3B/6.5B/\\13B/34B\end{tabular} &  Multiple & 100.9K & 4K & CLM & 1.6TB & \greencheck\\  
            & CodeShell~\citep{CodeShell2023} & 7B & - & 70.1K & 8K & CLM & 500B & \greencheck\\  
            & Lemur~\citep{xu2023lemur} & 70B & LLaMA2 & 32.0K & 4K & CLM & 90B & \greencheck\\  
            & DeepSeekCoder~\citep{deepseekcoder2023} & \begin{tabular}[c]{@{}c@{}}1.3B/5.7B/\\6.7B/33B\end{tabular} & - & 32.0K & 16K & FIM & 2T & \greencheck\\  
            & Symbol-LLM~\citep{xu2023symbolllm} & 7B/13B & LLaMA2 & 32.0K & 4K & CLM & 2.25GB & \greencheck\\ 
            & Stable Code~\citep{stable-code-3b} & 3B & - & 50.3K & 16K & FIM & 1.3T & \greencheck\\ 
            & DeciCoder~\citep{DeciFoundationModels} & 1B/6B & - & 49.2K & 2K & FIM & 446B & \greencheck\\ 
            & StarCoder2~\citep{lozhkov2024starcoder} & 3B/7B/15B & - & 49.2K & 16K & FIM & 900B/3TB & \greencheck\\ 
            & CodeGemma~\citep{team2024codegemma} & 2B/7B & Gemma & 250K & 8K & FIM & 1T & \greencheck\\ 
            & CodeStral~\citep{mistral2024codestral} & 22B & - & 32.0K & 32k & CLM+FIM & - & \greencheck\\ 
            & DeepSeekCoderV2~\citep{zhu2024deepseek} & 16B/236B & DeepSeekV2 & 100K & 128K & CLM+FIM & 10.2T & \greencheck\\ 
            & Crystal~\citep{tao2024crystal} & 7B & - & 32K & 2K & CLM & 1.4T & \greencheck\\ 
            & Yi-Coder~\citep{young2024yi} & 1.5B/9B & Yi & 64K & 128K & CLM & 2.4T & \greencheck\\ 
            & OpenCoder~\citep{huang2024ppencoder} & 1.5B/8B & - & 96.6K & 8K & CLM & 2.5T & \greencheck\\ 
		\bottomrule
		\end{tabular}
    \label{table:CodeLLM}
\end{table*}

%% file: tables/task-nl2code.tex
\begin{table*}[htb]
    \centering
    \caption{An overview of the representative NL2Code benchmarks categorized according to task purpose, along with the number of programming languages they cover and brief descriptions.
    The complete benchmarks are listed in Table~\ref{tab:nl2code_full}.}
    \begin{tabular}{l|l|c|c|l}
        \toprule[1.2pt]
        \textbf{Purpose} & \textbf{Dataset} & \textbf{Date} & \textbf{\# PLs.} & \textbf{Description}  \\
        \midrule
        \multirow{2}{*}{Open Domain} 
        & CONCODE~\citep{iyer2018mapping}~\href{https://github.com/sriniiyer/concode}{[link]} & 2018 & 1 & a dataset with over 1M examples consisting of Java classes\\
        & ODEX~\citep{wang2023odex}~\href{https://github.com/zorazrw/odex}{[link]} & 2023 & 1 & an open-domain execution-based NL to Python code generation dataset  \\
        \midrule
        \multirow{3}{*}{Code Exercise} 
        & HumanEval~\citep{chen2021evaluating}~\href{https://github.com/openai/human-eval}{[link]} & 2021 & 1 & a dataset of 164 handwritten programming problems with unit tests  \\
        & MBPP~\citep{austin2021program}~\href{https://github.com/google-research/google-research/tree/master/mbpp}{[link]} & 2021 & 1 & a dataset containing 974 short Python programs \\
        & BIG-Bench~\citep{srivastava2023bb,suzgun2023bbh}~\href{https://github.com/google/BIG-bench}{[link]} & 2023 & - & a benchmark containing over 12 tasks can be solved by coding \\ 
        \midrule
        \multirow{2}{*}{Competitions}
        & APPS~\citep{hendrycks2021apps}~\href{https://github.com/hendrycks/apps}{[link]} & 2021 & 1 & a benchmark including 10K less-restricted problems for code generation  \\
        & CodeContests~\citep{li2022alphacode}~\href{https://github.com/deepmind/code_contests}{[link]} & 2022 & 3 & a dataset specifically for competitive programming problems  \\
        \midrule
        \multirow{3}{*}{Multilingual}
        & MBXP~\citep{athiwaratkun2023mbxp}~\href{https://github.com/amazon-research/mbxp-exec-eval}{[link]} & 2023 & 12 & a benchmark to evaluate code generation for 12 programming languages \\
        & HumanEval-X~\citep{zheng2023codegeex}~\href{https://github.com/THUDM/CodeGeeX}{[link]} & 2023 & 4 & a benchmark of 164 code problems for evaluating multilingual models  \\
        & MultiPL-E~\citep{cassano2022multiple}~\href{https://github.com/nuprl/MultiPL-E}{[link]} & 2022 & 18 & a parallel, multilanguage benchmark for NL2Code generation \\       
        \midrule
        \multirow{3}{*}{Data Science} 
        & JuICe~\citep{agashe2019juice}~\href{https://github.com/rajasagashe/juice?tab=readme-ov-file}{[link]} & 2018 & 1 & a corpus of 1.5M examples with a curated test set of 3.7K instances  \\
        & DSP~\citep{chandel2022jupyt5}~\href{https://github.com/microsoft/DataScienceProblems}{[link]} & 2022 & 1 &  a collection of 1K problems curated from 306 pedagogical notebooks  \\
        & DS-1000~\citep{lai2022ds1000}~\href{https://ds1000-code-gen.github.io/}{[link]} & 2023 & 1 & a Python code generation benchmark of 1K data science problems  \\
        \midrule
        \multirow{3}{*}{Python Libs} 
        & PandasEval~\citep{zan2022cert}~\href{https://github.com/microsoft/PyCodeGPT}{[link]} & 2022 & 1 & a dataset consisting of 101 programming problems on Pandas library \\
        & NumpyEval~\citep{zan2022cert}~\href{https://github.com/microsoft/PyCodeGPT}{[link]} & 2022 & 1 & a dataset consisting of 101 programming problems on Numpy library  \\
        & TorchDataEval~\citep{zan2022private}~\href{https://github.com/microsoft/PyCodeGPT/tree/main/apicoder}{[link]} & 2022 & 1 & a dataset with 50 programming problems using the TorchData library \\
        \midrule
        \multirow{1}{*}{Multi-Turn} 
        & MTPB~\citep{nijkamp2022codegen}~\href{https://github.com/salesforce/CodeGen}{[link]} & 2023 & 1 & a benchmark containing 115 problem sets factorized into multi-turn prompts  \\
        \midrule
        \multirow{1}{*}{Command Line} 
        & NL2Bash~\citep{lin2018nl2bash}~\href{https://github.com/TellinaTool/nl2bash}{[link]} & 2018 & 1 & a corpus of 9K English-command pairs covering over 100 Bash utilities  \\
        \midrule
        \multirow{1}{*}{AI4Science} 
        & BioCoder~\citep{tang2023biocoder}~\href{https://biocoder-benchmark.github.io/}{[link]} & 2023 & 2 & a benchmark to evaluate LLMs in generating bioinformatics-specific code  \\
        \bottomrule[1.2pt]
    \end{tabular}
    \label{tab:nl2code_tasks}
\end{table*}

%% file: sections/5-reasoning.tex
\section{Synergies in Machine Intelligence}
\label{sec:reasoning}

Following our detailed exploration centered on code generation and comprehension,
in this section, 
we delve into some synergies with other aspects of machine intelligence.
Specifically, 
we will discuss from three perspectives:
new reasoning paradigms based on code generation, 
mathematical abilities enhanced by code training,
and the multi-dimensional capabilities of language models expanded through code.

\subsection{Binding Code Generation with LLM Reasoning}
\label{sec:pal}
Reasoning constitutes a long-standing task in the domain of machine intelligence~\citep{sun2023reasoning}.
The surge in LLM enthusiasm has brought to light approaches exemplified by chain-of-thought (CoT) prompting~\citep{wei2022chain}, 
proposed as a novel form of in-context learning~\citep{dong2023survey} wherein the exemplar contains the rationales instead of merely an answer.
These methodologies,
which encourage LLMs to generate a series of intermediate steps toward a final solution~\citep{nye2022show}, 
have made stunning progress across a wide spectrum of textual and numerical reasoning benchmarks~\citep{kojima2022zscot,wang2023sc,fu2023complexcot,wang2023cok,chu2023cotsurvey}.

\noindent\textbf{Unlocking a New Reasoning Paradigm.}
CoT-based approaches do not serve as a silver bullet for solving all reasoning problems.
Beyond factuality~\citep{wang2023cok} and hallucinations~\citep{yin2023selfaware,ji2024halu} issues in LLMs,
given the inherent limitations language models face with complex arithmetic operations and managing large numbers~\citep{nogueira2021investigating,qian2023limitations},
LLMs are susceptible to logical and arithmetic mistakes in the calculation phase, 
despite the problem decomposition being correct.
In light of this dilemma, 
code generation offers a potential pathway for disentangling computation from reasoning.
Program-Aided Large Language Models (PAL)~\citep{gao2022pal}\footnote{The idea of integrating LLMs with an external PL interface was proposed by \citet{gao2022pal} and \citet{chen2022program} within the same timeframe. Based on the descriptions adopted in the literature we surveyed, we use the terms PAL and PoT interchangeably in this paper.} 
and Program of Thoughts (PoT)~\citep{chen2022program} employ LLMs to comprehend natural language problems and synthesize programs as intermediate reasoning steps.
Crucially, 
they delegate the execution of the final solution to a symbolic solver
(includes but not limited to a Python interpreter~\citep{zhou2023solving}),
thereby resolving the computational limitations of LLMs.

This strategy of decoupling computation from reasoning and language understanding,
has achieved great success across a wide range of datasets involving mathematical reasoning~\citep{schubotz2018mathqa,cobbe2021gsm8k,hendrycks2021MATH}, 
symbolic reasoning~\citep{srivastava2023bb,suzgun2023bbh}, 
and semi-structured understanding~\citep{chen2021finqa,chen2022convfinqa}.
For instance, bolstered by PAL,
leveraging a CodeLLM like Codex has significantly outperformed the results obtained by larger models (\textit{e.g.}, PaLM/Minerva) on math word problems, 
BIG-Bench Tasks, 
and financialQA datasets using NL reasoning chains.
This approach has gradually emerged as a new reasoning paradigm for solving all numerical-related problems.

\begin{figure}[t]
   \begin{center}
   {\includegraphics[width=\linewidth]{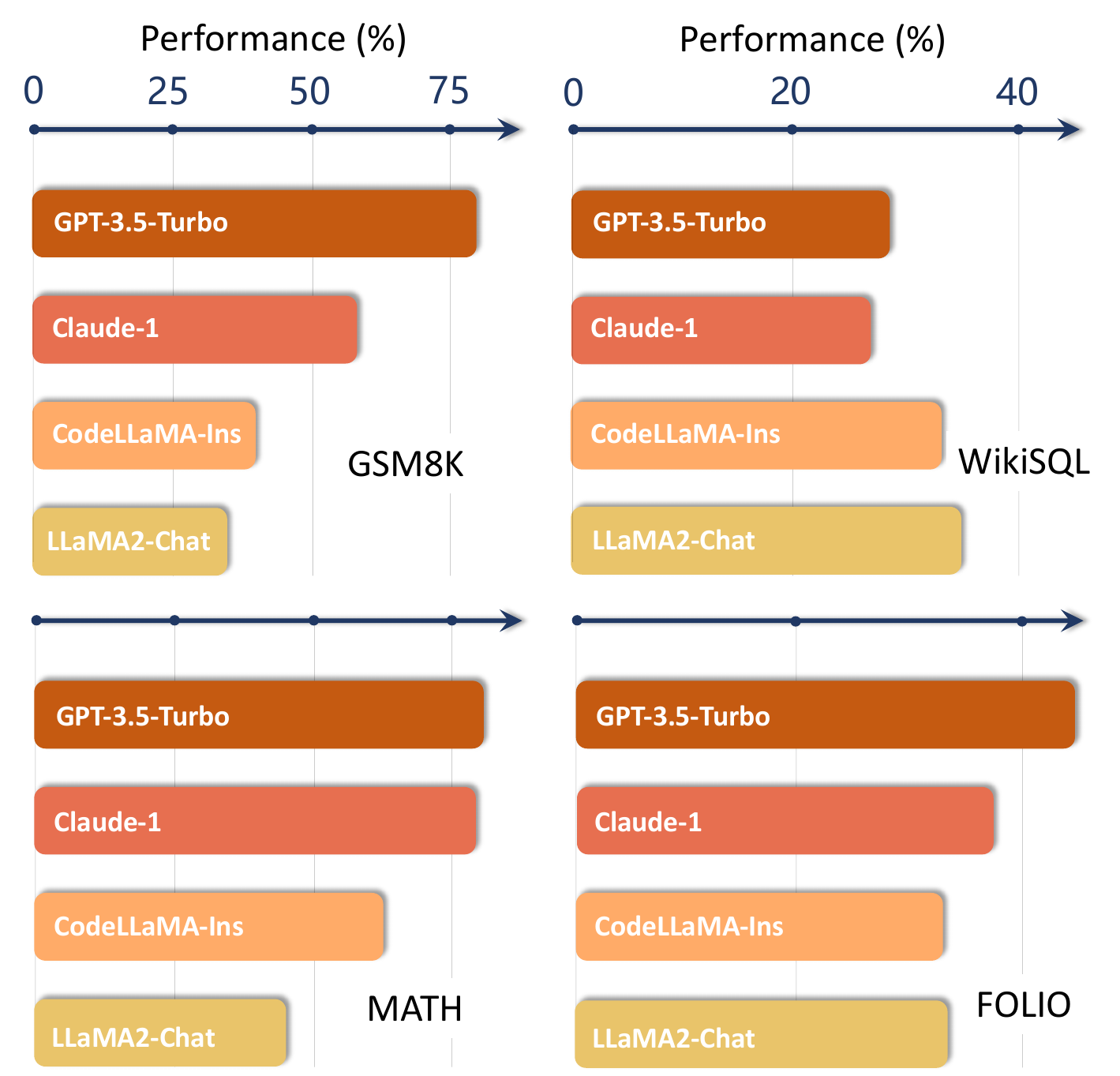}} 
   \end{center}
\caption{The performance of four LLMs using PAL on four different tasks.
Experimental details are shown in appendix~\ref{app:sources:pal}.
    }
\label{fig:pal_eval}
\vspace{-0.75em}
\end{figure}

\noindent\textbf{Laying Foundations for Broader Reasoning Scenarios.}
Emerging variations of PAL or code integration~\citep{wang2023mathcoder} into broader reasoning frameworks have also been developed.
For example, combining it with CoT to merge the strengths of both approaches through model selection offers enhanced flexibility and generalizability~\citep{zhao2023automatic}. 
Alternatively, 
integrating intermediate codes step with stochastic beam search to guide decoding presents another innovative direction~\citep{xie2023selfevaluation}.
Chameleon~\citep{lu2023chameleon} modularizes the function generation process and integrates it with other tools, 
such as web search engines, 
to tackle more complex reasoning scenarios.
For collaborative reasoning involving multi-model interactions, following the use of NL reasoning chains~\citep{yin2023eot}, code has also been leveraged as an alternative format for communication~\citep{sun2023corex,hong2023metagpt,chen2024natural}.
In addition to logic and arithmetic task~\citep{heyueya2023solving,xu2023are},
more recent research, Chain of Code~\citep{li2023coc}, 
encourages models to represent semantic tasks in the form of pseudocode.
This simple yet effective CoT extension enables models to explicitly identify and address undefined behaviors by simulating code execution.


Beyond offering performance gains in reasoning,
the deterministic execution of code also can be utilized for interpretability.
In pursuit of faithfulness, 
a notable example is Faithful CoT~\citep{lyu2023faithful}.
This approach converts a natural language query into a chain that interleaves natural language and a task-dependent symbolic language, 
such as Python or PDDL, helping us understand how the model arrives at the final answer. 

In sum,
integrating code generation with LLM reasoning
is a great breakthrough.
This new paradigm, which leverages models to interpret natural language and then generate programs as intermediate reasoning steps for execution, 
has transcended the entrenched limitations of traditional solutions.
Moreover, it lays the groundwork for exploring the synergy between generative models and neuro-symbolic AI in a broad range of scenarios~\citep{ye2023generating,lamalfa2024code,dinu2024symbolicai}.
Nevertheless, PAL is not a panacea and faces challenges such as model misinterpretation of questions and the fact that generated codes are not always error-free. 
Furthermore, the relationship between code utilization and enhancements in reasoning capabilities has not been fully elucidated~\citep{bi2023potwork}, making it an attractive and ongoing area of research. 

\input{tables/reasoning/reasoning.tex}

\subsection{Code Training Elicits Mathematical Capabilities}
Mathematics capability has historically been considered the Achilles' Heel of language models~\citep{luo2023wizardmath}.
However, with the advent of LLMs,
there has been a significant shift in this perception~\citep{ahn2024large}.  
Code training is believed to have played a pivotal role during this transformation.

\noindent\textbf{Unforeseen Benefits of Code Training.}
Before the buzz around LLMs, it was already discovered that language models trained on code could directly solve basic algebra~\citep{drori2021solvingla} or statistical problems~\citep{tang2021solving}. 
Early research at the onset of the LLM era find that \textit{code-davinci-002}, 
regardless of whether it utilizes chain-of-thought, significantly outperforms other models on mathematical tasks~\citep{qin2023chatgpt}. 
This observation has ignited the long-standing question among researchers: ``Does code training improve mathematical abilities?''
While the additional benefits from code training remain debatable, we collect some evidence to discuss the correlation between them.
\textit{Code-davinci-002} can easily demonstrate chain-of-thought capabilities in mathematical reasoning, achieving results far superior to those of \textit{text-davinci-001}, which did not undergo code training~\citep{wei2022chain}.
\citet{fu2023complexcot} also note that when the reasoning chain is longer,
\textit{code-davinci-002} is the best-performing model on mathematical benchmarks. 
Finally, the traditional next-token prediction objective usually captures ``local'' information,
while code often incurs longer dependencies and hierarchies,
such as referencing distant function definitions. 
This may indirectly cultivate the model's ability to understand more complex structures~\citep{fu2022gptroadmap}.

From the perspective of training, 
\citet{ma2023training} highlight that training with a mixture of code and text can significantly enhance LLM general reasoning capability, 
without almost any detriment to other aspects.
Recent data-centric research~\citep{toshniwal2024openmathinstruct1} has further demonstrated that the outcomes achieved from equivalent mathematical training on CodeLLMs (\textit{e.g.}, CodeLLaMA), 
markedly exceed the results compared to their base models or counterparts devoid of code training.
We left the further examination of this intriguing yet unverified hypothesis to future works.

\noindent\textbf{Building Math Models with Code.}
From a practical standpoint, 
CodeLLMs can serve as the foundation for developing sophisticated mathematical models. 
Llemma~\citep{azerbayev2024llemma} utilizes CodeLLaMA~\citep{roziere2023code} as the basis and is trained on a diverse mixture of math-related text and code to achieve remarkable mathematical capabilities, including the ability to use formal theorem provers.
Meanwhile, 
DeepSeekMath~\citep{shao2024deepseekmath} is developed based on DeepSeekCoder~\citep{deepseekcoder2023}, 
undergoes specialized training on web pages meticulously filtered for mathematical content. 
Both models exhibit superior performance on benchmarks such as MATH~\citep{hendrycks2021MATH},
STEM, and SAT, surpassing that of prior mathematics language models like Minerva~\citep{lewkowycz2022minerva}.
Beyond the unanimous choice of conducting continual pre-training on CodeLLMs,
comparative experiments from DeepSeek also suggest that code training can be a valuable prelude to math training, 
empirically showing the positive relationship between code training and mathematical capabilities.

Recently released InternLM-Math~\citep{ying2024internlmmath} incorporates code generated by proof assistants during its training process and innovatively interleaves coding processes within the problem-solving approach, 
presenting new state-of-the-art mathematical capabilities with the assistance of Python. 
Moreover, researchers have shown that code training enhances LLMs’ capabilities beyond mathematics, improving general problem-solving by fostering the ability to capture long-range dependencies and complex structures inherent in code~\citep{aryabumi2024code}.
To date,
code learning has been demonstrated to play a pivotal and positive role across various facets of reasoning. A deeper understanding of its impact in this domain remains an area for further investigation.


\subsection{Alternative Formats for Solving NLP Tasks}


\noindent\textbf{Information Extraction.}
Information extraction (IE) aims to extract structural knowledge (\textit{e.g.}, entities, relations, and events) from unstructured and/or semi-structured documents.
However, due to the tasks having diverse output forms and requiring complex decoding strategies to post-process them into valid structures, 
relying solely on the plain text output of LLMs proves challenging for efficient modeling~\citep{lu2022unified}.
In the context of generative IE~\citep{xu2023largeIE},
these semantic structures can be smoothly converted into structured code,
which plays a crucial role in processing various tasks in a unified schema~\citep{wang2023code4struct}.
CodeIE~\citep{li2023codeie} shows that formulating text-to-structure IE tasks into structure-to-structure code generation with ``code-style'' prompting leads to superior performance compared to previous NL approaches.
Additionally,
CodeKGC~\citep{bi2024codekgc} introduces a novel method in the construction of knowledge graphs by treating the generation of triples as code completion tasks and then develops schema-aware prompts to further leverage the structural knowledge inherent in code.
Code4UIE~\citep{guo2023retrievalaugmented} builds a framework for retrieval-augmented code generation, 
which utilizes Python classes to define various schemas under a unified format.
GoLLIE~\citep{sainz2024gollie} is developed through fine-tuning CodeLLaMA, 
by integrating Python code and comments into the input representation, 
thereby enhancing its zero-shot performance on unseen tasks.

\noindent\textbf{Code as Intermedia Representation.}
In addition to the numerical reasoning (discussed in Section~\ref{sec:pal}) and the aforementioned IE tasks, 
which can be formulated as code generation tasks in a relatively fixed paradigm, 
recent emerging research has applied code as a medium in more diverse scenarios, 
instead of text~\citep{xie2022uniskg}. 
Early in the rise of LLMs,
\citet{madaan2022language} highlight that CodeLLMs could effectively represent desired graph predictions as code for use in structured commonsense tasks. 
Utilizing CodeX by few-shot prompting, 
they achieve performance surpassing that of general language models fine-tuned on the target task.
Subsequently, 
this approach of constructing procedural processes based on code has also been applied to sequential decision making~\citep{logeswaran2023code}, 
story-based tasks~\citep{jiang2023transferring}, 
semantic parsing~\citep{bogin2023leveraging} and neurosymbolic understanding~\citep{dong2023corrpus}. 
For graph scenarios,
recently released InstructGraph~\citep{wang2024instructgraph} unifies the representation of graphs through a code-like format,
eliminating the need for external specific encoders in graph reasoning and generation.

Beyond scenarios oriented toward linguistic structures, 
modular approaches based on code can also be applied to understanding both textual and graphical information. 
For example, 
VisProg~\citep{gupta2022visual} generates Python-like modular programs that involve off-the-shelf models or functions to facilitate visual reasoning. 
ViperGPT~\citep{surís2023vipergpt} creates programs for visual queries that take images or videos as arguments for execution, rather than traditional end-to-end methods with limited interpretability and generalization. 
\citet{chen2023vistruct} develop code-vision representations to assist in capturing visual structural information of varying granularity.
More recently, 
\citet{sharma2024vision} use code to represent images for teaching models about more aspects of the visual world.
Moreover,
user queries received by robots can also be translated into executable actions through code generation, 
empowering intelligent robots with the capability for multi-step embodied reasoning~\citep{yoneda2023statler}. 
In sum,
integrating code promises to evolve into a novel paradigm for addressing diverse NLP tasks. 
However, this approach is not a free lunch,
since utilizing code as an intermediary can introduce overhead (\textit{e.g.}, defining a function) and consume more context windows.
Further explorations in this vein are ongoing.

\begin{TakeawayBox}
{\textit{Takeaways}}
\begin{itemize}[itemsep=2pt,topsep=3pt,parsep=0pt]
    \item[(1)] {The fusion of code generation and symbolic solvers with LLM reasoning represents a groundbreaking shift in tackling numerical tasks.
Replacing natural language with executable code as the medium for reasoning, 
it not only overcomes lingering computation limitations but also enhances model interpretability and generalizability.}
    \item[(2)] {Although a theoretical foundation has yet to be established, 
the capacity of code training to enhance the mathematical abilities of LLMs has been empirically demonstrated and is gradually being accepted within the community.}
    \item[(3)] {Adopting code as intermediate representations can significantly elevate the efficacy and versatility of tackling diverse NLP tasks.
By transcending traditional text-based processing, 
code-centric approaches, through the construction of a unified schema, 
can handle intricate and complex input and output forms that previous methods struggled with.}
\end{itemize}
\end{TakeawayBox}

%% file: tables/reasoning/reasoning.tex



%% file: sections/6-applications.tex
\section{Real-World Applications}
\label{sec:app}

Upon finishing our review of models, algorithms, and data,
we shift our focus to discussing their real-world applications and current developments.
Initially, our discourse centers around code processing and structured data, highlighting two main areas: (1) the direct application of language models for code in SE, 
aiming to bridge the gap between the NLP and SE communities~\citep{zhang2023unifying}; 
(2) the exploration of utilizing CodeLLMs to augment data science research. Subsequently, we delve into hybrid scenarios, encompassing 
(1) the construction of agents through the application of code intelligence and 
(2) an emerging domain that leverages code intelligence to support AI4Science research.

\subsection{Boosting Software Develop Workflows}
Neural code intelligence is redefining the landscape of software development, steering it towards unprecedented levels of accessibility and automation across various scenarios.

\noindent\textbf{Serving as Coding Assistants.}
\label{sec:app:coding}
It is an indisputable fact that coding assistants, 
exemplified by GitHub Copilot~\citep{copilot}, 
are fundamentally transforming the landscape of software development~\citep{maxim2022mlenhanced}. 
Earlier attempts,
such as aiXcoder\footnote{\url{https://github.com/aixcoder-plugin}} and Intellicode~\citep{Svyatkovskiy2020gptc}, 
have already demonstrated that deep learning models can assist with basic coding tasks.
With the advent of the LLM era, 
emerging products like Code Whisperer~\citep{codewhisper}, Tabnine~\citep{tabnine}, 
Coze\footnote{\url{https://coze.com/}},
and aiXcoder~\citep{jiang2024aixcoder7blightweighteffectivelarge}.

which are built on commercial models,
have begun to mature into established products.  
Concurrently, 
coding assistants based on open-source CodeLLMs are increasingly gaining popularity and are being utilized in multiple code downstream tasks. 
For instance, FauxPilot\footnote{\url{https://github.com/fauxpilot/fauxpilot}} and Starchat~\citep{tunstall2023starchatalpha}, 
respectively based on the CodeGen and BigCode models, 
can serve as locally hosted alternatives to Copilot.
These versatile tools can be used for code generation, completion, repair, and even predicting the time complexity~\citep{siddiq2023zero}.
Additionally, 
when paired with a code interpreter (\textit{e.g.}, ChatGPT Plugins\footnote{\url{https://openai.com/blog/chatgpt-plugins\#code-interpreter}}), chatbots are capable of aiding users in building temporary programs within conversations.
Furthermore,
interacting with users in the form of extensions has also started to become popular, 
with CodeGeeX~\citep{zheng2023codegeex} already adapting itself to IDEs such as Visual Studio Code and JetBrains. 
Empirical research has illuminated the profound impact of these AI coding tools on developer efficiency.
For instance, 
A prior study~\citep{ziegler2022productivity} has indicated that the majority of GitHub Copilot users have experienced enhanced programming efficiency and improved code quality when coding with it. Complementarily, 
\citet{peng2023impact} quantify this enhancement in productivity, 
demonstrating a notable acceleration in task completion when developers employ Copilot.


As interest grows, 
attention is also directed towards the multifaceted performance of these assistants. 
Beyond evaluating the generated code itself~\citep{barke2023grounded,nguyen2022empirical},
researchers have also recognized the influence of prompts on their operation~\citep{fagadau2024analyzing},
the robustness issues faced when tackling varying scenarios~\citep{mastropaolo2023robustness}, 
and their effectiveness across specific PLs~\citep{corso2024generating} or NLs~\citep{koyanagi2024exploring}.
Moreover,
the quality of generated codes~\citep{yeticstiren2023evaluating} is another aspect that should not be overlooked. 
\citet{fu2023security} have revealed common weakness enumeration and advocated for meticulous review processes to mitigate risks associated with automated code generation. 
So,
we need a dual focus on enhancing productivity while safeguarding against vulnerabilities in code generation.


\noindent\textbf{Streamlining Software Development.}
\label{sec:app:dev}
Compared to having models write or complete code of different granularities for you, 
allowing them to take over the entire software development process is also becoming a tangible reality.
ChatDev~\citep{qian2023communicative} represents a virtual ``software company'',
functioning through an array of intelligent agents assuming roles such as programmer, code reviewer, tester, and designer. 
Collectively, these agents establish a multi-agent ecosystem, facilitating comprehensive software development processes.
AgentCoder~\citep{huang2023agentcoder} also implements a multi-agent system, assigning different roles to models,
showing that role-playing strategy not only alleviates the need for manually crafted test cases but also achieves better outcomes than self-refinement methods~\citep{zhang2023self,olausson2023demystify,chen2023teaching}.

After that,
\citet{qian2023experiential} enrich these agents with experience, 
encouraging them to accrue shortcuts from previous experience to avoid inefficient attempts or repetitive errors. 
This strategy enables the agents to tackle software engineering tasks within their interactions more efficiently,
leading to improved automation and efficiency for unseen tasks.
Beyond software kernel design, graphical design also falls within the scope of interest~\citep{beltramelli2017pix2code}.
Newly released works like Design2Code~\citep{si2024design2code} and web2code~\citep{yun2024web2code} aim to automate front-end development by converting visual designs into code implementations.
More recently, SWE-Bench MM~\citep{yang2024swebenchmm} has started exploring visual software engineering tasks, such as UI design systems.

As for leveraging code intelligence within GitHub, 
CodeAgent~\citep{zhang2024codeagent} suggests the use of external tools (\textit{e.g.}, Format Checker) for repo-level code generation. 
Subsequently, autonomous AI software engineers like OpenDevin~\citep{openhands} began to flourish.
Meanwhile,
beyond using pre-defined function set~\citep{schick2023toolformer,qin2023toolllm},
GitAgent~\citep{lyu2023gitagent} autonomously integrates repositories in response to user queries, thereby expanding its toolkit. 
Regarding software maintenance, 
the quality of code documentation is directly linked to development efficiency~\citep{khan2023doc}. 
To automate this,
RepoAgent~\citep{luo2024repoagent} has been proposed for documentation generation. 
It utilizes AST analysis to understand code structure and discern reference relationships within files, 
providing a contextual perspective for LLMs to assist in identifying the functional semantics to support fine-grained code documentation generation.
For more reliable testing, Meta has introduced TestGen-LLM~\citep{alshahwan2024meta},
which aims to improve existing human-written tests for automated unit test generation. The majority of its improvements have landed in industrial production.
Meanwhile, Google has advocated for the automation of resolving review comments within daily development workflows~\citep{alexander2024resolving}.
Recently released MarsCode Agent\footnote{\url{https://www.marscode.com/}} integrates AI-powered tools into a cloud-based development environment, streamlining coding processes with features like code completion and automated bug fixing.




\subsection{Facilitating Data-Driven Decision-Making}
\label{sec:app:ds}


Neural code intelligence has emerged as a transformative force in data-driven decision-making by unlocking new potentials in more accessible database interactions and streamlining data science processes. 

\noindent\textbf{Democratizing Database Interactions.}    
Previous efforts primarily emphasize elaborate model design and optimization tailored for specific formal languages, which lack the innate capability to effectively adhere to instructions. Large language models implicitly contain extensive world knowledge, which brings some basic abilities to interact with users. However, the inherent autoregressive characteristics~\citep{radford2019language} of LLMs make it challenging for them to accurately retain and recall data, especially when facing large-scale private databases. Under such circumstances, it is necessary to equip LLMs with the capabilities to automatically interact with external databases. The ultimate challenge lies in the precise generation of formal calling languages (\textit{e.g.}, SQL) based on the given query, and it is in this aspect that CodeLLMs truly excel.

With the advent of CodeLLMs~\citep{roziere2023code,deepseekcoder2023},
recent endeavors have also focused on facilitating widespread access to database interactions, 
aiming to bridge the retrieval process~\citep{lin2020bridging}. 
Early attempt like Binder~\citep{cheng2023binding} leverages CodeX to generate the programming language to combine the external knowledge bases. \citep{sun2023sqlprompt} proposes the optimized prompt design to boost the performances of Text-to-SQL tasks.
In addition, SQL-PaLM~\citep{sun2023sqlpalm} targets powering off-the-shelf LLM with SQL-specific optimization. They represent two common types of practices in this direction: 1) prompting Code-LLMs; and 2) post-finetuning LLMs with SQL optimization.

To better measure the LLM performances in tackling database-related scenarios, the evaluation benchmarks are gradually improving~\citep{zhang2024benchmarking}. The overarching trend is shifting from static, single-turn interactions constrained by limited domains towards dynamic multi-turn conversations encompassing diverse domains. At the early stage, benchmarks, such as ATIS~\citep{dahl1994expand}, tackle domain-specific settings (\textit{e.g.}, flights) with a limited number of schema. 
Following them,
Spider~\citep{yu2018spider}, Sparc~\citep{yu2019sparc}, Cosql~\citep{yu2019cosql}, WikiSQL~\citep{zhong2017seq2sql} and WikiTableQA~\citep{panupong2015comp} etc. are proposed to cover plenty of domains and over thousands of schema. Powered by the advent of LLMs, some challenging and comprehensive benchmarks arouse wide interest. Recent work~\citep{li2023can} highly stresses pushing the boundaries of LLMs and making LLMs serve as a functional interface. InterCode-SQL~\citep{yang2023intercode} tackles the multi-turn interaction with the database, pushing LLM to more practical scenarios. To sum up, the potential of self-repairing abilities and multi-turn interactions with databases are highly valued in the current landscape.

\noindent\textbf{Accelerating Data Insight Discovery.}
Data science entails the extraction of insights from data~\citep{wang2021autods},
and has evolved to be a pivotal component of decision-making and knowledge discovery over the past few years~\citep{donoho2018ds}.
Earlier attempts in this domain included methods for querying tabular data in natural language~\citep{yin2016neural} or generating Pandas codes for data analysis~\citep{zan2022cert}.
Recent research has progressed to handling real-world data science notebooks,
which are more complex than mere code generation, 
as computational notebooks often mix codes, text, figures, and execution results~\citep{je2021reactive}.

Regarding models specifically designed for such scenarios,
after the initial construction of JupyT5~\citep{chandel2022jupyt5} as a data science assistant, 
PaChiNCo~\citep{yin2022nl2code} emerges as a 62B CodeLLM based on PaLM, 
specifically designed for Python data science. 
Beyond its substantial size, PaChiNCo is trained under massive multi-modal contexts, such as existing notebook cells, corresponding execution states, and previous interaction turns. These rich contents ensure the model aligns with the specific peculiarities of computational notebooks, accommodating more diverse elements.

As for benchmarks that more closely align with real-world applications, particularly those concerning data wrangling tasks to process raw data and exploratory data analysis, 
DS-1000~\citep{lai2022ds1000} stands out as an example.
It comprises high-quality problems sourced from StackOverflow, including use cases involving NumPy, Pandas, TensorFlow, PyTorch, etc.
In terms of evaluation, 
what sets it apart from conventional NL2Code tasks is not just the emphasis on functional correctness; 
it also imposes additional constraints, 
such as requiring the generated code to include specific APIs/keywords to ensure solutions are efficient and aligned with the query~\citep{wen2023grounding}.
ARCADE~\citep{yin2022nl2code} presents another challenging benchmark, with a focus on Pandas. 
Its cases are derived through repurposing high-quality portions of previous benchmarks~\citep{agashe2019juice,chandel2022jupyt5} and collecting interactions between professional data scientists and coding assistants. 
ARCADE features multiple rounds of code generation within the same notebook, emphasizing the iterative nature of real-world data science work.
Research in this domain continues to thrive, with scholars also focusing on handling sophisticated DataFrames~\citep{cao2023api}, 
tackling the challenges of complex data visualization~\citep{li2024visualization,yang2024matplotagent}, 
and addressing hallucinations in conversations~\citep{chopra2023conversational}.




\subsection{Building Code-Empowered Agents}

A perennial topic in machine intelligence is to develop agent systems~\citep{wang2024survey}, 
such as robots, 
that can perform complex tasks requiring interaction with real-world environments.
Recent breakthroughs in LLMs have notably enhanced the task-solving capabilities of AI agents~\citep{xu2024interactive}.
Typically focused on natural language generation, 
LLMs offer immense potential~\citep{ahn2022can, driess2023palm, huang2022language}, 
yet the inherent ambiguity of natural language can sometimes render the precision and efficiency of planning and interaction~\citep{hu2023toward}.
To address this, emerging research advocates the integration of code as the de facto standard of agent systems \citep{xi2023rise, sumers2024cognitive, durante2024agent, yang2024wizard, ma2024agentboard}.
The following discussions will delve into these advancements.

\noindent\textbf{Augmenting Robotics System.}
Building robots capable of manipulating objects to accomplish diverse tasks in physical environments poses a significant challenge.
Code generation enables robots to seamlessly combine environment perception \citep{liang2023code, huang2023voxposer, yoneda2023statler}, feedback loops \citep{singh2023progprompt, arenas2023prompt, jin2023robotgpt}, and parameterized actions \citep{parakh2023lifelong, huang2023instruct2act} into reasonable policy, effectively translating complex tasks into executable solutions.
Generating executable actions is fundamental for robots;
ProgPrompt~\citep{singh2023progprompt} pioneered the use of GPT-3 to generate code-based actions, offering better environmental grounding than free-form text.
Robot tasks often involve multiple sub-tasks and conditional loops \citep{chen2021evaluating}, where code-based planning naturally excels over natural language due to its structured advantage.
For instance, \citet{liang2023code} explored utilizing LLM to write robot policy code, incorporating complex feedback loops and primitive API calls.
Following studies concentrated on integrating visual input \citep{huang2023instruct2act, yang2023octopus, mu2024robocodex}, refining code through environmental feedback \citep{jin2023robotgpt, ha2023scaling} or reinforcement learning \citep{yang2023octopus}, and continually expanding agent's skill library \citep{parakh2023lifelong}.
Faced with difficulties in collecting human operation trajectories, a series of studies explored leveraging LLMs for code generation to automatically synthesize diverse robot data \citep{wang2024gensim, jin2023robotgpt}.
Additionally, \citet{ha2023scaling} proposed enhancing data quality by simultaneously generating robot operations and code snippets to verify task success.
Another line of research innovatively applies LLMs in reinforcement learning to program reward functions \citep{yu2023language, huang2023voxposer, ma2023eureka}, guiding robots with greater efficiency than human experts.
Text2Reward~\citep{xie2023text2reward} demonstrates that LLM can produce interpretable dense reward functions for robotic manipulation tasks and allow iterative refinement with human feedback. 

\noindent\textbf{Elevating Intelligent Automation.}
Beyond the extensively discussed robotics, the paradigm of interacting with the environment through code generation also plays a pivotal role in elevating the capabilities of various intelligent automation systems.
In digital device assistants, code serves as the natural choice for controlling interactions between the agent and digital environment \citep{zheng2023synapse, xu2023lemur}. 
\citet{gur2023real} designed a web agent that acts on websites via generated Python programs by PaLM \citep{chowdhery2022palm}.
CodeACT~\citep{wang2024executable} introduced unifying agent actions through code and integrated LLM with Python interpreter.
GAIA~\citep{mialon2023gaia} constructed a benchmark for general AI assistants, encompassing tasks involving LLMs answering questions through programming.
Recently released OS agent OS-Copilot~\citep{wu2024oscopilot} generates executable code for operating computers, enabling humans to interact with operating systems through natural language instruction.
Contemporary progress in gaming agents has explored the use of code for perceiving the environment and facilitating action~\citep{zhao2023see, tan2024towards}. 
For instance, 
Vogayer~\citep{wang2023voyager} employs programming for interactions with Minecraft and maintaining an ever-growing skill library of executable code,
aiming to foster embodied lifelong embodied agents.
Beyond these, code models are also widely applied in autonomous driving \citep{mao2023language, ma2023lampilot, ishida2024langprop}, automated data processing~\citep{qiao2023taskweaver, shi2024ehragent} and multimodal tasks \citep{surís2023vipergpt, gupta2023visual}.


Although code interfaces offer a superior method for planning and interacting with the environment for agents compared to natural language interfaces, these approaches still largely rely on predefined APIs or skills.
Intelligent systems require ongoing exploration to boost their generalization across diverse tasks in the real physical world.

\subsection{Advancing AI4Science Research}

The progress in scientific fields can be greatly facilitated by advancements in code generation and the use of symbolic languages, 
particularly in the domains of mathematical proof, chemistry, and biology. These advancements play a crucial role in driving innovation and pushing the boundaries of scientific understanding. 
The following paragraphs will outline these contributions in each domain.

\noindent\textbf{Automating Theorem Proving.} Formal theorem proving is a discipline that necessitates finding proofs for given conjectures articulated in structured, formal statements governed by the principles of logic. Traditional formal theorem proving called upon human experts to meticulously convert mathematical concepts into formal statements, which could then be verified using interactive theorem provers (ITP) like Isabelle~\citep{paulson1994isabelle} and Lean~\citep{de2015lean}. Each formal statement is similar to a code statement, and ITPs are ``compilers" specifically designed for math proofs.
This process is greatly accelerated by utilizing formal statements generators that generate single-step proofs that ITPs then verify.
A premise is often directly selected based on language modeling statistics~\citep{polu2020generative,jiang2021lisa,polu2022formal}, while  Leandojo~\citep{yang2023leandojo} uses a retrieval-augmentation generation pipeline where LLM directly generate proof step based on retrieved premises.
Each step is then checked for accuracy by ITPs before progressing to the next, with the ultimate objective being the completion of the proof. To optimize this procedure, sophisticated search algorithms are employed that identify and develop promising premises with the potential to culminate in successful proofs.~\citep{lample2022hypertree, wang2023dt}. A distinctive work in this field is AlphaGeometry~\citep{alphageometrytrinh2024},
which uses a language model to generate auxiliary construction for geometric proofs and then solve the newly constructed problem with symbolic solver~\citep{trinh2024solving}. 
Furthermore, recent research~\citep{jiang2022thor, jiang2022draft,zhao2023decomposing, first2023baldur} has explored the use of language models with potent code generation capabilities to directly produce entire proofs that could be further refined by ITPs.

Code generation has demonstrated its potential in proving existing mathematical theorems yet recent works demonstrate its potential in solving open mathematical questions. AlphaTensor \citep{fawzi2022discovering} tackles matrix decomposition by producing tensor decomposition statements and employing tree search to identify the most efficient solution. Similarly, FunSearch \citep{romera2024funsearch} addresses combinatorial problems such as cap sets and online bin packing through program generation and optimal selection.
 
\noindent\textbf{Catalyzing Biochemistry Discoveries.}
Code generation has emerged as a transformative force in biochemistry research, significantly enhancing the efficiency and effectiveness of scientific investigations in the field~\citep{hocky2022chem}. This can be largely attributed to the wealth of open-source tools and code packages now available, including tools for processing biochemical sequences~\citep{landrum2013rdkit, Jumper2021HighlyAP, M2021AccuratePO}, accessing databases~\citep{Apweiler2004UniProtTU, Finn2015ThePP, Kim2015PubChemSA}, and analyzing biochemical properties~\citep{Boeckmann2003TheSP, Zhu2022TorchDrugAP}. 
To this end, \citet{dias2023large} and \citet{bran2023chemcrow} pioneer in using LLMs to generate code APIs that use these tools for automating the chemical research pipeline. 
\citet{ma2023retrieved} have proposed retrieval APIs to expedite the analysis of protein sequences. \citet{tang2023biocoder} specifically designed to assess the capability of LLMs in generating bioinformatics code, marking a significant step towards standardized evaluation in this domain. In drug discovery, innovative approaches~\citep{ye2023drugassist, liu2023chatgpt, zheng2023chatgpt} to employ tool-using language models for automatically editing and optimizing molecular structures for therapeutic purposes -- a crucial phase in drug development. 
Beyond the generation of APIs for existing biochemical tools, 
\citet{Steiner2019OrganicSI} and \citet{rauschen2024universal} have also introduced novel chemical programming languages designed to automate the synthesis of chemical compounds. This progress represents a leap forward in our ability to program and execute complex synthetic chemistry with precision and scalability. 
Recently, 
SciCode~\citep{tian2024scicode} has emerged as an evaluation framework for LLMs’ scientific coding capabilities. It presents real-world challenges and facilitates the development of AI coding tools for tasks like protein analysis and biochemical data processing.

\begin{TakeawayBox}
{\textit{Takeaways}}
\begin{itemize}[itemsep=2pt,topsep=3pt,parsep=0pt]
    \item[(1)] {Coding assistants have revolutionized software engineering workflows by significantly enhancing programming efficiency and code quality. Further, the ongoing evolution towards fully automated software development ecosystems represents a leap towards utilizing intelligent code agents to alleviate humans from labor-intensive development tasks.}
    \item[(2)] {The evolution of the CodeLLMs has significantly broadened the scope of database interaction, thereby unlocking the potential of multi-turn retrieval in more generalized domains. Further,
the rise of code intelligence has also motivated more research into automating and accelerating real-world data science workflows.}
    \item[(3)] {The code-centric paradigm orchestrates perception, decision-making, action, and feedback for intelligent agents to tackle complex tasks in real-world environments. With their potent reasoning capacity and efficient interaction, these models establish a strong foundation for developing agent systems capable of navigating the highly variable physical world.}
    \item[(4)] Code models that can wield mathematical proof language and scientific tools have revolutionized the AI4Science field by advancing towards autonomous scientific discovery, matching human mathematicians in writing proofs, and chemists in analyzing compounds. The current languages used for AI4Science often differ from typical PLs; therefore, adapting code models to novel symbolic languages is essential for their further application in science.
\end{itemize}
\end{TakeawayBox}

%% file: sections/7-future.tex
\section{Opportunities and Future Directions}
\label{sec:discuss}

Thus far,
we have extensively reviewed and discussed the advancements in code intelligence, 
encompassing tasks, models, and applications,
aiming to provide a comprehensive view and bring interested researchers up to speed with this field.
In what follows,
we highlight several promising directions for future research that are ripe for contribution.

\subsection{Beyond Transformer Architecture}

Since the rise of transformer architecture~\citep{vaswani2017attention},
it has maintained a dominant position in NLP~\citep{lin2022transformer},
and most of the models we discuss in this paper are also based on this. 
Nowadays, 
with the emergence of diffusion models for controllable text generation generation~\citep{li2022diffusionlm,gong2023diffuseq,he2023diffusionbert}, 
it has also gradually been applied in the NL2Code field. 
CodeFusion~\citep{singh2023codefusion} has recently pioneered the application of diffusion models in code-related tasks, 
achieving stunning results on tasks for Python, Bash, and conditional formatting~\citep{singh2022cornet}, 
surpassing the performance of LLMs such as StarCoder and CodeT5+ with only a 75M size.
As we have discussed in Section~\ref{sec:code-embd}, 
utilizing specially tailored architectures or modeling for specific tasks separately remains a path worth exploring.

\subsection{Renaissance in Utilizing Code Features}
The most notable difference between code and natural language is its structured nature, 
which was widely leveraged in the pre-training~\citep{guo2021graphcodebert}, 
fine-tuning~\citep{zhu2022neural,chen2023pass}, 
evaluations~\citep{chirkova2022codebpe},
explainability~\citep{zhu2022catprobing} of models in the Pre-LLM era. 
Yet, in the current landscape dominated by LLMs,
the utilization of structural information has diminished. 
We believe the main reasons are the incompatibility of most LLM training pipelines with modalities other than text tokens, 
and the cost required to extract code features (such as data flow graphs and abstract syntax trees) from massive training data. 
We hold the view that finding appropriate strategies to integrate structural information into the CodeLLM training process is worth exploring. 
Beyond training,
as previously discussed in Section~\ref{sec:ptm:interpretable}
we believe that structural information can still help us better understand the behavior of models.
Recently we have also witnessed some researchers starting to utilize lexical properties for enhanced code generation~\citep{jain2023improving} or structural information to assess the impact of code complexities on program-aided reasoning~\citep{bi2023potwork}.
We are convinced that a revitalization of using code structure will significantly contribute to the ongoing advancement of code intelligence.

\subsection{Repo-Level Code Understanding and Generation}
Following the surge in the popularity of CodeLLMs,
their applications have primarily focused on individual functions or files.
However, 
in real-world software engineering, 
developers often need to consider the relationships between different files and functions within a code repository. 
When extending the tasks discussed in Section~\ref{sec:code-embd:task-overview} to broader scenarios, 
such as completing or repairing a code snippet within a repository~\citep{zhang2023repocoder,zhang2024codeagent},
The capability of models to invoke variables and functions from other files remains underexplored.
This may necessitate the integration of code understanding,
cross-file dependencies understanding, 
and retrieval-augmented generation~\citep{su2024arks}.
Also,
given the potential complexity of code repositories, 
this may require research into how to expand CodeLLMs' efficient context window~\citep{an2023leval} and their ability to understand cross-file dependencies. 
Therefore, 
repository-level code generation and understanding represent a direction of both practical and research significance, 
with production environments also serving as a sustainable and challenging testbed for future models.


\subsection{Towards Holistic and Reliable Evaluations}
\label{sec:future:eval}
Reliable and comprehensive benchmarking is a perennial topic in language model research, 
and the same applies to evaluations of language modes for code~\citep{lu2021codexglue,fu2023codeapex}.
Current mainstream evaluation methods primarily rely on code execution,
during which the security, diversity, and readability of the model-generated code are often overlooked~\citep{zhou2023codebertscore}. 
Additionally, researchers have pointed out that some benchmarks, represented by APPS~\citep{hendrycks2021apps} derived from competition platforms like Codeforces, 
are likely to have frequently appeared in public repositories~\citep{chen2021evaluating}, 
consequently leading to models ``remembering'' the potential solutions to these problems during the pre-training phase.  
The hand-written HumanEval dataset also faces inevitable data leakage or contamination issues~\citep{riddell2024quantifying} as the training corpus expands. 
In pursuit of more reliably evaluating CodeLLMs' performance, 
and to consider the naturalness, robustness, and lexical diversity of the generated code, 
fair and dynamic comprehensive benchmarking awaits further exploration.




\subsection{Interleaved Planning and Code-Driven Reasoning}

As discussed in Section~\ref{sec:pal}, Program-aided Language Models~\citep{gao2022pal} and Program of Thoughts~\citep{chen2022program} have demonstrated efficacy in numerical-related tasks, 
significantly outperforming some NL-centric paradigms~\citep{wei2022chain}. 
However, the success of such a paradigm hinges on the premise that a problem can be reliably decomposed into multiple lines of code for resolution. 
Compared to NL-centric reasoning, 
program-aided strategies are anticipated to push the boundaries of human intelligence. 
Mirroring the human approach to tackling complex challenges, more demanding problems  (\textit{e.g.}, Olympic competition problems~\citep{alphageometrytrinh2024}) require ongoing planning and reasoning. 
Each phase in such scenarios necessitates iterative planning, 
taking into account the problem at hand and the current state, 
while concurrently generating code to aid in the solution part. 
This synergistic blend of interleaved planning and program-aided reasoning provides a strategic advantage that is worthy of exploration.

\subsection{A Closer Look at Tokenizer Dynamics}
\label{sec:future:tokenizer}

Tokenization has historically played a pivotal role in language modeling~\citep{sennrich2016bpe},
where the choice of tokenizer can significantly influence a model's downstream performance~\citep{ali2023tokenizer} and multilingual capabilities~\citep{petrov2023language}. 
Recent studies have also identified its substantial impact on the generalization capabilities of LLMs~\citep{li2024evaluating,dagan2024getting}.
Unlike natural languages,
code operates under more rigid syntactic rules and structures, with meanings heavily reliant on the precision of these elements. 
Earlier research has uncovered that tokenization granularity exerts a non-trivial impact on code-related tasks~\citep{chirkova2022codebpe}.
So, we have reason to doubt that: when processing code with general tokenization methods like BBPE~\citep{wang2019neural}, 
as employed by models such as CodeLLaMA, 
there is a risk of disrupting this structure due to over-tokenization.
Hence, 
determining an effective approach to tokenization without compromising code structure and semantics poses a challenge.
One potential research direction involves developing tokenizers capable of recognizing and preserving code structures (discussed in Section~\ref{sec:code-embd:struct} )--such as keywords, function definitions, and control flow statements—or exploring better strategies that grasp a deeper understanding of code semantics.




\subsection{Efficient Methods for CodeLLMs}
\label{sec:future:efficient}

As the development of CodeLLMs progresses,
beyond striving for performance enhancements, 
the pursuit of model efficiency emerges as another critical direction for in-depth investigation.
As for training, 
whether starting from scratch or conducting additional training on code using general LLMs, 
the process proves exorbitantly expensive.
Although parameter-efficient methods~\citep{ding2023delta} like prefix-based strategies~\citep{li2021prefix,liu2022ptuning,wang2023upet} and LoRA~\citep{hu2022lora} significantly reduce resource consumption, 
applying them directly to language models for code~\citep{shamil2022petcode} has been shown to noticeably impact the performance of code-related tasks~\citep{zou2023comprehensive}.
Thus, 
identifying new efficient training techniques for code is worth exploring, 
and it necessitates finding the most suitable strategies and trade-offs between cost and performance across models of different scales~\citep{zhuo2024astraios}.
For deployment, 
due to the prevalent state-of-the-art CodeLLMs being powerful yet cumbersome, 
researchers have begun to emphasize the application~\citep{wang2023towardslow} and optimization of CodeLLMs for running within resource-constrained environments. 
As discussed in Section~\ref{sec:codellm:model},
this particularly includes enabling these models to function offline on consumer-level devices without the reliance on GPUs~\citep{stable-code-3b, DeciFoundationModels}.
Additionally, 
to satisfy more diverse demands,
CodeLLMs are faced with the strategic decision of scaling up~\citep{wu2024llamapro,lin2024scaling} or scaling out~\citep{jiang2024mixtral}.
Alternatively, 
acceleration can also be achieved through methods such as structured pruning~\citep{xia2023sheared} and distillation~\citep{chen2023personalized}
akin to approaches employed in general LLMs.


\subsection{Further Expansion of Multilingual Capability}
``Nobody should call themselves a professional if they only knew one language.'' — Bjarne Stroustrup.
Multilingualism has long been a staple in NLP research,
with LLMs proving capable of holding multilingual capabilities~\citep{pires2019multilingual, yuan2023multilingual}.
However, 
in the realm of code intelligence, the exploration of mastering multiple programming languages is a relatively recent development.
Although there has been research targeting multilingual scenarios~\citep{zheng2023codegeex},
significant performance variations across different programming languages by CodeLLMs are evident from various leaderboards (\textit{e.g.}, Big Code Models Leaderboard\footnote{\url{https://huggingface.co/spaces/bigcode/bigcode-models-leaderboard}}). 
This disparity is largely attributed to the uneven distribution of corpora which is dominated by popular languages like Python and Java.
Consequently, collecting more premium data for less prevalent languages (\textit{e.g.}, TypeScript, Kotlin, Scala), 
exploring data augmentation~\citep{zhuo2023source},
conducting specialized training, 
distilling language agnostic representation~\citep{huang2023program}
or even transferring knowledge between different languages~\citep{sun2023tr,cassano2024knowledge,mastropaolo2023transfer}, 
presents a worthwhile direction for research. 
Furthermore, 
multilingual models have also been shown to be more robust to prompt perturbation and excel in code summarization~\citep{athiwaratkun2023mbxp},
potentially contributing to an enhanced overall capability of CodeLLMs.
It’s also noteworthy that current mainstream multilingual benchmarks~\citep{cassano2022multiple} are mostly conversions from Python datasets, 
overlooking the unique characteristics of different languages.
Hence, developing new benchmarks that cater to specific language features is another avenue worth exploring.












\subsection{Copyright Challenges Faced by Coding Assistants}

With the expansion of the open-source community and the rise of coding assistant tools (discussed in Section~\ref{sec:app:coding}),
a series of ethical and security concerns regarding the distribution of source code have arisen, 
such as unauthorized use of copyrighted code, 
distributions without proper licenses, 
or exploitation of code for malicious purposes.
To circumvent potential legal and copyright disputes,
one viable strategy is the watermarking of source code protected by licenses such as GPL.
Furthermore, 
the use of copyrighted content detection in CodeLLMs' training corpora~\citep{sun2022coprotector} can also safeguard against the unintended output of such code.
Notably, 
watermarking and detection for code diverge from that for natural language due to the impact of obfuscation on the semantics of variable names, presenting an intriguing area for further investigation.
Unlike watermarking~\citep{yang2024watermarking} and content detection~\citep{duarte2024decop} applied to natural language, code presents unique challenges. 
As previously discussed,
the semantics of variable names in code can be significantly affected by obfuscation, 
making it a more challenging and worthwhile direction for further investigation.

%% file: sections/8-conclusion.tex
\section{Conclusion}
\label{sec:conclusion}

In this paper,
we present a systematic review of the entire evolutionary trajectory of code intelligence,
offering a comprehensive examination from the nascent application of deep neural networks on source code to the breakthroughs made in the LLM era.
Throughout this process, 
we have delved into the interconnections among research across different periods, 
engaging in detailed discussions and analyses centered on the paradigm shifts in models, tasks, and applications.
Moreover,
we explore the synergies between code learning and other facets of machine intelligence, 
along with both long-standing and emergent applications in the real world.
Bearing in mind the developmental pathway we have witnessed and insights garnered from our discussions,
we also identify several promising directions for future research in code intelligence.
Given the concurrent advancements in language models and the evolving needs of software development,
we are confident that this domain will continue to flourish in the forthcoming years.
It is our aspiration that the literature review, discussions, experiences, and resources provided in this survey paper will boost the community's future research.

%% file: appendices/resources.tex
\section{Additional Resources}

\subsection{Reading Lists}

In the \href{https://github.com/QiushiSun/Awesome-Code-Intelligence}{Awesome-Code-Intelligence} project related to this paper, 
we have prepared a curated reading list for our readers, encompassing a diverse array of research areas within the realm of code intelligence.

\subsection{Recreating the Figures}
\label{app:trees}

We used the template from \href{https://github.com/Mooler0410/LLMsPracticalGuide}{LLMsPracticalGuide}  maintained by \citet{yang2023harnessing} to construct Figure~\ref{fig:codelm-tree}.
Our primary criteria for determining the placement of models on specific branches of the tree were based on the release times of the models mentioned in Sections~\ref{sec:codeptms} and Section~\ref{sec:codellms},
as well as the relationships between the models. Relevant resources will be uploaded to our project for further reference and exploration.

%% file: appendices/more-benchmarks.tex
\section{More Benchmarks}

Given the constraints on space, 
Table~\ref{tab:code_tasks} and Table~\ref{tab:nl2code_tasks} feature only a subset of the representative tasks in code downstream applications and NL2Code.
To offer a more thorough overview, 
Table~\ref{tab:code_tasks_full_p1}, \ref{tab:code_tasks_full_p2}, \ref{tab:code_tasks_full_p3} and \ref{tab:nl2code_full} systematically present a more exhaustive review on datasets, organized by each specific task category.

\section{Detailed of CodePTM Training Objects}

Due to space limitations, we adopt abbreviations for the training objectives of CodePTMs in the overview presented in Table~\ref{table:CodePTMs},
section~\ref{sec:codeptm_models}. 
Here we provide detailed explanations for them in Table~\ref{tab:codeptm_objs}.

\section{Related Research}

Before the rise of employing transformer-based models in code intelligence, 
\citet{han2021comp} evaluates the performance of eight code embedding models on classic code-related tasks.
During the period when pre-trained language models are dominant in NCI research,
\citet{wu2022survey} categorize CodeLMs from the perspectives of code structures.
\citet{xu2022survey} revisit the training processes and downstream applications of CodePTMs, 
and \citet{xu2022systematic} conduct a systematic evaluation of mainstream models in this sphere.
Meanwhile, several studies~\citep{niu2022dlcode,niu2023empirical} have discussed the application of CodePTMs from the perspective of software engineering.
Amidst the burgeoning trend of LLMs, \citet{zan2023large} provides an empirical summary of the field's development from the standpoint of NL2Code.
\citet{hou2023large} and \citet{she2023pitfalls} discuss the application of CodeLLMs in SE, along with the opportunities and challenges.
In recent studies, 
\citet{zhang2023unifying} conduct a retrospective study of language models for coding from the unified viewpoints of NLP and SE.
\citet{wan2023deep} delve into an exploration of benchmarks and toolkits for neural code intelligence.
For the synergies between code learning and AI agents, 
\citet{yang2024wizard} investigate the role of code corpora in augmenting the capabilities of LLM-based agents.

\section{Data Sources}

\subsection{Evaluations}
\label{app:sources:pal}

The primary data sources for Figure~\ref{fig:nl2code-eval} are BigCode's Evaluation Harness~\citep{bigcode-eval} and the BigCode Models Leaderboard.
Similar evaluation results can also be found at OpenCompass~\citep{wu2023opencompass}. 
As for the results reported in the Figure~\ref{fig:pal_eval}, 
we access GPT-3.5-Turbo and Claude-1 through paid APIs. For open-source LLMs, we utilize CodeLLaMA-Instruct-13B and LLaMA2-Chat-13B under the greedy search setting. Both GSM8K and MATH evaluations are conducted with 3-shot promptings from \citet{gao2022pal}. WikiSQL~\citep{zhong2017seq2sql} and FOLIO~\citep{han2022folio} are evaluated under the 1-shot prompting, modified from ~\citet{xu2023symbolllm}.

\subsection{Publication Trends Data and Milestones}
 
The construction of Figure~\ref{fig:trend} was achieved through the use of the ArXiv advanced search\footnote{\url{https://arxiv.org/search/advanced}}, with the calculation based on an exact match by querying a group of keywords (\textit{e.g.}, code representation, code generation) in the title or abstract. 
Detailed information on this is provided within the \href{https://github.com/QiushiSun/Awesome-Code-Intelligence}{Awesome-Code-Intelligence} project.

As for the selection of milestones featured in Figure~\ref{fig:timeline}, Qiushi Sun, Zhirui Chen, and Zhangyue Yin each selected what they considered to be representative milestones based on their modeling approaches,
release time, 
the citations of papers/preprints, 
and their influence within the community. The final selection was made by identifying the intersection of their choices.

Beyond statistics, we can observe that the publication venues for research papers on code intelligence have undergone a significant transformation over the past decades.
In earlier research, publications were primarily confined to ACM/IEEE Conferences like ASE and ICSE, or journals like IEEE TSE.
However, with the rise of the neural approach, 
research began shifting towards machine learning conferences, 
such as NeurIPS.
Subsequently,
with the rapid advances in language modeling, 
code-related topics have also become frequent subjects at NLP conferences, exemplified by the *ACL Conferences.
These changes in publication venues reflect a blurring of the boundaries between previously distinct research areas, 
demonstrating the interdisciplinary nature of code intelligence research today.

\section*{Acknowledgement}

Thanks to everyone who has contributed to this paper.
We extend our gratitude to Xuesong Lu for meticulously reviewing our paper and providing helpful feedback, 
as well as assisting us in identifying and incorporating previously overlooked literature into our study.
Valuable suggestions during the paper revision process are also provided by Zichen Ding, Jingyang Gong, and Yichao Du, for which we extend our thanks.
We warmly welcome the community to utilize all resources provided in this survey for research, educational, knowledge-sharing, and domain background introduction purposes.
\textit{The authors of this survey paper retain the copyright of the figures/tables included herein, 
and any use of these materials for publication purposes requires authorization from the survey authors.}

\section*{Author Contributions}

Writing this comprehensive survey and continuously updating its contents is no easy job.
We are deeply grateful to the authors for their support and dedication to this work. Their contributions to this paper are as follows:

\begin{itemize}
    \item Project concept and leadership: Qiushi Sun, Xiang Li, and Zhiyong Wu.
    \item Paper Writing: Qiushi Sun, Fangzhi Xu, Chang Ma, Kanzhi Cheng, Zhirui Chen, Qipeng Guo, and Lingpeng Kong.
    \item Paper Revising: Qiushi Sun, Chang Ma, Zhirui Chen, Chengcheng Han, Renyu Zhu, Lingpeng Kong, Fei Yuan, Qipeng Guo, Pengcheng Yin, Xipeng Qiu, Xiang Li, and Xiaoli Li.
    \item Experiments: Fangzhi Xu and Qiushi Sun.
    \item GitHub project maintenance: Shuai Yuan, Zhirui Chen, Qiushi Sun, Jianing Wang, Kanzhi Cheng, Fangzhi Xu, and Zhangyue Yin.
    \item Strategic advice: Pengcheng Yin, Xipeng Qiu Xiang Li, and Xiaoli Li.
\end{itemize}

\section*{Document History}

\begin{itemize}
\item First release on March 21, 2024: the initial version.
\item Update on May 9, 2024: minor updates on formatting and typos, available at GitHub.
\item Update on June 23, 2024: revise Figure~\ref{fig:codelm-tree} and Section~\ref{sec:codellm:model}, add more related models/benchmarks.
\item Update on August 31, 2024: add more related models, minor updates on typos.
\item Update on October 25, 2024: add more related models, and updates on coding agents.
\item Update on November 1, 2024: add more related models, benchmarks, methods, and author contribution list.
\item Update on Jan 25, 2025: add a new section~\ref{sec:preference} on preference optimization for code generation, update the formatting of the takeaway sections in each part, add more CodeLMs \& benchmarks, and revise Figure~\ref{fig:codelm-tree}.
\end{itemize}

\input{tables/complete-benchmarks}

\input{tables/codeptm-objects}

%% file: tables/complete-benchmarks.tex
\begin{landscape}

\begin{table}[ht]

\footnotesize
    \caption{A more comprehensive collection of code-related benchmarks extended from Table~\ref{tab:code_tasks}. ``\# PLs'' denotes the number of programming languages each benchmark covers.}
    \adjustbox{width=\textwidth+5cm,center}{
    \begin{tabular}{l|l|c|c|l|c}
        \toprule[1.2pt]
        \textbf{Task} & \textbf{Dataset} & \textbf{Date} & \textbf{\# PLs} & \textbf{Description}  & \textbf{Eval. Metric} \\
        \midrule
          \multirow{3}{*}{Clone Detection}
        & POJ-104~\citep{mou2016convolutional}~\href{https://sites.google.com/site/treebasedcnn/}{[link]} & 2014 & 1 & a program classification dataset of 52K C/C++ programs  & Acc. \\
        & BigCloneBench~\citep{svajlenko2014towards}~\href{https://github.com/clonebench/BigCloneBench}{[link]} & 2015 & 1 & a clone detection dataset of 8M Java validated clones &  F1 score \\
        
        & CLCDSA~\citep{nafi2019CLCDSA}~\href{https://github.com/Kawser-nerd/CLCDSA}{[link]} & 2019 & 3 & a cross-language clone dataset dataset of more than 78K solutions & F1 score \\
        \midrule 
        \multirow{15}{*}{Defect Detection}
        & CGD~\citep{DBLP:conf/ndss/LiZXO0WDZ18}~\href{https://github.com/CGCL-codes/VulDeePecker}{[link]} & 2018 & 2 & a dataset focuses on two types of vulnerabilities in C/C++&  F1/Precision\\

        & Draper VDISC~\citep{DBLP:conf/icmla/RussellKHLHOEM18}~\href{https://osf.io/d45bw/}{[link]} & 2018 & 2 &  a vast dataset of open-source functions in C/C++&  F1\\

        & SySeVR~\citep{li2021sysevr}~\href{https://github.com/SySeVR/SySeVR}{[link]} & 2018 & 2 &   a dataset of 126 types of vulnerabilities in C/C++&  F1\\        
        
        & Devign~\citep{zhou2019devign}~\href{https://github.com/epicosy/devign}{[link]} & 2019 & 1 & a dataset of vulnerable C functions&  F1/Acc. \\

        &GREAT~\citep{hellendoorn2019global}~\href{https://github.com/VHellendoorn/ICLR20-Great}{[link]} & 2019 & 1 & a dataset extracted from the ETH Py150 dataset&  Acc. \\

        &MVD~\citep{zou2019mu}~\href{https://github.com/muVulDeePecker/muVulDeePecker}{[link]} & 2020 & 2 &  a dataset that contains 181K pieces of code from 33K C/C++ programs&  F1 \\

        &ReVeal~\citep{chakraborty2021deep}~\href{https://github.com/VulDetProject/ReVeal}{[link]} & 2020 & 1 &   a vulnerability dataset curated from two real-world projects (Chromium and Debian) &  F1 \\

        &BigVul~\citep{fan2020ac}~\href{https://github.com/ZeoVan/MSR_20_Code_vulnerability_CSV_Dataset}{[link]} & 2020 & 2 &   a large C/C++ code vulnerability dataset from open-source Github projects &  F1 \\  

        &D2A~\citep{DBLP:conf/icse/ZhengPLBEYLMS21}~\href{https://github.com/IBM/D2A}{[link]} & 2021 & 2 &    A dataset built for AI-based vulnerability detection methods &  F1 \\  

        &PyPIBugs~\citep{allamanis2021self}~\href{https://github.com/microsoft/neurips21-self-supervised-bug-detection-and-repair}{[link]} & 2021 & 1 &  A dataset retrieved form the most downloaded packages in the Python package index &  F1 \\  

        & CVEfixes~\citep{DBLP:conf/promise/BhandariNM21}~\href{https://github.com/secureIT-project/CVEfixes}{[link]} & 2021 & 27 &  a automate-collected dataset from Common Vulnerabilities and Exposures records &  F1 \\  
        
        & CrossVul~\citep{nikitopoulos2021crossvul}~\href{https://zenodo.org/records/4734050}{[link]} & 2021 & $>$ 40 & a dataset of 27K files containing vulnerabilities & F1/Acc. \\
        & DiverseVul~\citep{chen2023diversevul}~\href{https://github.com/wagner-group/diversevul}{[link]} & 2023 & 2 & a dataset of 19K vulnerable C/C++ functions and 330K nonvulnerable functions& F1/Acc. \\

        & VulnPatchPairs~\citep{risse2023limits}~\href{https://github.com/niklasrisse/LimitsOfML4Vuln}{[link]} & 2023 & 1 & a dataset contains 26.2K C functions& F1/Acc. \\

        & VulBench~\citep{gao2023far}~\href{https://github.com/Hustcw/VulBench}{[link]} & 2023 & 1 & a dataset offers a blend of CTF challenges and  real-world CVE vulnerabilities& F1 \\
        \midrule
       
        \multirow{24}{*}{Code Repair}
        & Defects4J~\citep{}~\href{https://github.com/rjust/defects4j}{[link]} & 2014 & 1 & a database and extensible framework providing real Java bugs& Mean \\
        
        & ManyBugs~\citep{DBLP:journals/tse/GouesHSBDFW15}~\href{https://github.com/squaresLab/ManyBugs}{[link]} & 2015 & 1 & a benchmark consisting of 1K defects in 15 C programs& Acc. \\        
        
        & BugAID~\citep{DBLP:conf/sigsoft/HanamBM16}~\href{http://salt.ece.ubc.ca/software/bugaid/}{[link]} & 2016 & 1 & a benchmark build through mining 105K commits from 134 JavaScript projects& Acc. \\           
        & DeepFix~\citep{gupta2017deepfix}~\href{https://bitbucket.org/iiscseal/deepfix/src/master/}{[link]} & 2017 & 1 & a set of 6971 erroneous C programs written by students for 93 programming tasks  &  Acc. \\

        &  Codeflaws~\citep{DBLP:conf/icse/TanYYMR17}~\href{https://github.com/codeflaws/codeflaws}{[link]} & 2017 & 1 & a collection of C programs with 3.9K defects& Acc. \\           
         & QuixBugs~\citep{prenner2022can}~\href{https://github.com/jkoppel/QuixBugs}{[link]} & 2017 & 2 & a multi-language benchmark (40 bugs in both Python and Java) & Pass Rate \\

         &  Bugs.jar~\citep{DBLP:conf/msr/SahaLLYP18}~\href{https://github.com/bugs-dot-jar/bugs-dot-jar}{[link]} & 2018 & 1 & a benchmark consisting of 1.1K bugs and patches & Acc. \\

         &  Bears~\citep{DBLP:conf/wcre/DelfimUMM19}~\href{https://github.com/bears-bugs/bears-benchmark}{[link]} & 2019 & 1 & an extensible bug benchmark for automatic repair studies in Java & Acc. \\

         &  BugsJS~\citep{DBLP:conf/icst/GyimesiVSMBF019}~\href{https://www.bugswarm.org/}{[link]} & 2019 & 1 & a benchmark of 453 real JavaScript bugs & Acc. \\

         &  BugSwarm~\citep{DBLP:conf/icse/DmeiriTWBLDVR19}~\href{https://www.bugswarm.org/}{[link]} & 2019 & 2 & a benchmark of 3K fail-pass pairs, in Java and Python & Acc. \\

         & ManySStuBs4J~\citep{karampatsis2020often}~\href{https://zenodo.org/records/3653444}{[link]} & 2019 & 1 &  a dataset of 153K single statement bugfix changes mined from Java projects & Acc. \\

         & Refactory~\citep{hu2019re}~\href{https://github.com/githubhuyang/refactory}{[link]} & 2019 & 1 &  a dataset consists of almost 1.8K real-life incorrect Python program submissions & Acc. \\  

         & Review4Repair~\citep{DBLP:journals/infsof/HuqHHMIA22}~\href{https://github.com/Review4Repair/Review4Repair}{[link]} & 2020 & 1 &  a dataset consists of 55K code reviews and associated code changes & Acc. \\ 

         & BugsInPy~\citep{DBLP:conf/sigsoft/WidyasariSLQPTT20}~\href{https://github.com/soarsmu/BugsInPy}{[link]} & 2020 & 1 &  a benchmark consists of 493 real bugs from 17 real-world Python programs & Acc. \\

         & TFix~\citep{berabi2021tfix}~\href{https://github.com/eth-sri/TFix}{[link]} & 2021 & 1 &  a benchmark consists of 52 different error types reported by a popular static analyzer & Acc. \\  

         & Megadif~\citep{monperrus2021megadiff}~\href{https://github.com/ASSERT-KTH/megadiff}{[link]} & 2021 & 1 &  A dataset of 600K java source code changes categorized by different size. & Acc. \\ 

         & SSB/TSSB~\citep{DBLP:conf/msr/RichterW22}~\href{https://github.com/cedricrupb/tssb3m}{[link]} & 2022 & 1 &  a collection of over 9M/3M general single statement. & Acc. \\          

         & FixJS~\citep{csuvik2022fixjs}~\href{https://github.com/AAI-USZ/FixJS}{[link]} & 2022 & 1 &  a dataset containing bug-fixing information of ~2M commits. & Acc. \\

         &  TypeBugs~\citep{oh2022pyter}~\href{https://github.com/kupl/PyTER}{[link]} & 2022 & 1 &  a benchmark dedicated to repairing type errors in Python. &  Precision \\

        & xCodeEval~\citep{khan2023xcodeeval}~\href{https://github.com/ntunlp/xCodeEval}{[link]} & 2023 & 11 &  an executable dataset of 450K small buggy/fixed program pairs & Pass Rate\\

        & RunBugRun~\citep{prenner2023runbugrun}~\href{https://github.com/giganticode/run_bug_run}{[link]} & 2023 & 8 &  an executable multilingual benchmark consisting of 25M document-level coding examples & Acc.\\ 

        &  HumanEvalPack~\citep{muennighoff2023octopack}~\href{https://github.com/bigcode-project/octopack}{[link]} & 2023 & 6 & expanding the HumanEval benchmark to 3 coding tasks across 6 languages  & Acc. \\        
         
        & ErrorCLR~\citep{han2023errorclr}~\href{https://dl.acm.org/doi/abs/10.1145/3135932.3135941}{[link]} & 2023 & 2 & a multilingual benchmark of similar buggy programs & Pass Rate \\

        \bottomrule[1.2pt]
    \end{tabular}
    }
    \label{tab:code_tasks_full_p1}
\end{table}
\end{landscape}

\begin{landscape}

\begin{table}[ht]

\footnotesize
    \caption{A more comprehensive collection of code-related benchmarks extended from Table~\ref{tab:code_tasks}. ``\#PLs.'' denotes the number of programming languages each benchmark covers (Cont'd). MRR (Mean Reciprocal Rank) indicates the average rank of correct answer choices. FRank (First Rank) measures the proportion of correct answers ranked first. CR (Compilation Rate) denotes the percentage of code snippets that compile successfully. NIM (Next Identifier Match) is the accuracy in predicting the next identifier or variable name. ISM (Identifier Sequence Match) measures the coherence of generated identifier sequences. PM (Prefix Match) means the alignment of generated code snippets with expected prefixes.}
    \adjustbox{width=\textwidth+5cm,center}{
    \begin{tabular}{l|l|c|c|l|c}
        \toprule[1.2pt]
        \textbf{Task} & \textbf{Dataset} & \textbf{Date} & \textbf{\# PLs} & \textbf{Description}  & \textbf{Eval. Metric} \\
        \midrule
        \multirow{3}{*}{Code Search}
        & CodeSearchNet~\citep{husain2019codesearchnet}~\href{https://github.com/github/CodeSearchNet}{[link]} & 2019 & 6 & a multilingual dataset of 6M functions and query-like natural language  & MRR\\
        & AdvTest~\citep{lu2021codexglue}~\href{https://github.com/microsoft/CodeXGLUE/tree/main/Text-Code/NL-code-search-Adv}{[link]} & 2021 & 1 & a Python code search dataset filtered from CodeSearchNet & MRR\\
        & WebQueryTest~\citep{lu2021codexglue}~\href{https://github.com/microsoft/CodeXGLUE/tree/main/Text-Code/NL-code-search-WebQuery}{[link]} & 2021 & 1 & a test set of Python code search including 1K query-code pairs & Acc. \\
        \midrule
        \multirow{8}{*}{Code Translation}
        & GeeksforGeeks~\citep{roziere2020unsupervised}~\href{https://github.com/wasiahmad/SumGenToBT}{[link]} & 2020 & 3 & a test set composed of 852 parallel functions for code translations & Acc./BLEU \\
        
        & CodeTrans~\citep{lu2021codexglue}~\href{https://github.com/microsoft/CodeXGLUE/tree/main/Code-Code/code-to-code-trans}{[link]} & 2021 & 2 & a C\#/Java code translation dataset collected from several public repos & Acc./BLEU/CodeBLEU \\

        & Avatar~\citep{DBLP:conf/acl/AhmadTCC23}~\href{https://github.com/wasiahmad/AVATAR}{[link]} & 2021 & 2 & a collection of 9.5K programming problems and solutions written in Java and Python & Acc./BLEU/CodeBLEU \\
        
        & CoST~\citep{zhu2022cost}~\href{https://github.com/reddy-lab-code-research/MuST-CoST}{[link]} & 2022 & 7 & a multilingual Code Snippet Translation dataset & BLEU/CodeBLEU \\
        
        & XLCoST~\citep{DBLP:journals/corr/abs-2206-08474}~\href{https://github.com/reddy-lab-code-research/XLCoST}{[link]} & 2022 & 7 &  a benchmark containing fine-grained parallel data from 7 programming languages. & BLEU/CodeBLEU \\

        & xCodeEval~\citep{khan2023xcodeeval}~\href{https://github.com/ntunlp/xCodeEval}{[link]} & 2023 & 11 &  an executable multilingual benchmark consisting of 25M document-level coding examples & Pass Rate\\

        & G-TransEval~\citep{DBLP:conf/kbse/JiaoYLQGS23}~\href{https://github.com/polyeval/g-transeval}{[link]} & 2023 & 5 &  a benchmark suite of 400 code translation pairs between 5 PLs, categorized into 4 levels & Acc./BLEU/CodeBLEU\\
        & CodeTransOcean~\citep{yan2023codetransocean}~\href{https://github.com/WeixiangYAN/CodeTransOcean}{[link]} & 2023 & 45 & a large-scale code translation benchmark with three novel multilingual datasets & EM/BLEU/CodeBLEU \\

        \midrule
        \multirow{10}{*}{Code Retrieval}

        &  StaQC~\citep{Yao2018Staqc}~\href{https://github.com/LittleYUYU/StackOverflow-Question-Code-Dataset}{[link]} & 2018 & 2 & a dataset of around 148K Python and 120K SQL domain question-code pairs & MRR \\

        &   DeepCS~\citep{Gu2018DeepCS}~\href{https://github.com/guxd/deep-code-search}{[link]} & 2018 & 1 & a large scale codebase collected from GitHub & FRank/SuccessRate/ Precision \\

        & CoNaLa~\citep{yin2018mining}~\href{https://conala-corpus.github.io/#dataset-information}{[link]} & 2018 & 1 &  a dataset automatically crawled from Stack Overflow & BLEU \\

        & CodeSearchNet~\citep{husain2019codesearchnet}~\href{https://github.com/github/CodeSearchNet}{[link]} & 2019 & 6 & a multilingual dataset of 6M functions and query-like natural language  & MRR\\

        & CosBench~\citep{Yan2020CosBench}~\href{https://conala-corpus.github.io/#dataset-information}{[link]} & 2020 & 1 & a dataset that consists of 1K projects and 52 code-independent natural-language queries & Precision/Precision/MRR \\

        & SO-DS~\citep{Heymen2020SODS}~\href{https://zenodo.org/records/4001602}{[link]} & 2020 & 1 & a dataset that consists of 12K snippets of Python & MRR/Recall \\

        & FB-Java~\citep{Ling2021FBJava}~\href{https://github.com/ryderling/DGMS}{[link]} & 2020 & 1 & a dataset consisting of 24K repositories with 4.6M functions in Java & MRR/SuccessRate\\

        & AdvTest~\citep{lu2021codexglue}~\href{https://github.com/microsoft/CodeXGLUE/tree/main/Text-Code/NL-code-search-Adv}{[link]} & 2021 & 1 & a Python code search dataset filtered from CodeSearchNet & MRR \\

        & WebQueryTest~\citep{lu2021codexglue}~\href{https://github.com/microsoft/CodeXGLUE/tree/main/Text-Code/NL-code-search-WebQuery}{[link]} & 2021 & 1 & a test set of Python code search of 1K query-code pairs & Acc. \\

        & CoSQA~\citep{Huang2021CosQA}~\href{https://github.com/Jun-jie-Huang/CoCLR}{[link]} & 2021 & 1 & a dataset of web queries for code search and question answering. & Acc. \\
        \midrule
        \multirow{6}{*}{Code Completion}
            
        & GitHub Java Corpus~\citep{allamanis2013mining}~\href{https://groups.inf.ed.ac.uk/cup/javaGithub/}{[link]} & 2013 & 1 &  a giga-token corpus of Java code from a wide variety of domains & EM/Acc./Edit sim \\
        & Py150~\citep{raychev2016py150}~\href{https://www.sri.inf.ethz.ch/py150}{[link]} & 2016 & 1 & a corpus of Python programs from GitHub & EM/Acc./Edit sim \\
        & JS150~\citep{raychev2016py150}~\href{https://dl.acm.org/doi/abs/10.1145/3022671.2984041}{[link]} & 2016 & 1 & a corpus of JavaScript programs from GitHub & EM/Acc./Edit sim \\

        & DotPrompts~\citep{Agrawal2023DotPrompts}~\href{https://github.com/microsoft/monitors4codegenl}{[link]} & 2023 & 1 &  a dataset of real-world open-source Java projects completion with their environments & CR/NIM/ISM/PM\\  
        & LCC~\citep{guo23longcoder}~\href{https://github.com/microsoft/CodeBERT}{[link]} & 2023 & 3 &  a benchmark that focuses on code completion with long code context &  EM/Edit Sim  \\

        & RepoBench~\citep{liu2023repobench}~\href{https://github.com/Leolty/repobench}{[link]} & 2023 & 2 &  a benchmark tailored for evaluating repository-level code autocompletion systems &  EM/Edit Sim  \\

        \midrule
        \multirow{14}{*}{GitHub} 
        & unamed~\citep{Jiang_McMillan2017}~\href{https://notredame.app.box.com/s/wghwpw46x41nu6iulm6qi8j42finuxni}{[link]} & 2017 & 1 & a dataset of 509K labeled diff files of Java &  Acc.  \\
        
        & CommitGen~\citep{loyola2017commitgen}~\href{http://github.com/epochx/commitgen}{[link]} & 2017 & 4 & a multilingual dataset collected from open source projects on Github  & BLEU \\

        & NNGen~\citep{Liu2018NNGen}~\href{https://github.com/Tbabm/nngen}{[link]} & 2018 & 1 & a cleaner version of CommitGen  & BLEU \\

        & PtrGNCMsg~\citep{Liu2019PtrGNCMsg}~\href{https://zenodo.org/records/2542706#.XECK8C277BJ}{[link]} & 2019 & 1 & a dataset of diffs and manual commit messages from Java projects in GitHub & BLEU/ROUGE \\

        & CoDiSum~\citep{Xu2019CoDiSum}~\href{https://github.com/SoftWiser-group/CoDiSum}{[link]} & 2019 & 1 & a cleaner version of \citet{Jiang_McMillan2017}'s dataset & BLEU/METEOR\\

        & ATOM~\citep{Liu2022ATOM}~\href{https://zenodo.org/records/4066398}{[link]} & 2019 & 1 & a dataset dataset crawled from 56 popular Java repositories & BLEU  \\       
        & CommitBERT~\citep{jung2021commitbert}~\href{https://github.com/graykode/commit-autosuggestions}{[link]} & 2021 & 6 & a multilingual dataset of code modification with commit messages in Github & BLEU \\

        & MCMD~\citep{tao2021cmgmodel}~\href{https://github.com/DeepSoftwareAnalytics/CommitMsgEmpirical/tree/main/dataset}{[link]} & 2021 & 5 & a large-scale, information-rich, and multi-language commit message dataset & B-Moses/B-Norm/B-CC \\

        & CoRec~\citep{Wang2021CoRec}~\href{https://zenodo.org/records/3828107}{[link]} & 2021 & 1 & a large-scale dataset crawled from 10K popular Java repositories in Github  & BLEU \\

        & ExGroFi~\citep{Wang2023ExGroFi}~\href{https://zenodo.org/records/3828107}{[link]} & 2023 & 1 & a dataset anchored on combining correlated commits and issues & BLEU/ROUGE/CIDEr \\

        & CommitChronicle~\citep{Eliseeva2023CommitChronicle}~\href{https://github.com/JetBrains-Research/commit_message_generation}{[link]} & 2023 & 20 & a dataset containing 10.7M commits across 20 programming languages & B-Norm/Edit Sim/EM \\

        & SWE-bench~\citep{jimenez2023swebench}~\href{https://www.swebench.com/}{[link]} & 2023 & 1 & a dataset of 2.2K software engineering problems with pull requests & Resolve Rate/Recall \\ 
        
        & CommitBench~\citep{schall2024commitbench}~\href{https://zenodo.org/records/10497442}{[link]} & 2024 & 6 & a reproducible and privacy- and license-aware benchmark for commit message generation & BLEU/METEOR/ROUGE \\ 
        & DevBench~\citep{li2024devbench}~\href{https://github.com/open-compass/DevBench}{[link]} & 2024 & 4 &  a benchmark to evaluate LLMs across various stages of the software development lifecycle &  pass@k  \\
        
        \bottomrule[1.2pt]
    \end{tabular}
    }
    \label{tab:code_tasks_full_p2}
\end{table}
\end{landscape}

\begin{landscape}

\begin{table}[ht]
    \caption{A more comprehensive collection of code-related benchmarks extended from Table~\ref{tab:code_tasks}. ``\# PLs'' denotes the number of programming languages each benchmark covers (Cont'd).}
    \adjustbox{width=\textwidth+5cm,center}{  
    \begin{tabular}{l|l|c|c|l|c}
        \toprule[1.2pt]
        \textbf{Task} & \textbf{Dataset} & \textbf{Date} & \textbf{\# PLs} & \textbf{Description} & \textbf{Eval. Metric}   \\     
       
        \midrule
        \multirow{5}{*}{Code Summarization} 
        & CODE-NN~\citep{iyer2016summarizing}~\href{https://github.com/sriniiyer/codenn}{[link]} & 2016 & 6 & (title, query) pairs from StackOverflow & BLEU \\

        & DeepCom~\citep{hu2018deep}~\href{https://github.com/xing-hu/DeepCom}{[link]} & 2018 & 1 & a large-scale Java corpus built from 9.7K open source projects from GitHub & BLEU \\
        
        & TL-CodeSum~\citep{ijcai2018TLCodeSum}~\href{https://github.com/xing-hu/TL-CodeSum}{[link]} & 2018 & 6 & a dataset of 69K pairs of code and summary  & BLEU\\
        & CodeSearchNet~\citep{husain2019codesearchnet}~\href{https://github.com/github/CodeSearchNet}{[link]} & 2019 & 6 & a multilingual dataset of 6M functions and query-like natural language & MRR \\
        &  HumanEvalPack~\citep{muennighoff2023octopack}~\href{https://github.com/bigcode-project/octopack}{[link]} & 2023 & 6 & a benchmark expanding HumanEval to 3 coding tasks across 6 languages & BLEU \\
        \midrule
        \multirow{3}{*}{Code Debug} 
       
         & QuixBugs~\citep{ye2021quixbugs}~\href{https://github.com/jkoppel/QuixBugs}{[link]} & 2017 & 2 & a multi-language benchmark (40 bugs in both Python and Java) & Pass Rate \\
         & EvalGPTFix~\citep{zhang2023critical}~\href{https://github.com/hitz-zentroa/lm-contamination}{[link]} & 2023 & 1 & a benchmark containing 151 pairs of bugs and fixes in Java & Pass Rate \\
         &DebugBench~\citep{tian2024debugbench}~\href{https://github.com/thunlp/DebugBench}{[link]} & 2024 & 3 & a debugging benchmark containing 4K instances of four major bug categories &  Pass Rate \\

        \midrule
        \multirow{3}{*}{Question Answering} 

         & CodeQA~\citep{liu2021codeqa} & 2021 & 2 & a free-form QA dataset for Java/Python code comprehension   & BLEU/EM/F1\\
         & CodeQueries~\citep{sahu2024codequeries} & 2024 & 1 & a dataset to evaluate models to answer semantic queries over python code &  EM/Acc./Recall\\
         & InfiCoder-Eval~\citep{li2024inficodereval} & 2024 & 1 & a large-scale freeform QA benchmark for code collected from Stack Overflow &  Customized  \\

        \midrule
        \multirow{1}{*}{Code Editing} 
         &  EditEval~\citep{hu2023instructcoder} & 2023 & 1 &  a human-written execution-based benchmark of 194 code editing tasks & Acc. \\

        \midrule
        \multirow{8}{*}{Text2SQL} 
         &ATIS~\citep{dahl1994expand}~\href{https://github.com/howl-anderson/ATIS_dataset}{[link]} & 1994 & 1 & a dataset containing 3K utterances reserved for testing  &  Acc. \\ 
         &WikiTQ~\citep{panupong2015comp}~\href{https://github.com/salesforce/WikiSQL}{[link]} & 2015 & 1 & a dataset of 22K complex questions on Wikipedia tables &  Acc.  \\         &WikiSQL~\citep{zhong2017seq2sql}~\href{https://github.com/salesforce/WikiSQL}{[link]} & 2017 & 1 & a dataset of 80K hand-annotated questions and SQL queries from Wikipedia  &  Acc./F1 \\          
         & Spider~\citep{yu2018spider}~\href{https://github.com/taoyds/spider}{[link]} & 2018 & 1 & a dataset consisting of 10K questions and 5.6K unique complex SQL queries &  Acc./F1  \\ 
         & SParC~\citep{yu2019sparc}~\href{https://github.com/taoyds/sparc?tab=readme-ov-file#sparc-cross-domain-semantic-parsing-in-context}{[link]} & 2019 & 1 & a dataset consisting of 4.2K coherent question sequences  & Acc. \\ 
         &Cosql~\citep{yu2019cosql}~\href{https://github.com/taoyds/sparc?tab=readme-ov-file#sparc-cross-domain-semantic-parsing-in-context}{[link]} & 2019 & 1 & a dataset consisting  of 30K+ turns and 10K+ annotated SQL queries &  Acc.\\ 
         &Squall~\citep{DBLP:conf/emnlp/ShiZBDL20}~\href{https://github.com/tzshi/squall}{[link]} & 2020 & 1 & a dataset of 11K English questions with manually created SQL equivalents &  Acc.\\
         &KaggleDBQA~\citep{DBLP:conf/acl/LeePR20}~\href{https://aka.ms/KaggleDBQA}{[link]} & 2021 & 1 &  a cross-domain evaluation dataset of real Web databases &  Acc. \\

        \bottomrule[1.2pt]
    \end{tabular}
    }
    \label{tab:code_tasks_full_p3}
\end{table}
\end{landscape}

\begin{landscape}

\begin{table}[ht]
    \centering
    \footnotesize
    \caption{More NL2Code benchmarks extended from Table~\ref{tab:nl2code_tasks}. FGA indicates the F1-score of group accuracy.}
    \adjustbox{width=\textwidth+6.5cm,center}{
    \begin{tabular}{l|l|c|c|l|c}
        \toprule[1.2pt]
        \textbf{Purpose} & \textbf{Dataset} & \textbf{Date} & \textbf{\# PLs.} & \textbf{Description} & \textbf{Eval. Metric} \\
        \midrule
        \multirow{2}{*}{Open Domain} 
        & CONCODE~\citep{iyer2018mapping}~\href{https://github.com/sriniiyer/concode}{[link]} & 2018 & 1 & a dataset with 100K examples consisting of Java classes from online code repositories & EM/BLEU \\
        & ODEX~\citep{wang2023odex}~\href{https://github.com/zorazrw/odex}{[link]} & 2023 & 1 & an open-domain execution-based natural language to Python code generation dataset & pass@k \\
        \midrule            
        \multirow{10}{*}{Code Exercise} 
        & HumanEval~\citep{chen2021evaluating}~\href{https://github.com/openai/human-eval}{[link]} & 2021 & 1 & a dataset of 164 handwritten programming problems with unit tests & pass@k \\
        & MBPP~\citep{austin2021program}~\href{https://github.com/google-research/google-research/tree/master/mbpp}{[link]} & 2021 & 1 & a dataset containing 974 short Python programs & pass@k \\
        & MathQA-Python~\citep{austin2021program}~\href{https://arxiv.org/abs/2108.07732}{[link]} & 2021 & 1 & a Python version of the MathQA benchmark contains 23K problems & Acc. \\   
        & AixBench~\citep{Hao2022AixBench}~\href{https://github.com/aixcoder-plugin/nl2code-dataset}{[link]} & 2022 & 1 & a code generation benchmark dataset & Pass Rate \\ 
        & BIG-Bench~\citep{srivastava2023bb,suzgun2023bbh}~\href{https://github.com/google/BIG-bench}{[link]} & 2023 & - & a benchmark containing over 12 tasks can be solved by coding & -  \\ 
        & CoderEval~\citep{Yu2023CoderEval}~\href{https://github.com/CoderEval/CoderEval}{[link]} & 2023 & 2 & a dataset consisting of 230 Python and 230 Java code generation tasks & Pass Rate \\ 
        &  CodeApex~\citep{fu2023codeapex}~\href{https://github.com/APEXLAB/CodeApex}{[link]} & 2023 & 1 &  a bilingual programming evaluation benchmark for large language models & Acc. \\
        & ClassEval~\citep{du2023classeval}~\href{https://github.com/FudanSELab/ClassEval}{[link]} & 2023 & 1 &  a handcrafted benchmark for evaluating class-level code generation & pass@k \\
        & OOP~\citep{wang2024oop}~\href{https://github.com/alphadl/OOP-eval}{[link]} & 2024 & 1 &  an OOP-focused benchmark, with 431 Python programs that encompass OOP concepts like classes & pass@k \\
        & NCB~\citep{zhang2024naturalcodebench}~\href{https://github.com/THUDM/NaturalCodeBench}{[link]} & 2024 & 2 &  402 high-quality python and java problems, selected from natural user queries from coding services & pass@k \\
&MHPP~\citep{dai2024mhpp}~\href{https://github.com/SparksofAGI/MHPP}{[link]} & 2024 & 1 & a dataset of 140 Python problems designed to assess LLMs' advanced code generation abilities. & pass@k \\

        \midrule 
        \multirow{3}{*}{Competitions}
        & APPS~\citep{hendrycks2021apps}~\href{https://github.com/hendrycks/apps}{[link]} & 2021 & 1 & a benchmark including 10K problems for Python code generation  & Acc. \\
        & CodeContests~\citep{li2022alphacode}~\href{https://github.com/deepmind/code_contests}{[link]} & 2022 & 3 & a dataset of competitive programming problems & pass@k \\
        & TACO~\citep{li2023taco}~\href{https://github.com/FlagOpen/TACO}{[link]} & 2023 & 1 &a dataset on algorithmic code generation, consisting of 25K training problems and 1K testing problem& pass@k \\
        \midrule
        \multirow{6}{*}{Multilingual}
        & MultiPL-E~\citep{cassano2022multiple}~\href{https://github.com/nuprl/MultiPL-E}{[link]} & 2022 & 18 & a parallel, multilanguage benchmark for natural-language-to-code generation & pass@k \\   
        & MCoNaLa~\citep{Wang2023Mconala}~\href{https://github.com/zorazrw/multilingual-conala}{[link]} & 2022 & 1 & a benchmark for code generation from multiple natural languages  & BLEU\\ 
        & MBXP~\citep{athiwaratkun2023mbxp}~\href{https://github.com/amazon-research/mbxp-exec-eval}{[link]} & 2023 & 12 & a benchmark to evaluate code generation models in over 10 programming languages & pass@k \\
        & MathQA-X~\citep{athiwaratkun2023mbxp}~\href{https://github.com/amazon-research/mxeval}{[link]} & 2023 & 3 & a dataset transpiled from the original Python dataset MathQA  & Pass Rate/Acc. \\ 
        & xCodeEval~\citep{khan2023xcodeeval}~\href{https://github.com/ntunlp/xCodeEval}{[link]} & 2023 & 11 &  an executable dataset of 450K small buggy/fixed program pairs & Pass Rate \\    
        & HumanEval-X~\citep{zheng2023codegeex}~\href{https://github.com/THUDM/CodeGeeX}{[link]} & 2023 & 4 & a benchmark of 164 code problems for evaluating multilingual models &  pass@k \\
        \midrule
    
        \multirow{6}{*}{Data Science} 
        & JuICe~\citep{agashe2019juice}~\href{https://github.com/rajasagashe/juice?tab=readme-ov-file}{[link]} & 2018 & 1 & a corpus of 1.5M examples with a curated test set of 3.7K instances & EM/BLEU \\
        & PlotCoder~\citep{chen-2021-plotcoder}~\href{https://github.com/Jungyhuk/plotcoder}{[link]} & 2021 & 1 & a dataset of plot samples extracted from the original dev and test splits of JuICe & Acc. \\ 
        & DSP~\citep{chandel2022jupyt5}~\href{https://github.com/microsoft/DataScienceProblems}{[link]} & 2022 & 1 &  a collection of 1.1K problems curated from 306 pedagogical notebooks & pass@k \\
        & ExeDS~\citep{huang2022exeds}~\href{https://github.com/Jun-jie-Huang/ExeDS}{[link]} & 2022 & 1 & a dataset of 534 problems for execution evaluation for data science code generation tasks & (Code)BLEU/EM  \\
        & DS-1000~\citep{lai2022ds1000}~\href{https://ds1000-code-gen.github.io/}{[link]} & 2023 & 1 & a code generation benchmark with 1K data science problems spanning 7 Python libraries & pass@k  \\
        & DAEval~\citep{hu2024infiagentdabench}~\href{https://github.com/InfiAgent/InfiAgent}{[link]} & 2024 & 1 & a dataset comprising 257 data analysis questions derived from 52 CSV files  & Pass Rate \\ 
        \midrule
        \multirow{3}{*}{Python Libs} 
        & PandasEval~\citep{zan2022cert}~\href{https://github.com/microsoft/PyCodeGPT}{[link]} & 2022 & 1 & a dataset consisting of 101 programming problems on Pandas library & pass@k \\
        & NumpyEval~\citep{zan2022cert}~\href{https://github.com/microsoft/PyCodeGPT}{[link]} & 2022 & 1 & a dataset consisting of 101 programming problems on Numpy library  & pass@k \\
        & TorchDataEval~\citep{zan2022private}~\href{https://github.com/microsoft/PyCodeGPT/tree/main/apicoder}{[link]} & 2022 & 1 & a dataset with 50 programming problems using the TorchData library & pass@k \\
        \midrule
        \multirow{1}{*}{Multi-Turn} 
        & MTPB~\citep{nijkamp2022codegen}~\href{https://github.com/salesforce/CodeGen}{[link]} & 2023 & 1 & an open benchmark consisting of 115 diverse problem sets that are factorized into multi-turn prompts & Pass Rate \\
        \midrule
        \multirow{1}{*}{Command Line} 
        & NL2Bash~\citep{lin2018nl2bash}~\href{https://aclanthology.org/L18-1491/}{[link]} & 2018 & 1 & a corpus of 9K English-command pairs, covering over 100 unique Bash utilities &  Acc. \\
        \midrule
        \multirow{2}{*}{AI4Science} 
        & BioCoder~\citep{tang2023biocoder}~\href{https://biocoder-benchmark.github.io/}{[link]} & 2023 & 2 & a benchmark developed to evaluate large language models in generating bioinformatics-specific code & pass@k \\
        & SciCode~\citep{tian2024scicode} & 2024 & 1 & a scientist-curated benchmark containing 80 real scientific coding problems & pass@k \\
        \midrule
        \multirow{2}{*}{Education} 
        & StudentEval~\citep{Babe2023StudentEval}~\href{https://huggingface.co/datasets/wellesley-easel/StudentEval}{[link]} & 2023 & 1 & a benchmark containing 1.7K prompts for 48 problems, written by 80 students &  pass@1 \\
        & COJ2022~\citep{han2023errorclr}~\href{https://github.com/DaSESmartEdu/ErrorCLR.}{[link]} & 2023 & 1 & a dataset containing 5.9K C programs with semantic errors submitted to 498 different assignments & F1  \\
        \midrule
        \multirow{2}{*}{Log Parsing} 
        & LogHub\textsuperscript{1}~\citep{Zhu2019LogHUb}~\href{https://github.com/logpai/logparser/tree/main/data#loghub_2k}{[link]} & 2019 & - &  a dataset of logs produced by 16 different systems & Acc.  \\
        & LogHub\textsuperscript{2}\citep{Jiang2023LogHub}~\href{https://github.com/logpai/LogPub}{[link]} & 2023 & - &  a suite of 14 datasets, each averaging 3.6M log lines  & FGA \\

        
        \midrule
        \multirow{1}{*}{Hardware Design}
        &  VerilogEval~\citep{Liu2023Verilogeval}~\href{https://github.com/NVlabs/verilog-eval}{[link]} & 2023 & 1 &  a dataset consists of 156 problems from the Verilog instructional website HDLBits & BLEU/pass@k \\

        \bottomrule[1.2pt]
    \end{tabular}
    }
    \label{tab:nl2code_full}
\end{table}
\end{landscape}

%% file: tables/codeptm-objects.tex
\begin{landscape}

\begin{table}[ht]
    \centering
    \caption{Detailed Pre-training Objectives of CodePTMs listed in Table~\ref{table:CodePTMs}. 
    }
    \adjustbox{width=\textwidth+5cm,center}{
    \begin{tabular}{l|l|l}
        \toprule[1.2pt]
        \textbf{Model}
        & \textbf{Abbr.}  & \textbf{Description} \\
        \midrule
        \multirow{2}{*}{CuBERT~\citep{kanade2020learning}}
        & MLM & \textbf{M}asked \textbf{L}anguage \textbf{M}odeling \\
        & NSP & Given two sentences, \textbf{P}redict whether one sentence is the \textbf{N}ext \textbf{S}entence of the other~\citep{devlin2018bert}. \\ 
        \midrule 
        \multirow{2}{*}{CodeBERT~\citep{feng2020codebert}}
        & MLM & \textbf{M}asked \textbf{L}anguage \textbf{M}odeling\\
        & RTD & \textbf{D}etect the \textbf{T}okens that are randomly \textbf{R}eplaced in a sentence~\citep{clark2020electra}.  \\
        \midrule
        \multirow{3}{*}{GraphCodeBERT~\citep{guo2021graphcodebert}}
        & MLM & \textbf{M}asked \textbf{L}anguage \textbf{M}odeling\\
        & Edge Pred. & (Edge Prediction) Mask direct edges that connect sampled nodes in data flow, and predict these masked edges.\\
        & Node Align. & (Node Alignment) Mask edges between a variable in data flow and code tokens, and predict which token the variable in data flow is identified from.\\
        \midrule
        \multirow{4}{*}{SynCoBERT~\citep{wang2022syncobert}}
        & MMLM & \textbf{M}ulti-Modal \textbf{M}asked \textbf{L}anguage \textbf{M}odeling which jointly models NL, PL, and AST (CFG).\\
        & IP & \textbf{P}redict the type of all code tokens as either \textbf{I}dentifier or non-identifier.\\
        & TEP & Mask edges in the AS\textbf{T} and ask the model to \textbf{P}redict these \textbf{E}dges.\\
        & MCL & \textbf{M}ulti-Modal \textbf{C}ontrastive \textbf{L}earning \\
        
        \midrule
        \multirow{3}{*}{CODE-MVP~\citep{wang2022codemvp}}
        & MCL & \textbf{M}ulti-Modal \textbf{C}ontrastive \textbf{L}earning \\
        & FGTI & Traverse the AST and use the type checker to \textbf{I}nfer \textbf{F}ine-\textbf{G}rained identifier \textbf{T}ypes.\\
        & MMLM & \textbf{M}ulti-Modal \textbf{M}asked \textbf{L}anguage \textbf{M}odeling \\
        \midrule
        \multirow{1}{*}{SCodeR~\citep{li2022soft}}
        & - & Soft-Labeled Contrastive Pre-training that uses relevance scores between different samples as softlabels to learn function-level code representation. \\
        \midrule
        \multirow{3}{*}{DISCO~\citep{ding2022disco}}
        & MLM & \textbf{M}asked \textbf{L}anguage \textbf{M}odeling \\
        & NT-MLM & Local AST \textbf{N}ode \textbf{T}ype-MLM which learns to recover the masked AST type. \\
        & CLR & \textbf{C}ontrastive \textbf{L}earning \\
        \midrule
        \multirow{3}{*}{PLBART~\citep{ahmad2021unified}}
        & - & Random tokens are sampled and replaced with a mask token from the input sequence. \\
        & - & Random tokens are deleted from the input sequence. \\
        & - & A number of text spans are sampled and replaced with a single mask token. \\
        \midrule
        \multirow{4}{*}{CodeT5~\citep{wang2021codet5}}
        & MSP & Randomly mask spans with arbitrary lengths and \textbf{P}redict these \textbf{M}asked \textbf{S}pans combined with some sentinel tokens at the decoder. \\
        & IP & \textbf{P}redict the type of all code tokens as either \textbf{I}dentifier or non-identifier.\\
        & MIP & \textbf{M}ask all \textbf{I}dentifiers in the PL segment and \textbf{P}redict them.\\
        & - & Bimodal Dual Generation\\
        \midrule
        \multirow{1}{*}{PyMT5~\citep{clement2020PyMT5}}
        & MSP & Randomly mask spans with arbitrary lengths and \textbf{P}redict these \textbf{M}asked \textbf{S}pans combined with some sentinel tokens at the decoder. \\
        \midrule
        \multirow{5}{*}{UniXcoder~\citep{guo2022unixcoder}}
        & MLM & \textbf{M}asked \textbf{L}anguage \textbf{M}odeling \\
        & ULM & \textbf{U}nidirectional \textbf{L}anguage \textbf{M}odeling\\
        & MSP & Randomly mask spans with arbitrary lengths and \textbf{P}redict these \textbf{M}asked \textbf{S}pans combined with some sentinel tokens at the decoder. \\
        & MCL & \textbf{M}ulti-modal \textbf{C}ontrastive \textbf{L}earning\\
        & CMG & (\textbf{C}ross-\textbf{M}odal \textbf{G}eneration) Generate code comment which describes the function of the code.\\
        \midrule
        \multirow{1}{*}{NatGen~\citep{chakraborty2022natgen}}
        & - &  Naturalizing pre-training which forces the model to generate correct semantically equivalent natural code that is just what a human originally wrote. \\ 
        \midrule
        \multirow{2}{*}{TreeBERT~\citep{jiang2021treebert}}
        & TMLM & \textbf{T}ree \textbf{M}asked \textbf{L}anguage \textbf{M}odeling for masking the nodes in the AST and the tokens in the code snippet.\\
        & NOP & Randomly exchange the positions of some nodes in the AST path, and \textbf{P}redict whether the \textbf{O}rder of \textbf{N}odes in the AST is correct.\\
        \midrule
        \multirow{2}{*}{ERNIE-Code~\citep{chai2023erniecode} }
        & SCLM & (\textbf{S}pan-\textbf{C}orruption \textbf{L}anguage \textbf{M}odeling) Corrupts 15\% of the original NL/PL input tokens and predict the corrupted span on the target side.\\
        & PTLM & (\textbf{P}ivot-based \textbf{T}ranslation \textbf{L}anguage \textbf{M}odeling) Concatenate parallel source-target sentences and learn to predict the corrupted target language.\\
        \midrule
        \multirow{2}{*}{CodeExecutor~\citep{liu2023codeexecutor}}
        & - & Code Execution which improves the model ability to understand and execute code. \\
        & - & Curriculum Learning which starts from easy instances and then gradually handles harder ones. \\
        \midrule
        \multirow{1}{*}{LongCoder~\citep{guo23longcoder}}
        & CLM & \textbf{C}ausal \textbf{L}anguage \textbf{M}odeling \\
        \midrule
        \multirow{3}{*}{\makecell[l]{GPT-C~\citep{Svyatkovskiy2020gptc} \\ CodeGPT~\citep{lu2021codexglue} \\ PyCodeGPT~\citep{zan2022cert} }}
        & \multirow{3}{*}{CLM} & \multirow{3}{*}{\textbf{C}ausal \textbf{L}anguage \textbf{M}odeling} \\
        & & \\
        & & \\
        \bottomrule[1.2pt]
    \end{tabular}
    }
    \label{tab:codeptm_objs}
\end{table}

\end{landscape}